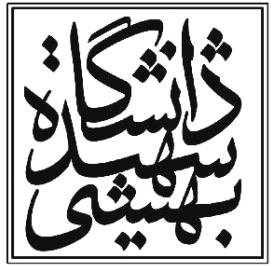

# دانشگاه شهید بهشتی

**دانشکده علوم**
**گروه فیزیک**

پایان نامه کارشناسی ارشد
گرانش

**موضوع پایان نامه**
## جرم ویریال در کیهان شناسی DGP

**استاد راهنما**
حمیدرضا سپنجی

**استاد مشاور**
مهرداد فرهودی

**نگارنده**
سید شهاب الدین شهیدی شادکام

تابستان ۱۳۸۸

## چکیده

در این پایان نامه ما در مورد مسئله جرم ویریال در مدل DGP صحبت خواهیم کرد. مدل DGP یک نوع مدل جهان شامه ای است که در فواصل خیلی بیشتر از مقیاس فاصله $r_0$، که بصورت نسبت مقیاس پلانک در شامه به مقیاس پلانک در توده تعریف می شود، تصحیحاتی در گرانش معمول اینشتینی ایجاد می کند. بعد اضافه در این مدل غیر فشرده و نامتناهی است. کنش این مدل شامل کنش اینشتین- هیلبرت در پنج بعد و یک جمله القا شده چهار بعدی است که تضمین می کند گرانش در روی شامه وجود داشته باشد. مهمترین کاربرد این مدل این است که می تواند شتاب تند شونده جهان در زمان حال را بدون استفاده از هر نوع انرژی تاریک توضیح دهد. جرم ویریال یکی از مسائل مهم در کیهان شناسی است که بوسیله ماده تاریک می توان آن را توضیح داد. این مسئله از آنجا ناشی می شود که دو راه معمولی که برای اندازه گیری جرم یک خوشه کهکشانی استفاده می شود به جواب یکسانی ختم نمی شوند. یکی از این دو راه اندازه گیری جرم تک تک کهکشان ها و جمع کردن آنها با هم، و دیگری استفاده از قضیه ویریال است. تقریبا در تمام خوشه های کهکشانی جرم ویریال ۲۰ تا ۳۰ برابر بیشتر از جرم مشاهده شده در خوشه کهکشانی است. در این پایان نامه نشان خواهیم داد که مسئله جرم ویریال را می توان بوسیله مدل DGP توضیح داد. این کار با تعریف یک جمله جرمی جدید که نماینده بعد اضافه در مدل DGP است انجام می پذیرد.

**کلمات کلیدی**: بعد اضافه، جرم ویریال، کیهان شناسی DGP

# فهرست

# فهرست شکل ها

# فهرست جداول

# فصل ۱

# مقدمه

ثابت کیهان شناسی یکی از مهمترین و سخت ترین مسائل در فیزیک ذرات و کیهان شناسی محسوب می شود [۲۶]. مسئله اولیه کوچک بودن ثابت کیهان شناسی نسبت به مقیاس بنیادی فیزیک ذرات بوده است اما با کشف اینکه جهان با شتاب تند شونده منبسط می شود، مسئله قدری پیچیده تر شد [۲۷]. گرانش معمول اینشتینی در چهار بعد بدون ثابت کیهان شناسی پیش بینی می کند که شتاب جهان کند شونده باشد (مقدمه فصل ۳). مسئله انبساط تندشونده جهان را می توان با یک ثابت کیهان شناسی خیلی کوچک حل کرد. برای حل این مشکل تلاش های بسیاری انجام شده است که مهمترین آنها معرفی انرژی تاریک و مدل های گرانش تصحیح شده است. تعریف غالب برای مسئله انبساط تندشونده جهان این است که تصحیحاتی روی گرانش معمول اینشتینی اعمال کنیم که بتواند شتاب تندشونده جهان را بدون استفاده از ثابت کیهان شناسی در فواصل زیاد و در زمان حال[1] نشان دهد. یکی از معروف ترین مثال ها برای حل این مسئله مدل DGP است که در این پایان نامه به آن خواهیم پرداخت. در مدل های انرژی تاریک که ثابت کیهان شناسی نیز یکی از کاندیداهای آن است، قانون گرانش همان قانون گرانش اینشتینی است اما محتوای انرژی ـ تکانه سیستم را با معرفی یک میدان یا ذره جدید که بطور ضعیف با متریک فضا جفت شده است اصلاح می کنیم.

در مدل های گرانش تصحیح شده فرض می کنیم که در فواصل کیهانی قانون گرانش حاکم بر جهان با قانون گرانش معمول اینشتینی تفاوت دارد. یکی از مهمترین شاخه های این نوع مدل ها نظریه های جهان شامه ای است. این نظریه ها بر این پایه استوار هستند که جهان دارای ابعادی بیش از چهار است (که آن را جهان توده می نامیم) و جهان چهار بعدی ما (که آن را جهان شامه می نامیم) به صورت مرزی برای جهان توده است. البته در اولین تلاش ها برای ایجاد این نوع مدل ها فرض نمی کرده اند که

---
[1] Late time

۱

شامه یک ابرسطح برای توده است. اولین مدل جهان شامه ای در سال ۱۹۹۸ برای حل مسئله سلسله مراتب[2] توسط نیما ارکانی ـ حامد، S. Dimopoulos و G. Dvali ارائه شد [۷] که در آن پیشنهاد شده بود که بعد فضای توده حداقل ۶ باشد. مسئله سلسله مراتب از این حقیقت ناشی می شود که مقیاس انرژی پلانک بسیار بیشتر از مقیاس انرژی ضعیف است و این باعث می شود که قدرت نیروی گرانشی خیلی کمتر از سایر نیروها باشد. شش بعدی بودن فضای توده به همراه فشرده بودن ابعاد اضافه به ما این امکان را می دهد که با انتخاب اندازه بعد اضافه درحد میلیمتر بتوانیم مقیاس پلانک در شش بعد را هم مرتبه با مقیاس ضعیف قرار دهیم. اینکه اندازه بعد اضافه در حد میلیمتر است یعنی باید بتوانیم اثرات ابعاد اضافه را در برهم کنش های گرانشی در این فواصل مشاهده کنیم، و این کار به انرژی نیاز دارد که بزودی در LHC [3] قابل دسترس خواهد بود.

این که دو بعد اضافه در این مدل وجود داشته باشد و اینکه این ابعاد اضافه فشرده نیز باشند ناراحتی هایی را برای یک فیزیکدان ایجاد می کند. برای حل مسئله سلسله مراتب فقط با استفاده از یک بعد اضافه به دو شامه همزمان احتیاج داریم [۸] که یکی به عنوان جهان چهار بعدی ما و با تنش مثبت، و دیگری با تنش منفی باشد. در این حالت توده باید حول شامه دارای تقارن $Z_2$ بوده و هندسه آن باید از نوع AdS[4] باشد تا هم محبوس بودن ذرات مدل استاندارد را بر روی شامه و هم جایگزیده بودن گرویتون در اطراف شامه را تضمین کند. اگر شامه با تنش منفی را به بینهایت میل دهیم [۸] به یک مدل با بعد اضافه غیر فشرده می رسیم که در حد فواصل زیاد به گرانش معمول اینشتینی کاهش می یابد. تصحیحات این مدل (که آن را مدل RS II می نامند) بر گرانش اینشتینی در فواصل بسیار کم اتفاق می افتد و لذا نمی تواند شرایطی برای حل انبساط تند شونده جهان در فواصل کیهانی ارائه دهد. در پی یافتن مدل های شامه ای که بتواند تصحیحات گرانشی در فواصل کیهانی ایجاد کند مدل های متنوعی ارائه شده است که می توان به مدل های گرانش جرم دار [۲۸ و ۲۹]، مدل GRS، مدل DGP و اخیراً مدل ghost condensation اشاره کرد. مدل GRS در واقع اصلاح مدل RS با اضافه کردن یک شامه دیگر است [۱۸]. شامه سوم که در طرف دیگر شامه با تنش مثبت قرار گرفته است و تنش منفی دارد باعث می شود که تصحیحات گرانشی هم در فواصل خیلی کم و هم در فواصل خیلی زیاد اتفاق بیافتد. در فواصل میانی گرانش معمول اینشتینی قابل بازیابی خواهد بود. این مدل دارای ناپایداری ghost است به این معنی که مدهایی در نظریه وجود دارند که جمله انرژی جنبشی برای آنها با علامت مخالف

---

[2] Hierarchy problem
[3] Large Hadron Collider
[4] Anti de Sitter

۲

ظاهر می شود. و این باعث ناپایداری خلاء می شود که خلق ذره ghost را همراه با ماده معمولی ممکن می سازد. با اینکه بعد اضافه در این مدل نامتناهی است اما وجود سه شامه در مدل باعث شد که این مدل مقبولیت چندانی بدست نیاورد.

به دنبال یافتن نظریه ای شامه با یک بعد اضافه که در آن نا متناهی باشد و بتواند تصحیحات گرانشی در فواصل کیهانی ایجاد کند، G. Dvali و G. Gabadadze و M. Porrati در سال ۲۰۰۰ مدلی ارائه دادند که در آن علاوه بر احراز شرایط بالا لازم نیست هندسه توده از نوع AdS باشد. نبودن هندسه توده باعث می شود که گرویتون در اطراف شامه جایگزیده نشود. برای حل این مشکل در این مدل گرانش را بصورت دستی در چهار بعد وارد نظریه می کنیم. این کار با اضافه کردن کنش چهار بعدی اینشتین ـ هیلبرت در چهار بعد به کنش پنج بعدی مدل انجام می شود. تغییرات در قانون گرانش در این مدل با معرفی یک مقیاس فاصله $r_0$ انجام می شود بطوریکه برای $r \ll r_0$ گرانش به طور مؤثری چهار بعدی است اما در حد $r \gg r_0$ قانون گرانش پنج بعدی می شود. در فصل ۳ خواهیم دید که چهار بعدی بودن مدل در حد $r \ll r_0$ به معنی بازیابی گرانش اینشتین در این حد نیست. در واقع پارامتر جدیدی در مدل ارائه خواهیم داد که در حد $r \ll r_0 \ll r_\star$ گرانش بصورت مدل برنز- دیکی در چهار بعد با پارامتر $\omega = 0$ در می آید و در حد $r \ll r_\star$ گرانش معمول اینشتینی قابل بازیابی است. البته برای اینکه مدل ارائه شده با مشاهدات کیهان شناسی همخوانی داشته باشد مجبوریم که مقیاس $r_0$ را در حد شعاع هابل کنونی جهان در نظر بگیریم، اما با این مقدار زیاد $r_0$ نیز می توان تصحیحاتی در قانون گرانش حتی در ابعاد منظومه شمسی انتظار داشت. این مدل نیز دارای ناپایداری ghost می باشد اما شاید مهمترین نتیجه این مدل این است که می تواند انبساط تند شونده جهان را در زمان حال و بدون در نظر گرفتن ثابت کیهان شناسی توضیح دهد. در مطالعاتی که در مورد کیهان شناسی این مدل انجام شده [۱۱] این نتیجه بدست آمده است که معادله فریدمن در این مدل دو شاخه دارد که با اضافه شدن $\pm \dfrac{H}{r_0}$ در معادله ظاهر می شود. شاخه مربوط به علامت مثبت همان شاخه کیهان شناسی معمولی یعنی جهان FRW است، اما شاخه مربوط به علامت منفی شاخه ای است که شتاب تندشونده جهان را تضمین می کند. در واقع در حد $t \to \infty$ ، پارامتر هابل $H$ به عدد ثابت $\dfrac{1}{r_0}$ میل می کند. این که توانسته ایم شتاب تندشونده جهان را با استفاده از مدل DGP توضیح دهیم به این معنی است که دیگر نیازی به استفاده از انرژی تاریک برای حل این مسئله نداریم. این توانایی شاید ما را به این فکر بیاندازد که مدل DGP بتواند مسائلی را که ما برای توضیح آنها ناگزیر به معرفی ماده تاریک هستیم را نیز بدون استفاده از ماده تاریک توضیح دهد.

۳

اولین مسئله ای که به خلق ماده تاریک منجر شد منحنی چرخش کهکشان ها[5] بود. بعد از آن مسائلی از قبیل مسئله جرم ویریال نیز بوسیله این ماده توضیح داده شد.

در این پایان نامه ما سعی خواهیم کرد که مسئله جرم ویریال را با استفاده از مدل DGP توضیح دهیم. مسئله جرم ویریال یکی از مسائل قدیم در کیهان شناسی است، و از این حقیقت ناشی می شود که اندازه گیری مستقیم جرم یک خوشه کهکشانی بصورت جمع جرم تک تک اعضای آن با محاسبه همین جرم با استفاده از قضیه ویریال همخوانی ندارد. در واقع در اکثر مواقع جرم ویریال در حدود ۲۰ تا ۳۰ برابر بیشتر از جرم اندازه گیری شده ی خوشه کهکشانی است. همان طور که گفته شد می توان این مسئله را با استفاده از فرض وجود  یک ماده ناشناخته در خوشه کهکشانی که تنها می تواند برهم کنش گرانشی داشته باشد توضیح داد. اما مدل DGP همان طور که در انتهای فصل ۳ بدست خواهیم آورد یک جمله جدید در سیستم معادلات تعریف می کند که صرفاً از هندسه بعد اضافه می آید و می تواند جای ماده تاریک را بگیرد. نشان خواهیم داد که جمله مربوط به چگالی انرژی برای این ماده هندسی که باید در مؤلفه ($00$) معادلات اینشتین ظاهر شود صفر است.

این پایان نامه بصورت زیر ساماندهی شده است. در فصل دوم مدل های شامه ای را مرور خواهیم کرد. این مرور بر پایه سیر تاریخی پیشرفت این نظریه تا ارائه مدل DGP است. مدل ADD که صرفاً پیشنهاد می کند با اضافه کردن حداقل دو بعد فضایی به نظریه می توان مسئله سلسله مراتب را توضیح داد بدون اینکه خللی در مشاهدات امروز ما از چهار بعد ایجاد کند در بخش۲.۲ توضیح داده شده است. مدل RS I که این ابعاد را به یک کاهش می دهد، و مدل RS II که مسئله مهمتر فشرده بودن بعد اضافه را حل می کند با این شرط که متریک فضای توده از نوع AdS بوده و توده حول شامه از تقارن $Z_2$ برخوردار باشد، در بخش ۲.۳ توضیح داده شده است. مدل GRS تعمیمی از مدل RS است که از سه شامه استفاده می کند اما محدودیت مدل های قبل که تصحیحات گرانشی را فقط در فواصل کوتاه ایجاد می کردند برطرف می کند، به این صورت که این بار تصحیحات گرانشی در هر دو رژیم فواصل خیلی کم و خیلی زیاد (نسبت به دو مقیاس فاصله در مدل- به فصل ۲ مراجعه کنید) ظاهر می شود. این مدل در بخش ۲.۴ توضیح داده شده است. در فصل ۳ به توضیح مدل DGP می پردازیم. کیهان شناسی در مدل DGP دارای دو شاخه برای معادله فریدمن است که یکی مربوط به جهان معمول FRW و دیگری مربوط به جهانی با شتاب تند شونده است. این که چگونه مدل DGP شتاب تندشونده جهان را در زمان حال توضیح می دهد در این فصل بحث خواهد شد. همچنین در مورد اینکه

---

[5] Galaxy rotation curves

۴

مدل DGP چگونه به گرانش معمول اینشتینی کاهش می یابد در بخش ۳.۶ بحث خواهیم کرد. در واقع مشاهده خواهیم کرد که این مدل نیز دارای دو مقیاس فاصله ای $r_0$ و $r_\star$ است که اگر فاصله از $r_\star$ بسیار کمتر باشد به جوابهای نسبیت عام معمول خواهیم رسید. در حد $r_0 \ll r$ نظریه گرانش کاملاً پنج بعدی خواهد بود و این محدوده ای است که برای مطالعات کیهان شناسی بسیار کارساز است. این مطلب را که مقیاس فاصله $r_0$ باید هم مرتبه با پارامتر هابل باشد در بخش ۳.۵ توضیح خواهیم داد. در بخش ۳.۷ معادلات DGP را روی شامه و برای یک جهان ایستا با تقارن کروی بدست خواهیم آورد. فصل ۴ مربوط به مسئله جرم ویریال است. در این فصل رابطه جرم ویریال را با جرم مشاهده شده برای خوشه کهکشانی بدست می آوریم و نشان می دهیم که با در نظر گرفتن این واقعیت که جرم ویریال ۲۰ تا ۳۰ برابر بیشتر از جرم مشاهده شده است جرم ویریال هم مرتبه با جرم تعریف شده بوسیله  بعد اضافه در مدل DGP است. در بخش ۴.۴ تخمینی برای جرم حاصل از بعد اضافه $\mathcal{N}(r)$ ارائه خواهیم کرد و نشان خواهیم داد که این مقدار می تواند مقدار محاسبه شده برای جرم ویریال را توضیح دهد. در بخش ۴.۵ نیز رابطه پراکندگی شعاعی سرعت خوشه های کهکشانی را با پارامتر های مسئله بدست می آوریم.

# فصل ۲

# مقدمه ای بر مدل های شامه ای

## ۲.۱ مقدمه

در این فصل در مورد مدل های شامه ای صحبت خواهیم کرد. مطالب این فصل را از لحاظ تاریخی مرتب کرده ایم. ایده اصلی برای شامه در نظریه ریسمان ایجاد شده است. ما در اینجا از زمانی شروع می کنیم که فیزیکدانان سعی کردند برای حل مسائل کیهان شناسی از مدل های جهان شامه ای استفاده کنند. دربخش ۲.۲ مدل ADD را به طور مختصر توضیح خواهیم دارد. بخش ۲.۳ مربوط به مدل شامه ای RS است که شاید مهمترین این نوع نظریه ها باشد. در این بخش اندکی در مورد چیدمان مدل صحبت خواهیم کرد و نقص های این مدل را توضیح می دهیم. مدل های جدیدتر جهان شامه ای در واقع به منظور برطرف کردن این نواقص ایجاد شده اند. در بخش ۲.۴ در مورد مدل GRS صحبت می کنیم که برای حل این مشکل در نظریه RS بوجود آمده است که تصحیحات گرانشی فقط در فواصل کوتاه رخ می دهد. مطالعه در مورد مدل DGP را که مهمترین تعمیم مدل RS است به فصل بعد موکول خواهیم کرد. از دیگر تصحیحاتی که بر مدل های جهان شامه ای ایجاد شده است می توان به مدل های گرویتون جرم دار و مدل ghost condensation اشاره کرد که تصحیحات گرانشی را برای فواصل کیهانی ایجاد می کنند. در این پایان نامه ما به این مدل ها نخواهیم پرداخت برای مطالعات بیشتر می توانید به مراجع [۲۸ و ۲۹] و [۳۳] مراجعه کنید.

## ۲.۲ مدل ADD

این مدل که ما آن را مدل ADD می نامیم در سال ۱۹۹۸ برای توضیح مسئله سلسله مراتب توسط نیما ارکانی ـ حامد، S.Dimopoulos و G. Dvali ارائه شد. مسئله سلسله مراتب ناشی می شود از این حقیقت که قدرت نیروی گرانشی نسبت به بقیه نیروها بسیار کمتر است. برای توضیح بیشتر باید خاطر نشان کنیم که دو مقیاس بنیادی انرژی در طبیعت وجود دارد که یکی مقیاس الکتروضعیف $m_{EW} \sim 10^3 GeV$ و دیگری مقیاس پلانک $M_{Pl} = G_N^{-1/2} \sim 10^{18} GeV$ است. تفاوت اساسی بین این دو مقیاس این است که با اینکه برهم کنش های ضعیف در فواصل نزدیک به $m_{EW}^{-1}$ آزمایش شده اند، نیروهای گرانشی در فواصل $M_{Pl}^{-1}$ تاکنون آزمایش نشده اند. گرانش تنها می تواند در بازه حدوداً یک سانتی متری بطور دقیق اندازه گیری شود. تعبیر ما از اینکه $M_{Pl}$ یک مقیاس بنیادی انرژی- به این معنی که در آن انرژی برهم کنش های گرانشی قدرتمند می شوند- است بر این پایه استوار است که گرانش در بازه ای در اندازه $10^{33}$ ، یعنی از مقدار فاصله ای که ما می توانیم گرانش را بیازماییم، یک سانتی متر، تا طول پلانک، $10^{-33} cm$، نیاز به اصلاح ندارد. در این مدل فرض شده است که تنها یک مقیاس برای انرژی وجود دارد و آن مقیاس الکتروضعیف است، و این مقیاس در مورد قدرت نیروی گرانشی نیز صادق خواهد بود. برای انجام این کار فرض کنید که علاوه بر چهار بعد معمول در فیزیک، $n$ بعد اضافه فضایی و فشرده نیز وجود داشته باشد. فرض کنید شعاع ابعاد اضافه $R$ باشد. برای اینکه بتوانیم گرانش را با بقیه نیروها یکی کنیم فرض می کنیم که مقیاس پلانک در $4+n$ بعد $M_{Pl(4+n)}$ با مقیاس الکتروضعیف هم مرتبه باشند. طبق قانون گؤس پتانسیل وارد بر دو ذره با جرم $m$ که در فاصله $r \ll R$ از یکدیگر قرار گرفته اند در $4+n$ بعد برابر است با

$$V(r) \sim \frac{m_1 m_2}{M_{Pl(4+n)}^{n+2}} \frac{1}{r^{n+1}} \quad (r \ll R). \qquad (۲.۲.۱)$$

از طرف دیگر اگر این دو ذره در فاصله $R \ll r$ از یکدیگر قرار گرفته باشند ابعاد اضافه نمی تواند در پتانسیل بین این دو ذره نقشی داشته باشد لذا قانون پتانسیلی بین آن دو باید همان قانون $\frac{1}{r}$ معمول در چهار بعد باشد

$$V(r) \sim \frac{m_1 m_2}{M_{Pl(4+n)}^{n+2} R^n} \frac{1}{r} \quad (r \gg R). \qquad (۲.۲.۲)$$

۷

از این رابطه می توانیم مقیاس مؤثر پلانک را برای چهار بعد بدست آوریم

$$M_{Pl}^2 \sim M_{Pl(4+n)}^{n+2} R^n. \qquad (۲.۲.۳)$$

اگر قرار دهیم $M_{Pl(4+n)} \sim m_{EW}$ و بخواهیم $R$ را طوری بدست بیاوریم که مقدار $M_{Pl}$ با مقدار مشاهده شده همخوانی داشته باشد بدست می آوریم

$$R \sim 10^{\frac{30}{n}-17} cm \times \left(\frac{1 TeV}{m_{EW}}\right)^{1+\frac{2}{n}}. \qquad (۲.۲.۴)$$

برای $n=1$ بدست می آوریم $R \sim 10^{13} cm$ و این نتیجه ایجاب می کند که بتوانیم انحرافات از گرانش نیوتنی را در فواصلی در حد منظومه شمسی مشاهده کنیم، لذا فرض $n=1$ از نظر تجربی رد می شود. برای $n$ های بزرگتر از ۲ به راحتی این انحرافات از چشم ما دور خواهد ماند. برای مثال در مورد $n=2$ محدوده $R$ بین ۱۰۰ میکرون تا یک میلیمتر خواهد بود و این فاصله ای است که می توان آن را در LHC آزمود و به دنبال اثرات بعد اضافه گشت. با اینکه گرانش در فواصل کمتر از یک میلی متر اصلاً آزمایش نشده است اما نیروهای مدل استاندارد تا فواصل مقیاس ضعیف به دقت آزمایش شده اند و اثری از بعد اضافه در آنها دیده نشده است. لذا ذرات مدل استاندارد نمی توانند آزادانه در ابعاد اضافه منتشر شوند و باید در چهار بعد جایگزیده باشند. و از آنجایی که فرض کرده بودیم که مقیاس ضعیف تنها مقیاس بنیادی برای فواصل کم است لذا جهان چهار بعدی ما باید دارای ضخامتی در حدود $m_{EW}^{-1}$ باشد. پس تنها میدانی که می تواند در ابعاد اضافه منتشر شود گرانش خواهد بود. ایده های مختلفی برای جایگزیده کردن ذرات مدل استاندارد در چهار بعد وجود دارد که شاید مهمترین آنها استفاده از نظریه ریسمان باشد. در اینجا ما از توضیح بیشتر در مورد جایگزیده کردن ذرات مدل استاندارد روی شامه صرف نظر خواهیم کرد. از نتایج این مدل می توان به تصحیح گرانش در فواصل کوتاه اما تغییر نکردن آن در فواصل زیاد اشاره کرد، چیزی که در مدل DGP بصورت برعکس اتفاق می افتد اشاره کرد (برای توضیح بیشتر به فصل بعد مراجعه کنید). همچنین یکی از مهمترین فرضیات در این مدل این است که ابعاد اضافه فشرده و دارای اندازه متناهی هستند، و این فرضی است که در مدل های بعد از آن تصحیح شده است. این فرض باعث می شود که همان طور که در (۲.۲.۳) دیدیم مقیاس پلانک در چهار بعد از طریق حجم این ابعاد اضافه به مقیاس پلانک در $4+n$ بعد ارتباط پیدا کند.

۸

$$M_{Pl}^{2} \sim M_{Pl(4+n)}^{n+2} V_n. \qquad (2.2.5)$$

## 2.3 مدل RS

با اینکه مدل ADD توانست مسئله سلسله مراتب را بین مقیاس ضعیف و مقیاس پلانک بر طرف کند اما این مدل یک مسئله دیگر به وجود آورد و آن مسئله سلسله مراتب جدیدی بین مقیاس ضعیف و مقیاس فشرده سازی[6] $\mu_c \sim \dfrac{1}{V_n^{1/n}}$ است. از این رو L.Randall و R.Sundrum به دنبال جایگزینی برای مدل ADD، مدل جدیدی شامل یک بعد اضافه و دو ابر سطح یکی با تنش مثبت و شامل ذرات مدل استاندارد و دیگری با تنش منفی ارائه کردند. در این مدل که ما آن را مدل RS I می نامیم خواهیم دید که متریک فضای شامه توسط یک ضریب تابیدگی[7] به متریک فضای توده مرتبط می شود

$$ds^{2} = e^{-2kr_c\phi}\eta_{\mu\nu}dx^{\mu}dx^{\nu} + r_c^{2}d\phi^{2}, \qquad (2.3.1)$$

که در آن $k$ هم مرتبه با مقیاس پلانک است، $x^{\mu}$ نشان دهنده مختصات چهار بعد شامه است و $0 \leq \phi \leq \pi$ مختصات بعد اضافه است که بازه متناهی دارد و اندازه اش توسط ثابت $r_c$ تعیین می شود. از آنجایی که در متریک یک تابع نمایی قرار گرفته است، لذا برای حل مسئله سلسله مراتب برخلاف مدل ADD احتیاجی نداریم که اندازه بعد اضافه ($r_c$) خیلی بزرگ باشد. این یکی از تفاوت های اساسی این مدل با مدل ADD است که حل مسئله سلسله مراتب به تولید مسئله جدیدی نمی انجامد. در واقع تفاوت بین مقیاس پلانک در پنج بعد و مقیاس فشرده سازی $\mu_c$ از مرتبه ۱۰ است. تفاوت شاید مهمتر این دو نظریه در این است که در اینجا فقط از یک بعد اضافه استفاده می شود. در این مدل نشان داده می شود که متریک بالا حل معادلات اینشتین برای دو شامه با ثابت کیهان شناسی مناسب است. این دو شامه در $\phi = 0$ و $\phi = \pi$ قرار گرفته اند. هم چنین فرض می کنیم که بعد اضافه حول شامه دارای تقارن $Z_2$ باشد یعنی دو نقطه $(x, \phi)$ و $(x, -\phi)$ را یکی در نظر می گیریم. متریک روی دو شامه بصورت زیر خواهند بود

---

[6] Compactification scale
[7] Warp factor

۹

$$g^{vis}_{\mu\nu}(x^{\mu}) \equiv G_{\mu\nu}(x^{\mu}, \phi = \pi),$$
$$g^{hid}_{\mu\nu}(x^{\mu}) \equiv G_{\mu\nu}(x^{\mu}, \phi = 0),$$
(۲.۳.۲)

که در آن $G_{MN}$ متریک فضای پنج بعدی است. $vis$ به معنای قابل مشاهده و $hid$ به معنای پنهان به ترتیب مربوط به شامه های با تنش مثبت و منفی هستند. اگر کنش مسئله را بصورت زیر تعریف کنیم

$$S = S_{gravity} + S_{vis} + S_{hid},$$
$$S_{gravity} = \int dx^4 \int_{-\pi}^{\pi} d\phi \sqrt{-G} \{-\Lambda + 2M^3 R\},$$
$$S_{vis} = \int dx^4 \sqrt{-g_{vis}} \{\mathcal{L}_{vis} - V_{vis}\},$$
$$S_{hid} = \int dx^4 \sqrt{-g_{hid}} \{\mathcal{L}_{hid} - V_{hid}\},$$
(۲.۳.۳)

که در آن ثابت های انرژی خلاء (کیهان شناسی) را از ماده معمولی جدا کرده ایم، معادلات اینشتین در پنج بعد بصورت زیر در می آیند [۳۵]

$$\sqrt{-G}\left(R_{MN} - \frac{1}{2}G_{MN}R\right) = -\frac{1}{4M^3}[\Lambda\sqrt{-G}G_{MN} + V_{vis}\sqrt{-g_{vis}}g^{vis}_{\mu\nu}\delta^{\mu}_M\delta^{\nu}_N\delta(\phi-\pi)$$
$$+ V_{hid}\sqrt{-g_{hid}}g^{hid}_{\mu\nu}\delta^{\mu}_M\delta^{\nu}_N\delta(\phi)].$$
(۲.۳.۴)

با انتخاب متریک توده بشکل

$$ds^2 = e^{-2\sigma(\phi)}\eta_{\mu\nu}dx^{\mu}dx^{\nu} + r_c^2 d\phi^2,$$
(۲.۳.۵)

و قرار دادن آن در معادلات اینشتین بدست می آوریم

$$\frac{6\sigma'^2}{r_c^2} = \frac{-\Lambda}{4M^3},$$
(۲.۳.۶)

$$\frac{3\sigma''}{r_c^2} = \frac{V_{hid}}{4M^3 r_c}\delta(\phi) + \frac{V_{vis}}{4M^3 r_c}\delta(\phi-\pi).$$
(۲.۳.۷)

از حل معادله اول می رسیم به

$$\sigma = r_c |\phi| \sqrt{-\frac{\Lambda}{24M^3}},  \qquad (2.3.8)$$

که از آن نتیجه می گیریم $\Lambda < 0$. با مشتق گیری دوباره از (2.3.6) و مقایسه آن با (2.3.7) مشاهده می کنیم که ثابت های کیهان شناسی و انرژی خلاء باید توسط یک ثابت به هم مربوط شوند

$$V_{hid} = -V_{vis} = 24M^3 k, \qquad \Lambda = -24M^3 k^2, \qquad (2.3.9)$$

هم چنین معادله (2.3.8) نشان می دهد که میان دو شامه، متریک توده باید از نوع AdS باشد. همچنین ثابت پلانک بوسیله فرمول زیر به اندازه بعد اضافه مربوط خواهد شد

$$M_{Pl}^2 = M^3 r_c \int_{-\pi}^{\pi} d\phi\, e^{-2kr_c|\phi|} = \frac{M}{k}\left[1 - e^{-2kr_c\pi}\right]. \qquad (2.3.10)$$

حل مسئله سلسله مراتب به این صورت انجام می شود که مشاهده می کنیم هر پارامتر جرم $m_0$ روی شامه که توسط نظریه تعیین می شود بوسیله فرمول

$$m \equiv e^{-kr_c\pi} m_0, \qquad (2.3.11)$$

به جرم فیزیکی مرتبط می شود به این معنی که قدرت آن بوسیله یک ضریب قابل تنظیم کاهش می یابد.

در مدل RS I بعد اضافه فشرده و از اندازه متناهی بود به همین منظور این دو دانشمند مدل دیگری بر پایه مدل اول ارائه دادند که بعد اضافه در آن غیر فشرده باشد. در این مدل که آن را RS II می نامیم یک بعد اضافه غیر فشرده وجود دارد و شامل یک شامه با تنش مثبت است. مقیاس پلانک در این مدل بوسیله انحنای ابعاد بالا تعیین می شود. همان طور که در مدل RS I گفته شد متریک بین دو شامه از نوع $AdS$ است. اگر $r_c$ را به بینهایت میل دهیم (که این معادل است با این که شامه با تنش منفی را به بینهایت ببریم) بعد اضافه بصورت نیمه بینهایت در می آید. نامتناهی شدن بعد اضافه به خاطر شرط $Z_2$ بودن فضای توده

(اکنون حول تک شامه مسئله) محقق خواهد شد. در این مدل نیز انحرافات از گرانش معمول اینشتینی فقط در فواصل کوتاه اتفاق می افتد.

## ۲.۴ مدل GRS

همان طور که در بخش قبل دیدیم مدل RS II توانست مدل ADD را به مدلی شامه ای با بعد اضافه غیر فشرده ارتقاء دهد. اما در هر دوی این مدل ها تصحیحات گرانشی برای فواصل خیلی کوتاه اتفاق می افتد. در سال ۲۰۰۰ میلادی R. Gregory، V. Rubakov و M. Sibiryakov مدلی بر پایه مدل RS با سه شامه ارائه دادندکه در فواصل بسیار زیاد هم تاثیرات بعد پنجم را بروز می دهد [۱۸]. این مدل از یک شامه با تنش مثبت $\sigma > 0$ و (به خاطر حفظ تقارن $Z_2$) دو شامه با تنش $-\frac{\sigma}{2}$ که با فاصله مساوی در دو طرف شامه اول روی بعد اضافه قرار گرفته اند تشکیل شده است. همچنین فرض می کنیم که ذرات مدل استاندارد روی شامه با تنش مثبت قرار داشته باشند. ثابت کیهان شناسی توده بین شامه ها، مانند مدل RS، منفی است اما در دو طرف شامه های با تنش منفی صفر است. می توان نشان داد که معادلات اینشتین در پنج بعد جوابی برای پیکربندی بالا دارد که متریک آن بصورت

$$ds^2 = a^2(z)\eta_{\mu\nu}dx^\mu dx^\nu - dz^2, \qquad (\text{۲.۴.۱})$$

نوشته می شود که در آن

$$a(z) = \begin{cases} e^{-kz} & 0 < z < z_c \\ e^{-kz_c} \equiv a_- & z > z_c \end{cases}. \qquad (\text{۲.۴.۲})$$

ثابت $k$ با $\sigma$ و $\Lambda$ بصورت زیر ارتباط دارد

$$\sigma = \frac{3k}{4\pi G_{bulk}}, \qquad \Lambda = -6k^2, \qquad (\text{۲.۴.۳})$$

که در آن $G_{bulk}$ ثابت نیوتن در پنج بعد است. همان طور که می بینیم متریک در طرف راست شامه با تنش منفی تخت است و بین دو شامه بصورت $AdS$ می باشد. طرف چپ فضا نیز تقارن آینه ای طرف راست خواهد بود. این پس زمینه دارای دو مقیاس طولی است $k^{-1}$ و $\zeta_c \equiv k^{-1}e^{kz_c}$. اگر ما به حالتی نگاه کنیم که $z_c$ به اندازه کافی بزرگ باشد، این دو ثابت تفاوت زیادی با هم خواهند داشت یعنی $k^{-1} \ll \zeta_c$. خواهیم دید که اگر $k^{-1} \ll r \ll \zeta_c (k\zeta_c)^2$ باشد گرانش بطور مؤثر چهار بعدی است، اما در دو حالت $r \ll k^{-1}$ و $(k\zeta_c)^2 \zeta_c \gg r$ گرانش پنج بعدی خواهد بود. برای اینکه رفتار پتانسیل گرانشی را در این مدل ببینیم اختلال های حول متریک پس زمینه (۲.۴.۱) را مطالعه می کنیم. ما از پیمانه گؤسی نرمال $g_{zz} = -1$ و $g_{z\mu} = 0$ استفاده خواهیم کرد. فرض کنید متریک اختلالی به صورت زیر باشد

$$ds^2 = a^2(z)\eta_{\mu\nu}dx^\mu dx^\nu + h_{\mu\nu}(x,z)dx^\mu dx^\nu - dz^2. \tag{۲.۴.۴}$$

برای $h_{\mu\nu}$ نیز پیمانه بی رد و متعامد[8] $h_\mu^\mu = 0$ و $h_{\nu;\mu}^\mu = 0$ را انتخاب می کنیم. معادلات خطی برای تمام مؤلفه های $h_{\mu\nu}$ بصورت زیر خواهد شد

$$\begin{cases} h'' - 4k^2 h - \dfrac{1}{a^2}\Box^{(4)} h = 0 & 0 < z < z_c \\ h'' - \dfrac{1}{a_-^2}\Box^{(4)} h = 0 & z > z_c \end{cases}, \tag{۲.۴.۵}$$

که در آن اندیس مولفه ها را حذف کرده ایم. اگر فرض کنیم $h = \psi(z)e^{ip_\mu x^\mu}$ که $p^2 = m^2$ مشاهده می کنیم که $\psi(z)$ از معادله

$$\begin{cases} \psi'' - 4k^2\psi + \dfrac{m^2}{a^2}\psi = 0 & 0 < z < z_c \\ \psi'' + \dfrac{m^2}{a_-^2}\psi = 0 & z > z_c \end{cases}, \tag{۲.۴.۶}$$

پیروی می کند. حل این معادلات به جواب زیر می انجامد

---
[8] Transverse and traceless

۱۳

$$\psi_m = \begin{cases} C_m \left[ N_1\left(\frac{m}{k}\right) J_2(m\zeta) - J_1\left(\frac{m}{k}\right) N_2(m\zeta) \right] & for \quad 0 < z < z_c \\ A_m \cos\left[\frac{m}{a_-}(z - z_c)\right] + B_m \sin\left[\frac{m}{a_-}(z - z_c)\right] & for \quad z > z_c \end{cases}, \qquad (2.4.7)$$

که در آن $J$ و $N$ توابع بسل هستند و $\zeta = \frac{1}{k} e^{kz}$ . برای بدست آوردن ثابت ها میتوانیم از معادلات اتصال Israel استفاده نماییم که برای مسئله ما بصورت زیر هستند

$$\begin{cases} \psi' + 2k\psi = 0 & at \quad z = 0 \\ [\psi'] - 2k\psi = 0 & at \quad z = z_c \end{cases}, \qquad (2.4.8)$$

که در آن $[\psi']$ ناپیوستگی مشتق $\psi$ نسبت به بعد اضافه $z$ در $z_c$ است. با استفاده از این معادلات $A_m$ و $B_m$ بصورت زیر خواهند شد

$$A_m = C_m \left[ N_1\left(\frac{m}{k}\right) J_2(m\zeta_c) - J_1\left(\frac{m}{k}\right) N_2(m\zeta_c) \right],$$
$$B_m = C_m \left[ N_1\left(\frac{m}{k}\right) J_1(m\zeta_c) - J_1\left(\frac{m}{k}\right) N_1(m\zeta_c) \right]. \qquad (2.4.9)$$

ثابت $C_m$ از طریق شرط بهنجارش بدست می آید. برای حد $m\zeta_c \gg 1$ ثابت $C_m$ بصورت زیر است

$$C_m^{\ 2} = \frac{m}{2k}\left[ J_1^{\ 2}\left(\frac{m}{k}\right) + N_1^{\ 2}\left(\frac{m}{k}\right) \right]^{-1}, \qquad (2.3.10)$$

و برای حد $m\zeta_c \ll 1$ بصورت

$$C_m^{\ 2} = \frac{\pi}{(k\zeta_c)^3}\left( 1 + \frac{4}{(m\zeta_c)^2 (k\zeta_c)^4} \right)^{-1}, \qquad (2.4.11)$$

است. پتانسیل گرانشی بصورت زیر محاسبه می شود

$$V(r) = G_{bulk} \int_0^\infty dm \frac{e^{-mr}}{r} \psi_m^2(z=0), \qquad (2.4.12)$$

که برای حد $r \ll k^{-1}$ بصورت

$$V(r) = \frac{G_{bulk}}{kr^3} = \frac{G_N}{r} \frac{1}{k^2 r^2}, \qquad (2.4.13)$$

بدست می آید که در آن $G_N = G_{bulk} k$. همان طور که انتظار داشتیم این حل به شکل قانون نیوتن برای پنج بعد است. برای حد مخالف یعنی $r \gg k^{-1}$ داریم

$$V(r) = \frac{G_N}{r} \frac{2}{\pi} \int_0^\infty dx \frac{e^{-\left(\frac{2r}{r_c}\right)x}}{x^2 + 1} = \frac{2G_N}{\pi r} \left[ Ci\left(\frac{2r}{r_c}\right) \sin\left(\frac{2r}{r_c}\right) - Si\left(\frac{2r}{r_c}\right) \cos\left(\frac{2r}{r_c}\right) \right], \qquad (2.4.14)$$

که در آن $x = \frac{mr_c}{2}$ است و

$$r_c = \zeta_c (k\zeta_c)^2 \equiv k^{-1} e^{3kz_c}. \qquad (2.4.15)$$

اگر $r \ll r_c$ باشد پتانسیل بصورت معمول نیوتنی خود $V(r) = \frac{G_N}{r}$ در می آید، اما اگر $r \gg r_c$ بدست می آوریم

$$V(r) = \frac{G_{bulk} r_c}{\pi r^2}, \qquad (2.4.16)$$

که باز به فرم قانون نیوتن در پنج بعد است. یکی از نقایص این مدل وجود ghost است. همان طور که در مقدمه گفته شد ghost یک ناپایداری برای خلاء است که اجازه می دهد خلق ذرات همراه با ghost آنها اتفاق بیافتد. این ناپایداری بصورت یک جمله انرژی جنبشی با علامت اشتباه در معادلات ظاهر می شود. مدل های متنوعی برای برداشتن دو قید

- محدود بودن تصحیحات گرانشی به فواصل کم
- فشرده بودن بعد اضافه

ارائه شده است. برای مثال می توان به [۲۲] اشاره کرد که در آن یک مدل با بعد اضافه فشرده معرفی شده است که گرانش در فواصل بسیار زیاد پنج بعدی است. اما شاید مهم ترین اقدام در جهت رفع این دو قید توسط G. Dvali، G. Gabadadze و M. Porrati در مقاله معروف [۱] داده شده باشد. در این مدل تصحیحات گرانشی در فواصل زیاد اتفاق می افتد و بعد اضافه غیر فشرده است. این کار توسط اضافه کردن کنش چهار بعدی اینشتین به کنش پنج بعدی مسئله انجام می شود. در فصل بعد بطور مفصل در مورد این مدل و نتایج آن بحث خواهیم کرد.

# فصل ۳

# مدل DGP

## ۳.۱ مقدمه

یکی از مشاهدات اخیر در کیهان شناسی منبسط شدن شتابدار جهان است که با اندازه گیری انتقال به سرخ ابر نو اختر نوع Ia بدست آمده است [۶] و این مسئله ای است که نسبیت عام اینشتین از توضیح آن باز می ماند. در نسبیت عام اینشتین با اعمال متریک FRW به معادله زیر می رسیم

$$3\frac{\ddot{a}}{a} = -4\pi(\rho + 3p). \qquad (۳.۱.۱)$$

بوضوح برای ماده باریونیک که برای آن چگالی انرژی و فشار مثبت است، شتاب جهان باید منفی باشد. برای حل این مسئله دو راه کلی وجود دارد. راه اول که بیشتر مورد علاقه کیهان شناسان است این است که فرض کنیم معادلات اینشتین همچنان درست است و مسئله بالا (یا مسائل مشابه آن) را با در نظر گرفتن یک منبع ماده یا میدان جدید که تاکنون آن را در نظر نگرفته ایم حل کنیم. در گرانش اینشتینی انرژی خلاء یا ثابت کیهان شناسی با ایجاد دافعه گرانشی می تواند انبساط شتابدار جهان را توضیح دهد. گونه های دیگری از انرژی خلاء مانند quintessence نیز وجود دارند که این مدل ها را از سطح یک مدل چگالی انرژی به سطح یک پتانسیل چگالی انرژی برای یک میدان دینامیکی ارتقاء می دهد. به تمامی این نوع مدل ها، **مدل های انرژی تاریک** می گوییم. در این روش معادله فریدمن را بصورت زیر می نویسیم

$$H^2 = \left(\frac{\dot{a}}{a}\right)^2 = \frac{2\Lambda}{3} + \frac{2A_D}{a^3} + \frac{2A_B}{a^3} + \frac{2A_R}{a^4} + \frac{k}{a^2}, \qquad (۳.۱.۲)$$

۱۷

که در آن $\Lambda$ ثابت کیهان شناسی است و بوسیله $\Lambda = 8\pi G \rho_v$ به چگالی انرژی خلاء مربوط می شود، و $A_R, A_D, A_B$ به ترتیب ثابت های مربوط به ماده غیر نسبیتی (باریونی)، ماده تاریک و تابش هستند. $k$ انحنای فضایی است که به ترتیب مقدارهای $-1, 1, 0$ را برای جهان های تخت، بسته و باز می پذیرد. در جهان در حال انبساط به زمانی می رسیم که جمله اول در (۳.۱.۲) بر جمله های دیگر غالب می شود. حل معادله (۳.۱.۲) در حد $t \to \infty$ بصورت زیر است

$$a(t) = A_V f(t), \qquad f(t) = \begin{cases} \cosh\left(\dfrac{t}{A_V}\right) & k = 1 \\ \sinh\left(\dfrac{t}{A_V}\right) & k = -1, \\ \exp\left(\dfrac{t}{A_V}\right) & k = 0 \end{cases} \qquad (۳.۱.۳)$$

که در آن $A_V = (8\pi G \rho_V / 3)^{1/2}$. بوضوح این حل نشان دهنده انبساط تند شونده جهان است.

راه دوم این است که به دنبال نظریه ای بر پایه نسبیت عام اینشتین بگردیم که بتواند انبساط تند شونده جهان را پیش بینی کند. نظریه DGP چنین نظریه ای است. این نظریه یک مدل شامه ای است که بعد اضافه در آن تخت و نامتناهی است. بر خلاف نظریه های شامه ای دیگر مانند ADD و RS این نظریه در فواصل کوتاه به یک نظریه چهار بعدی تبدیل می شود اما در فواصل زیاد پتانسیل پنج بعدی خواهد شد. در بخش ۳.۶ مشاهده خواهیم کرد که این مدل در حد فاصله های کم ابتدا به نظریه اسکالر – تانسوری برنز – دیکی تبدیل می شود و اگر باز هم به فواصل کوتاه تر برویم می توانیم گرانش معمول اینشتین را بازیابی کنیم.

## ۳.۲ مدل DGP

مدل DGP در سال ۲۰۰۰ میلادی توسط M.Porrati, G.Gabadadze, G.Dvali ارائه شد [۱]. در این مدل توده پنج بعدی است و شامه به عنوان یک ابر سطح از فضای توده در نظر گرفته می شود. بر خلاف مدل RS که با در نظر گرفتن

متریک AdS گرانش اطراف شامه جایگزیده می شود، در این مدل فرض می کنیم که فضای توده تخت است و متریک آن را مینکوفسکی در نظر می گیریم. همان طور که در [۹] بررسی شده است، گرانش در مدل های شامه ای فقط وقتی اطراف شامه جایگزیده می شود که

۱. توده حول شامه تقارن $Z_2$ داشته باشد.

۲. هندسه توده از نوع $AdS$ باشد.

برای حل این مشکل، گرانش را بصورت کنش چهار بعدی اینشتین- هیلبرت به کنش پنج بعدی مدل DGP می افزاییم. کنش کامل این مدل بصورت زیر است

$$S = \frac{m_4^3}{2}\int d^5x \sqrt{-g}\,^5R + \frac{m_3^2}{2}\int d^4x \sqrt{-q}\,^4R + S_m(q_{\mu\nu}) + S_B(g_{AB}),  \qquad (۳.۲.۱)$$

که در آن $g_{AB}$ ($q_{\mu\nu}$) متریک توده (شامه)، $^5R$ ($^4R$) اسکالر ریچی در پنج (چهار) بعد و $m_4^3$ ($m_3^2$) مقیاس پلانک برای توده (شامه) هستند. دو جمله آخر در (۳.۲.۱) کنش های مربوط به ماده در شامه و توده هستند. در این مدل $m_4^3$ و $m_3^2$ مستقل از هم هستند (در حالت کلی این دو کمیت به هم وابسته اند). همچنین $q_{\mu\nu}$ متریک القایی روی شامه است و بصورت زیر تعریف می شود

$$q_{AB}(x) = g_{AB}(x, y=0),  \qquad (۳.۲.۲)$$

که در آن شامه را در $y=0$ قرار داده ایم. در حد $m_4^3 \to 0$ با فرض اینکه $m_3^2$ متناهی باشد کنش (۳.۲.۱) توصیف کننده گرانش در چهار بعد روی شامه است. از طرف دیگر در حد $m_3^2 \to 0$ و با فرض متناهی بودن $m_4^3$ این کنش توصیف کننده گرانش در توده پنج بعدی است. قبل از بدست آوردن معادلات DGP در مورد منشأ پیدا شدن جمله چهار بعدی در کنش

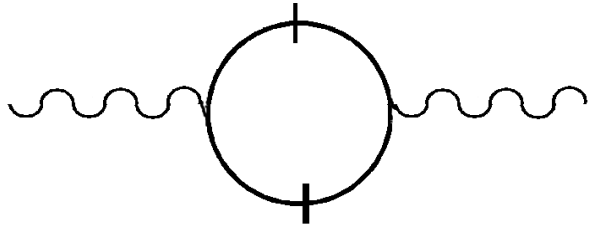

شکل (۳.۱) نمودار یک حلقه ای برای تولید اسکالر ریچی در ۴ بعد. خطوط مواج گرویتون ها را مشخص می کند، خطوط پر اسکالر های جرم دار یا فرمیون ها را مشخص می کند و خطوط عمودی روی انتشارگر های فرمیون/اسکالر نشان دهنده جرم دار بودن آنهاست. [برگرفته از مرجع

[G. Dvali, G. Gabadadze and M. Porrati, Phys. Lett. B **485** (2000) 208

(۳.۲.۱) صحبت می کنیم. مهم ترین احتمال به این حقیقت برمی گردد که در بعضی حالات خاص (که در زیر به یکی از آنها می پردازیم) ماده جایگزیده روی شامه از طریق تصحیحات حلقه ای[9] می تواند یک جمله جنبشی چهار بعدی برای گرویتون ها ایجاد کند. برای توضیح بیشتر این مطلب فرض کنید که میدان های ماده، روی شامه مقید شده اند. تانسور انرژی ـ تکانه ماده بصورت زیر نوشته می شود

$$T_{AB} = \begin{pmatrix} T_{\mu\nu}(x)\delta(y) & 0 \\ 0 & 0 \end{pmatrix}. \tag{۳.۲.۳}$$

در نتیجه لاگرانژی برهم کنش برای این ماده جایگزیده، با اختلالات پنج بعدی متریک بصورت $h_{AB}(x,y) \equiv G_{AB}(x,y) - \eta_{AB}$ به جمله زیر کاهش می یابد

$$\mathcal{L}_{\text{int}} = \int dy\, h^{\mu\nu}(x,y) T_{\mu\nu}(x)\delta(y) = h^{\mu\nu}(x,0) T_{\mu\nu}(x), \tag{۳.۲.۴}$$

که در آن از متریک القایی روی شامه $g_{\mu\nu}(x) = \eta_{\mu\nu} + h_{\mu\nu}$ استفاده کرده ایم. به خاطر این برهم کنش می توانیم یک جمله جنبشی در چهار بعد برای $g_{\mu\nu}(x)$ تولید کنیم. برای مثال نمودار شکل (۳.۱) با اسکالرهای جرم دار [۱۵] یا فرمیون ها [۱۶و۱۷] که در حلقه می چرخند، می تواند جمله چهار بعدی زیر را در انرژی های پایین تولید کند

---

9 Loop corrections

۲۰

$$\int d^4x\, dy\, \delta(y)\sqrt{|g|}R. \qquad (3.2.5)$$

ثابت گرانش القا شده مربوط به این جمله می تواند توسط توابع همبستگی[10] تعیین شوند. برای مثال در یک چارچوب چهار بعدی [16و17] داریم

$$m_3^2 = \frac{i}{96}\int d^4x\, x^2 \left\{ \langle T\, S(x)S(0)\rangle - \langle S\rangle^2 \right\}, \qquad (3.2.6)$$

که در آن $S(x) = T_\mu^{\,\mu}$ رد تانسور انرژی ـ تکانه حالت چهار بعدی است که در حلقه می چرخد و $T$ ترتیب زمانی[11] است. ثابت کیهان شناسی القا شده در چهار بعد بصورت $\Lambda = \langle 0|T_\mu^{\,\mu}|0\rangle$ تعریف می شود. همچنین در این مدل از جملات مرتبه بالاتر ($\mathcal{O}(R^2)$) که به همین ترتیب می تواند در مدل ایجاد شود صرفنظر می کنیم.

وردش کنش (3.2.1) ما را به معادلات اینشتین در پنج بعد می رساند

$$m_4^3\left({}^5R_{AB} - \frac{1}{2}g_{AB}\,{}^5R\right) + m_3^2\left({}^4R_{\mu\nu} - \frac{1}{2}q_{\mu\nu}\,{}^4R\right)\delta^\mu_{\,A}\delta^\nu_{\,B}\delta(y) \qquad (3.2.7)$$
$$= \delta^\mu_{\,A}\delta^\nu_{\,B}\left(T_{\mu\nu} - m_3^2\Lambda q_{\mu\nu}\right)\delta(y) + \hat{T}_{AB},$$

که در آن $\hat{T}_{AB}\left(T_{\mu\nu}\right)$ تانسور انرژی ـ تکانه ی توده (شامه) است.

## 3.3 نتایج اولیه مدل

---

[10] Correlation functions

[11] Time ordering

در این بخش به عنوان یک مدل اسباب بازی مثالی می زنیم که ایده ای از رفتار تابع پتانسیل گرانشی در چهار و پنج بعد را به ما می دهد. این مدل شامل یک اسکالر پنج بعدی است. این ترفند را در ادامه بر روی مدل DGP اعمال خواهیم کرد. فرض کنید که کنش مدل بصورت زیر باشد

$$S = m_4^3 \int d^4 x dy \, \partial_A \Phi(x,y) \partial^A \Phi(x,y) + m_3^2 \int d^4 x dy \, \delta(y) \partial_\mu \Phi(x,0) \partial^\mu \Phi(x,0), \qquad (3.3.1)$$

که در آن برای اینکه راحت بتوانیم نتایج را با گرانش مقایسه کنیم از میدان اسکالر بهنجار و بی بعد $\Phi$ استفاده کرده ایم. توجه کنید که این نوع کنش وقتی بوجود می آید که مثلاً $\Phi$ در پنج بعد در پس زمینه ( $\chi = v \tanh(vy)$ ) قرار داشته باشد و بر هم کنش بین میدان ها بصورت $\left(\partial_A \Phi\right)^2 + \left(\partial_A \chi\right)^2 \left(\partial^A \chi\right)^2 - \left(\partial_A \Phi \partial^A \chi\right)^2 + \cdots$ باشد. با این فرضیات معادله حرکت $\Phi$ بصورت $2\partial_A \Phi \partial^A \Phi + \left(\chi'\right)^2 \partial_\mu \Phi \partial^\mu \Phi = 0$ در می آید که در آن پریم نشانه مشتق نسبت به y است و $\chi$ تابعی است که در $y=0$ ماکزیمم دارد. این معادله می تواند توسط کنش (3.3.1) مدل شود.

هدف ما این است که وابستگی برهم کنش ها به فاصله را تعیین کنیم. برای این کار باید تابع گرین تأخیری[12] مربوط به مسئله را بیابیم و سپس پتانسیل را محاسبه نماییم. معادله کلاسیکی تابع گرین بصورت زیر است

$$\left(m_4^3 \partial_A \partial^A + m_3^2 \delta(y) \partial_\mu \partial^\mu\right) G_R(x,y;0,0) = \delta^4(x) \delta(y), \qquad (3.3.2)$$

که در آن $G_R(x,y;0,0) = 0$ برای $x_0 < 0$. پتانسیل مربوط به اسکالر $\Phi$ در جهان چهار بعدی شامه بصورت زیر تعیین می شود

$$V(r) = \int G_R(t, \mathbf{x}, y=0; 0,0,0) dt, \qquad (3.3.3)$$

که در آن $r = \sqrt{x_1^2 + x_2^2 + x_3^2}$. برای حل معادله (3.3.2) تمام کمیت ها را تحت تبدیل فوریه نسبت به چهار_مختصه $x_\mu$ قرار می دهیم

---

[12] Retarded

$$G_R(x,y;0,0) \equiv \int \frac{d^4p}{(2\pi)^4} e^{ipx} \tilde{G}_R(p,y). \tag{3.3.4}$$

اگر به فضای اقلیدسی برویم معادله (3.3.2) بصورت زیر در می آید

$$\left[m_4^3\left(p^2 - \partial_y^2\right) + m_3^2 p^2 \delta(y)\right]\tilde{G}_R(p,y) = \delta(y), \tag{3.3.5}$$

که در آن $p^2$ نشان دهنده مربع چهار-تکانه اقلیدسی است. جواب این معادله با در نظر گرفتن شرایط مرزی مناسب بصورت زیر است

$$\tilde{G}_R(p,y) = \frac{1}{m_3^2 p^2 + 2m_4^3 p} e^{-p|y|}, \tag{3.3.6}$$

که در آن $p = \sqrt{p^2}$. با استفاده از این حل و معادله (3.3.3) می توانیم فرمولی برای پتانسیل مربوط به این میدان اسکالر در شامه چهار بعدی بدست آوریم

$$V(r) = -\frac{1}{8\pi^2 m_3^2} \frac{1}{r} \left\{ \sin\left(\frac{r}{r_0}\right) Ci\left(\frac{r}{r_0}\right) + \frac{1}{2}\cos\left(\frac{r}{r_0}\right)\left[\pi - 2Si\left(\frac{r}{r_0}\right)\right] \right\}, \tag{3.3.7}$$

که در آن

$$Ci(z) \equiv \gamma + \ln(z) + \int_0^z \frac{(\cos(t)-1)}{t} dt,$$
$$Si(z) \equiv \int_0^z \frac{(\sin(t)-1)}{t} dt, \tag{3.3.7a}$$

هستند و $\gamma = 0.577$ ثابت اویلر-ماچرونی[13] است و مقیاس فاصله ای $r_0$ را بصورت زیر تعریف کرده ایم.

$$r_0 = \frac{m_3^2}{2m_4^3}. \tag{3.3.8}$$

---
[13] Euler-Masceroni

می توان با استفاده از این مقیاس فاصله ای در مورد رفتار تابع پتانسیل در فواصل کم و زیاد نسبت به $r_0$ صحبت کرد. در واقع همین مقیاس فاصله است که در مورد پتانسیل گرانشی در مدل DGP نیز استفاده می شود. در فواصل کوتاه وقتی که $r \ll r_0$ داریم

$$V(r) \simeq -\frac{1}{8\pi^2 m_3^2} \frac{1}{r} \left\{ \frac{\pi}{2} + \left[ -1 + \gamma + \ln\left(\frac{r}{r_0}\right) \right] \times \left(\frac{r}{r_0}\right) + \mathcal{O}(r^2) \right\}. \tag{3.3.9}$$

همان طور که می بینیم پتانسیل در فواصل کوتاه رفتار صحیح نیوتنی در چهار بعد را دارد، یعنی $\frac{1}{r}$، که البته با یک جمله دافعه لگاریتمی اصلاح شده است. در مورد فواصل زیاد وقتی که $r_0 \ll r$ رفتار پتانسیل بصورت زیر خواهد بود

$$V(r) = -\frac{1}{8\pi^2 m_3^2} \frac{1}{r} \left\{ \left(\frac{r_0}{r}\right) + \mathcal{O}\left(\frac{1}{r^2}\right) \right\}, \tag{3.3.10}$$

که بصورت $\frac{1}{r^2}$ تغییر می کند و مطابق قوانین نظریه پنج بعدی است. این نتایج مشابه نتایجی است که در فصل قبل بخش 2.4 توضیح داده شد [18]. حال آماده ایم تا به مسئله اصلی خود بپردازیم. خواهیم دید که رفتار پتانسیل گرانشی در مدل DGP همانند رفتار پتانسیل در این مدل اسباب بازی است. قبل از انجام این کار توجه کنید که ساختار تانسوری برای انتشارگر یک گرویتون بدون جرم در چهار بعد بصورت زیر است

$$\frac{1}{2}\eta^{\mu\alpha}\eta^{\nu\beta} + \frac{1}{2}\eta^{\mu\beta}\eta^{\nu\alpha} - \frac{1}{2}\eta^{\mu\nu}\eta^{\alpha\beta}. \tag{3.3.11}$$

اما در مورد یک گرویتون جرم دار در چهار بعد یا بطور معادل یک گرویتون بدون جرم در پنج بعد ساختار تانسوری انتشارگر بصورت

$$\frac{1}{2}\eta^{\mu\alpha}\eta^{\nu\beta} + \frac{1}{2}\eta^{\mu\beta}\eta^{\nu\alpha} - \frac{1}{3}\eta^{\mu\nu}\eta^{\alpha\beta}, \tag{3.3.12}$$

خواهد بود. همان طور که مشاهده می کنید تفاوت این دو در یک ضریب $\frac{1}{2}$ و $\frac{1}{3}$ است. ضریب $\frac{1}{3}$ برای توصیف مشاهدات مشکل ساز است. بدون توجه به اینکه جرم گرویتون در چهار بعد چقدر کم باشد، ساختارهای تانسوری بالا در مورد به دام انداختن نور پیش بینی های متفاوتی خواهند داشت [19و20] و ساختار (3.3.12) نتایج رصدی را ارضاء نخواهد کرد. در واقع هر نظریه ای با گرویتون جرم دار، بدون توجه به اینکه جرمش چقدر کم باشد، با این مشکل مواجه است چون در آنها یک ناپیوستگی در حدی که جرم را به صفر می بریم وجود خواهد داشت. این ناپیوستگی را «ناپیوستگی VDVZ» می نامیم[14]. برای توضیح بیشتر در مورد این ناپیوستگی ابتدا توجه کنید که نظریه نسبیت عام یک نظریه گرانشی است که با گرویتون بدون جرم با دو درجه آزادی (دو قطبش) سازگار است. برای توضیح گرانش با یک میدان تانسوری جرم دار، مجبور به نادیده گرفتن همودینامایی عام[15] معادلات خواهیم بود و این باعث می شود که گرویتون از پنج درجه آزادی برخوردار شود. پتانسیل گرانشی $h_{\mu\nu}$ (که آن را بصورت $h_{\mu\nu} = g_{\mu\nu} - \eta_{\mu\nu}$ تعریف می کنیم) تولید شده بوسیله یک منبع ایستای $T_{\mu\nu}$ برای یک گرویتون جرم دار با جرم $m$ حول پس زمینه تخت مینکوفسکی بصورت زیر است

$$h_{\mu\nu}^{massive}(q^2) = -\frac{8\pi}{m_3^2}\frac{1}{q^2+m^2}\left(T_{\mu\nu} - \frac{1}{3}\eta_{\mu\nu}T_\alpha^\alpha\right), \qquad (3.3.13)$$

که در آن $q^i$ مختصات تکانه در فضای سه بعدی تکانه است. بطور مشابه این پتانسیل برای نظریه معمول گرانش اینشتینی بصورت زیر خواهد بود

$$h_{\mu\nu}^{massless}(q^2) = -\frac{8\pi}{m_3^2}\frac{1}{q^2}\left(T_{\mu\nu} - \frac{1}{2}\eta_{\mu\nu}T_\alpha^\alpha\right). \qquad (3.3.14)$$

در حدی که جرم به صفر می رود، پنج درجه آزادی گرویتون جرم دار می تواند به یک تانسور بدون جرم (گرویتون)، یک بردار بدون جرم (گرویفوتون) و یک اسکالر بدون جرم تبدیل شود. این اسکالر به عنوان یک درجه آزادی اضافه برای همه رژیم های نظریه وجود خواهد داشت و یک نیروی جاذبه اضافه ایجاد خواهد کرد. در بخش 3.7 تعبیر این پتانسیل را در نظریه DGP خواهیم دید. در واقع این اسکالر مدل DGP را برای حد فواصل کوتاه به نظریه برنز – دیکی تبدیل می کند. پس نظریه گرانش

---

[14] Van Dam-Veltman-Zakharov discontinuity
[15] General Covariance

جرم دار حتی در حد جرم صفر نیز با نظریه گرانش معمول اینشتین متفاوت است. این تفاوت اساس فرمول بندی ناپیوستگی $VDVZ$ است. در بخش های بعد در مورد این ناپیوستگی صحبت خواهیم کرد. برای اینکه ببینیم ساختار تانسوری در مدل $DGP$ به چه صورتی است باید انتشارگر را در این مدل بیابیم. برای این کار نوسانات متریک را بصورت زیر تعریف می کنیم

$$G_{AB} = \eta_{AB} + h_{AB}. \tag{3.3.15}$$

همچنین ما برای توده از پیمانه هارمونیک استفاده خواهیم کرد

$$\partial^A h_{AB} = \frac{1}{2} \partial_B h_C^C. \tag{3.3.16}$$

براحتی می توان چک کرد که قرار دادن

$$h_{\mu 5} = 0, \tag{3.3.17}$$

با معادلات (3.3.1) سازگار است، لذا تنها مؤلفه های باقیمانده $h_{AB}$ عبارتند از $h_{\mu\nu}$ و $h_{55}$. در این پیمانه معادله اینشتین برای مؤلفه (55) بصورت زیر است

$$\partial_\mu \partial^\mu h_\nu^\nu = \partial_\mu \partial^\mu h_5^5, \tag{3.3.18}$$

که با استفاده از معادلات (3.3.16) و (3.3.17) می توان نوشت

$$\partial_A \partial^A h_\nu^\nu = \partial_B \partial^B h_5^5. \tag{3.3.19}$$

معادله اینشتین مربوط به مؤلفه های ($\mu\nu$) نیز بعد از اندکی جابجایی در جملات بصورت زیر در می آیند

$$\left( m_4^3 \partial_A \partial^A + m_3^2 \delta(y) \partial_\alpha \partial^\alpha \right) h_{\mu\nu}(x,y) = \left\{ T_{\mu\nu} - \frac{1}{3} \eta_{\mu\nu} T_\alpha^\alpha \right\} \delta(y) + m_3^2 \delta(y) \partial_\mu \partial_\nu h_5^5. \tag{3.3.20}$$

مشاهده می کنیم که ساختار این معادله گرویتون های جرم دار در چهار بعد یا به طور معادل گرویتون های بی جرم در پنج بعد را نشان می دهد. این معادله را می توان بصورت زیر نیز نوشت

۲۶

$$\left(m_4^3 \partial_A \partial^A + m_3^2 \delta(y) \partial_\alpha \partial^\alpha \right) h_{\mu\nu}(x,y)$$
$$= \left\{ T_{\mu\nu} - \frac{1}{2} \eta_{\mu\nu} T_\alpha^\alpha \right\} \delta(y) - \frac{1}{2} m_4^3 \eta_{\mu\nu} \partial_A \partial^A h_\alpha^\alpha + m_3^2 \delta(y) \partial_\mu \partial_\nu h_5^5, \quad (\text{۳.۳.۲۱})$$

که در آن طرف راست بصورت یک گرویتون بی جرم چهار بعدی (۳.۳.۱۱) و یک جمله اضافه مربوط به رد پتانسیل گرانشی $h_\alpha^\alpha$ نوشته شده است. از نقطه نظر گرانش چهار بعدی این مدل شامل یک گرویتون و یک اسکالر است. با فرایندی مشابه چیزی که در مدل اسباب بازی دنبال کردیم می توان $h_{\mu\nu}$ را در فضای تکانه بدست آورد

$$\tilde{h}_{\mu\nu}(p, y=0) = \frac{1}{m_3^2 p^2 + 2 m_4^3 p} \left[ \tilde{T}_{\mu\nu}(p^\lambda) - \frac{1}{3} \eta_{\mu\nu} \tilde{T}_\alpha^\alpha(p^\lambda) \right], \quad (\text{۳.۳.۲۲})$$

که در آن علامت مَد نشان دهنده کمیت تبدیل شده تحت تبدیل فوریه است. پتانسیل گرانشی برای یک میدان ماده به جرم $m$ در مقیاس های بسیار کوچکتر و بسیار بزرگتر از $r_0$ بصورت زیر خواهد بود

$$V_{grav} \sim -\frac{G_{brane} m}{r}, \qquad r \ll r_0, \quad (\text{۳.۳.۲۳a})$$

و

$$V_{grav} \sim -\frac{G_{bulk} m}{r^2}, \qquad r \gg r_0, \quad (\text{۳.۳.۲۳b})$$

که در آن تعریف کرده ایم

$$8\pi G_{brane} = m_3^{-2}, \qquad 8\pi G_{bulk} = m_4^{-3}. \quad (\text{۳.۳.۲۴})$$

پس رفتار پتانسیل در مقیاس های $r \ll r_0$ همان رفتار پتانسیل گرانشی نیوتنی در چهار بعد است و تغییر وقتی ایجاد می شود که به ناحیه $r_0 \ll r$ می رویم. در این حالت پتانسیل بصورت پنج بعدی رفتار خواهد کرد. در بخش ۳.۵ تخمینی برای مقدار $r_0$ ارائه می دهیم که بتواند همخوانی نظریه را با مشاهدات رصدی تضمین کند. برای اینکه پیش بینی های این نظریه با تجربیات ما

سازگار باشد نیاز داریم $r_0$ مقدار زیادی داشته باشد اما این بدین معنی است که $m_4$ عدد بسیار کوچکی باشد. برای مثال برای اینکه $r_0$ در حد شعاع هابل کنونی باشد یعنی $r_0 \sim H_0^{-1}$ باید داشته باشیم $m_4 \sim 100 MeV$. این مقدار کم $m_4$

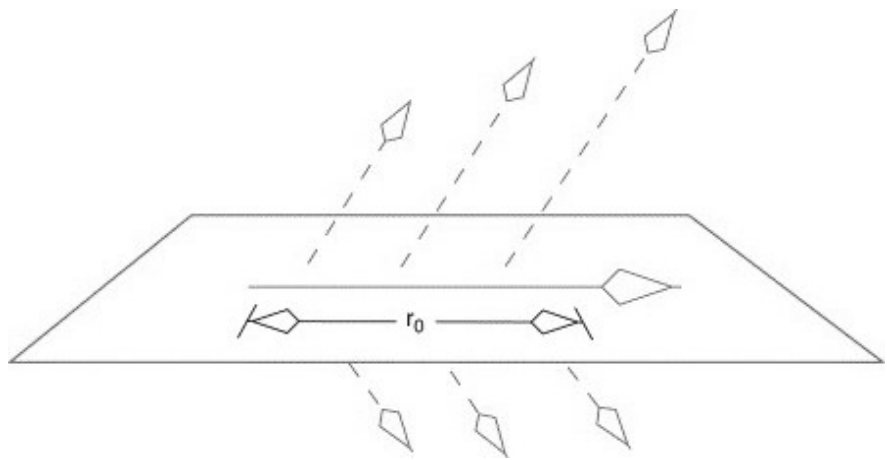

شکل (۳.۲) اگر فواصل خیلی کمتر از مقیاس $r_0$ باشد گرانش بطور مؤثری چهار بعدی است. اما اگر فواصل خیلی بیشتر از مقیاس $r_0$ باشد مقدار زیادی از انرژی گرویتون ها حین انتشار به بعد اضافه نفوذ کرده است و لذا گرانش پنج بعدی است. [برگرفته از مرجع

[A. Lue, Phys. Reports **423** (2006) 1

باعث می شود که ثابت گرانشی در توده بسیار زیاد باشد و این ایجاب می کند که نیروی گرانشی وارد بر یک منبع به جرم $m$ در توده بسیار قوی تر از نیروی گرانشی وارد بر آن در شامه باشد. محاسبات زیر به ما کمک می کند که بفهمیم مدل $DGP$ چگونه کار می کند و چرا پتانسیل گرانشی به این صورت عمل می کند. وقتی که $m_4 \gg m_3$ باشد یک نابرابری بزرگ بین انرژی نوسان های گرانشی در توده نسبت به شامه بوجود می آید. فرض کنید که یک میدان گرانشی با دامنه و اندازه معلوم داده شده باشد. می توانیم یک حساب سر انگشتی از انرژی های این مدل در توده و شامه ارائه دهیم.

$$E_{brane} \sim m_3^2 \int d^3x \, (\partial h)^2 \sim m_3^2 \times size, \qquad (3.3.25)$$

$$E_{bulk} \sim m_4^3 \int d^3x dy \, (\partial h)^2 \sim m_4^3 \times (size)^2 \sim E_{brane} \times \frac{size}{r_0}. \qquad (3.3.26)$$

اتفاقی که می افتد این است که با این که نوسانات گرانشی و میدان آزاد هستند که بدون هیچ محدودیتی در کل فضای پنج بعدی حرکت کنند اما دارای انرژی به مراتب کمتری در توده خواهند بود. یک موقعیت مشابه می تواند مطلب را روشن کند. فرض کنید که شامه یک ورقه فلزی غوطه ور در هوا باشد و فرض کنید که امواج صوتی نشان دهنده گرانش باشد. اگر یک نفر ضربه ای به ورقه فلزی بزند امواج صوتی آزادانه در فلز و همین طور در هوا منتشر خواهند شد اما انرژی این امواج در هوا بسیار پایین تر از انرژی آنها در فلز است. امواج صوتی در فلز بسیار آهسته ضعیف می شوند و لذا در فواصل کم طوری منتشر می شوند که گویی توده وجود ندارد. فقط وقتی که موج صوت فاصله بسیار طولانی را روی فلز می پیماید و به مقدار قابل توجهی ضعیف می شود ناظر روی موج صوتی می تواند بگوید که یک بعد اضافه وجود دارد، چون می بیند که مقداری از انرژی صوت در جای نامعلومی گم می شود (شکل 3.2) در نتیجه در فواصل کوتاه فیزیک صوت از بعد پایین است و در فواصل بلند از بعد بالا خواهد بود. مشابه این مورد در مدل DGP اتفاق می افتد [2].

## 3.4 کیهان شناسی DGP

یکی از مهمترین نتایج مدل DGP توضیح انبساط تندشونده جهان است که در این بخش به آن می پردازیم. کارهای اولیه در مورد کیهان شناسی مدل DGP حتی قبل از مقاله معروف DGP و در غالب نظریه های جهان شامه ای انجام شده است [12-14]. در اینجا ما خط سیر [11] را ادامه خواهیم داد که برای اولین بار نشان داد که چگونه از معادلات میدان DGP به حل های تند شونده برای انبساط جهان برسیم. برای بررسی کیهان شناسی در مدل DGP از المان طول تابع زمان زیر استفاده می کنیم

$$ds^2 = -N^2(\tau, y) d\tau^2 + A(\tau, y)^2 \delta_{ij} d\lambda^i d\lambda^j + B(\tau, y)^2 dy^2, \qquad (3.4.1)$$

که در آن $\tau$ زمان کیهانی است و $\lambda^i$ مختصات فضایی همراه[16] برای جهان قابل مشاهده ماست. $y$ مختصه بعد اضافه است. متریک مربوط به قسمت فضایی سه بعدی را تابع دلتای کرونکر اتخاذ کرده ایم چون به دنبال حل های کیهان شناسی برای یک فضای تخت هستیم. برای سادگی تمام ماده را روی شامه فرض می کنیم و توده را خالی در نظر می گیریم، لذا داریم

$$T_A^{\ B} = 0, \tag{3.4.2}$$

$$T^{\mu}_{\ \nu} = diag\,(-\rho, p, p, p). \tag{3.4.3}$$

همچنین فرض می کنیم ثابت کیهان شناسی صفر باشد. تانسور اینشتین در توده بصورت زیر خواهند بود

$$G^0_{\ 0} = \frac{3}{N^2}\left[\frac{\dot{A}}{A}\left(\frac{\dot{A}}{A}+\frac{\dot{B}}{B}\right)\right] - \frac{3}{B^2}\left[\frac{A''}{A}+\frac{A'}{A}\left(\frac{A'}{A}+\frac{B'}{B}\right)\right], \tag{3.4.4}$$

$$G^i_{\ j} = \frac{1}{N^2}\delta^i_{\ j}\left[\frac{2\ddot{A}}{A}+\frac{\ddot{B}}{B}-\frac{\dot{A}}{A}\left(\frac{2\dot{N}}{N}+\frac{\dot{A}}{A}\right)-\frac{\dot{B}}{B}\left(\frac{\dot{N}}{N}-\frac{2\dot{A}}{A}\right)\right]$$
$$-\frac{1}{B^2}\delta^i_{\ j}\left[\frac{N''}{N}+\frac{2A''}{A}+\frac{A'}{A}\left(\frac{2N'}{N}+\frac{A'}{A}\right)-\frac{B'}{B}\left(\frac{N'}{N}+\frac{2A'}{A}\right)\right], \tag{3.4.5}$$

$$G^5_{\ 5} = \frac{3}{N^2}\left[\frac{\ddot{A}}{A}-\frac{\dot{A}}{A}\left(\frac{\dot{N}}{N}-\frac{\dot{A}}{A}\right)\right] - \frac{3}{B^2}\left[\frac{A'}{A}\left(\frac{N'}{N}+\frac{A'}{A}\right)\right], \tag{3.4.6}$$

$$G^0_{\ 5} = 3\left[\frac{\dot{A}'}{A}-\frac{\dot{A}}{A}\frac{N'}{N}-\frac{\dot{B}}{B}\frac{A'}{A}\right], \tag{3.4.7}$$

که در آن نقطه نشانه مشتق گیری نسبت به زمان و پریم نشانه مشتق نسبت به بعد اضافه است. همچنین شامه را در $y = 0$ در نظر گرفته ایم. از آنجایی که می توان این انتخاب را به منزله انتخاب پیمانه مختصاتی در نظر گرفت لذا محدودیتی برای دینامیک شامه ایجاد نخواهد کرد. به خاطر تقارن $Z_2$ مولفه های متریک را می توان بصورت زیر نوشت

---

[16] comoving

۳۰

$$N = 1 \mp |y| \frac{\ddot{a}}{\dot{a}},$$
$$A = a \mp |y| \dot{a}, \qquad (3.4.8)$$
$$B = 1,$$

که در آن $a(\tau) = A(\tau, y = 0)$ همان ضریب مقیاس[17] چهار بعدی شامه و پارامتر باقیمانده مسئله است که باید معین شود. در آخر خواهیم دید که علامت های مثبت و منفی در (3.3.8) به ترتیب نمایانگر جهان در حال انبساط تند شونده و جهان $FRW$ هستند. برای بدست آوردن ضریب مقیاس $a(\tau)$ نیاز به شرایط مرزی داریم. این شرایط مرزی با انتگرال گیری از معادلات میدان نسبت به بعد اضافه و درست در نزدیکی شامه یا بطور معادل با شرایط اتصال Isreal بدست می آیند. در اینجا از راه کار اول استفاده می کنیم و بدست می آوریم

$$\left.\frac{N'}{N}\right|_{y=0} = -\frac{8\pi G}{3}\rho, \qquad (3.4.9)$$

$$\left.\frac{A'}{A}\right|_{y=0} = \frac{8\pi G}{3}(3p + 2\rho), \qquad (3.4.10)$$

که در آن G برابر است با [11]

$$G = \frac{4}{3} G_{brane}. \qquad (3.4.11)$$

با مقایسه این شرایط اتصال[18] با حل های (3.4.8) برای توده به یک جفت معادله برای ضریب مقیاس می رسیم که در واقع همان معادله فریدمن اصلاح شده برای مدل DGP و معادله بقای انرژی ـ تکانه هستند

$$H^2 \pm \frac{H}{r_0} = \frac{8\pi G}{3}\rho(\tau), \qquad (3.4.12)$$

$$\dot{\rho} + (3\rho + p)H = 0, \qquad (3.4.13)$$

---

[17] Scale factor
[18] Junction conditions

که در آن پارامتر هابل را بصورت زیر تعریف کرده ایم

$$H = \frac{\dot{a}}{a}.$$ (۳.۴.۱۴)

معادله بقای انرژی ـ تکانه (۳.۴.۱۳) در واقع همان معادله بقا در گرانش معمول اینشتینی است که با متریک $FRW$ بدست می آید. تغییر اساسی در معادله (۳.۴.۱۲) اتفاق افتاده است، و آن ظاهر شدن جمله جدید $\pm\frac{H}{r_0}$ در معادله فریدمن است. در بخش قبل دیدیم که در مدل $DGP$ گرانش در مقیاس های کوچک چهار بعدی و در مقیاس های بزرگ پنج بعدی خواهد بود.

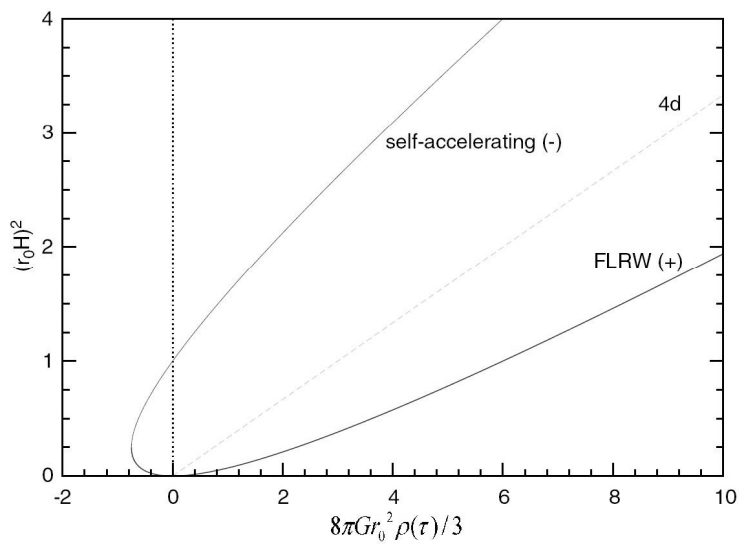

شکل (۳.۳) رسم نمودار $\left(r_0 H\right)^2$ بر حسب $\frac{r_0^2 \rho}{m_3^2}$. در این شکل بوضوح دو شاخه معادله (۳.۴.۱۲) از هم قابل تشخیص هستند. خطوط پر نشان دهنده جوابهای معادله (۳.۴.۱۲) هستند. خط نقطه چین نقطه چین حل معمول جهان $FRW$ است. [برگرفته از مرجع ۱ (2006) **423** A. Lue, Phys. Reports]

۳۲

در شکل (۳.۳) نمودار $\left(r_0 H\right)^2$ بر حسب $\dfrac{r_0^2 \rho}{m_3^2}$ کشیده شده است. مشاهده می کنیم که معادله فریدمن دو شاخه دارد که یکی مربوط به جهان $FRW$ و دیگری حل جهانی در حال انبساط تند شونده است. همان طور که گرانش در فواصل کم چهار بعدی و در فواصل زیاد پنج بعدی است پارامتر هابل نیز چنین تغییری خواهد کرد بطوری که در فواصل خیلی کمتر از $r_0$ که پارامتر هابل $H(\tau)$ مقدار بسیار بیشتری از $r_0^{-1}$ خواهد داشت سهم جمله $\pm\dfrac{H}{r_0}$ در مقایسه با $H^2$ در معادله (۳.۴.۱۲) بسیار ناچیز خواهد بود و معادله فریدمن بطور مؤثری چهار بعدی است، اما وقتی به محدوده $H(\tau) \sim r_0^{-1}$ می رسیم تغییرات در این معادله (که تحول $H(\tau)$ را مشخص می کند) مشهود خواهد بود. برای اینکه ببینیم شاخه پایینی معادله چگونه تند شونده شتاب جهان را توضیح می دهد فرض کنید که معادله حالت ماده بصورت

$$p = \alpha \rho, \qquad (۳.۴.۱۵)$$

باشد. معادله (۳.۴.۱۳) نتیجه می دهد

$$\rho = \dfrac{1}{a^{3+\alpha}}. \qquad (۳.۴.۱۶)$$

اگر $a \to \infty$ باشد بدست می آوریم $\rho \to 0$ و معادله (۳.۴.۱۲) نتیجه می دهد

$$H \to \dfrac{1}{r_0}. \qquad (۳.۴.۱۷)$$

اما این بدین معناست که $\ddot{a} > 0$ یعنی شتاب تندشونده جهان. تحول جهان به این صورت خواهد بود که وقتی جهان جوان است بصورت یک جهان $FRW$ با شتاب کند شونده منبسط می شود. با گذشت زمان و پخش شدن ماده در جهان مقدار پارامتر هابل کاهش می یابد. زمانی که پارامتر هابل به مقدار آستانه خود یعنی $r_0^{-1}$ می رسد این کاهش پایان می پذیرد و به یک مقدار ثابت میل می کند، حتی وقتی جهان از انرژی _ تکانه خالی شده است. برای اینکه بتوانیم تفاوت دو شاخه معادله (۳.۴.۱۲) را به تفصیل مطالعه کنیم ابتدا توجه کنید که می توان متریک توده (۳.۴.۱) را در یک هندسه مینکوفسکی نشاند[۲ و ۱۱]. با تعریف تبدیلات زیر متریک فضای توده بصورت

۳۳

$$ds^2 = -dT^2 + (dX^1)^2 + (dX^2)^2 + (dX^3)^2 + (dY^5)^2, \qquad (3.4.18)$$

در می آید

$$T = A(y,\tau)\left(\frac{\lambda^2}{4} + 1 - \frac{1}{4\dot{a}^2}\right) - \frac{1}{2}\int d\tau \frac{a^2}{\dot{a}^3}\frac{d}{d\tau}\left(\frac{\dot{a}}{a}\right)$$

$$Y^5 = A(y,\tau)\left(\frac{\lambda^2}{4} - 1 - \frac{1}{4\dot{a}^2}\right) - \frac{1}{2}\int d\tau \frac{a^2}{\dot{a}^3}\frac{d}{d\tau}\left(\frac{\dot{a}}{a}\right) \qquad (3.4.19)$$

$$X^i = A(y,\tau)\lambda^i$$

که در آن $\lambda^2 = \delta_{ij}\lambda^i\lambda^j$. برای اینکه به ساختار سراسری[19] مدل DGP پی ببریم توجه خود را به زمان های ابتدای جهان معطوف می کنیم. جهان را بصورت تابش غالب و با معادله حالت $p = \frac{1}{3}\rho$ در نظر می گیریم. معادلات (3.4.12) و (3.4.13) برای $H(\tau) \gg r_0^{-1}$ به حل $a(\tau) = \tau^{1/2}$ می انجامد. ساختار سراسری شامه با قرار دادن $y = 0$ در معادلات (3.4.19) بدست می آید

$$T = \tau^{\frac{1}{2}}\left(\frac{\lambda^2}{4} + 1 - \tau\right) - \frac{4}{3}\tau^{\frac{3}{2}},$$

$$Y^5 = \tau^{\frac{1}{2}}\left(\frac{\lambda^2}{4} - 1 - \tau\right) - \frac{4}{3}\tau^{\frac{3}{2}}, \qquad (3.4.20)$$

$$X^i = \tau^{\frac{1}{2}}\lambda^i.$$

معادله نقاطی که این معادلات را ارضاء می کنند بصورت

$$Y_+ = \frac{1}{4Y_-}\sum_{i=1}^{3}(X^i)^2 + \frac{1}{3}Y_-^3, \qquad (3.4.21)$$

---

[19] Global structure

۳۴

است که در آن تعریف کرده ایم $Y_\pm = \frac{1}{2}(T \pm Y^5)$. قابل توجه است که اگر فقط جمله اول را در عبارت بالا در نظر بگیریم این معادله همان مخروط نوری ایجاد شده از نقطه $0 = (T, X^i, Y^5)$ است. همچنین از معادله (۳.۴.۲۰) مشاهده می کنیم که

$$Y_- = \tau^{\frac{1}{2}}, \tag{۳.۴.۲۲}$$

یعنی $Y_-$ می تواند به عنوان یک مقیاس زمانی در نظر گرفته شود. Big bang مترادف با $\tau = 0$ یا همان خط $Y^5 = T$ است. در شکل (۳.۴) مخروط نوری مربوط به معادله (۳.۴.۲۱) رسم شده است. ناظر توده، شامه را به عنوان سطح یک ابر کره فشرده می بیند که از زمان انفجار بزرگ شروع به انبساط کرده است. با اینکه ناظر توده شامه را فشرده می بیند اما ناظر در شامه جهان خود را از نظر فضایی فشرده نخواهد دید. همزمان ناظر در توده مشاهده می کند که چگالی انرژی روی شامه در حال تغییر است اما ناظر روی شامه در هر برش زمانی جهان خود را از نظر فضایی همگن می بیند. در شکل (۳.۵) تصویر شماتیک این انبساط نشان داده شده است.

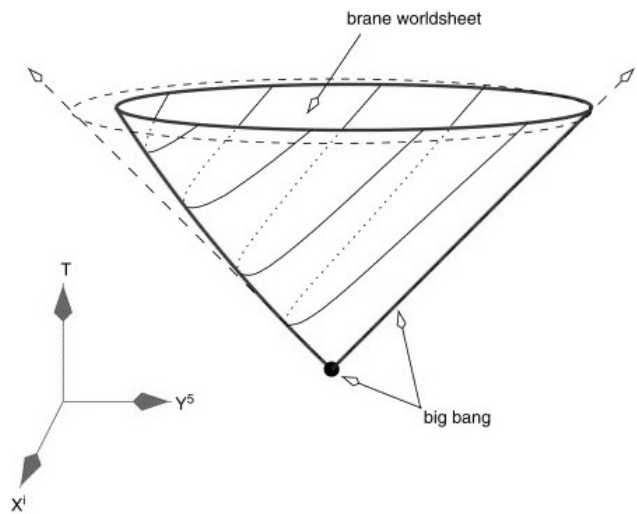

شکل (۳.۴) نمودار مربوط به معادله (۳.۴.۲۱). زمان در توده بصورت عمودی و رو به بالا نشان داده شده است. بقیه مختصات بصورت فشرده و عمود بر آن نشان داده شده اند. Big bang مترادف با خط $Y^5 = T$ است. خطوط نقطه دار مخروط نوری آینده برای رویداد در مبداء است که با یک نقطه بزرگ نشان داده شده است[برگرفته از مرجع ۱ (2006) 423 A. Lue, Phys. Reports].

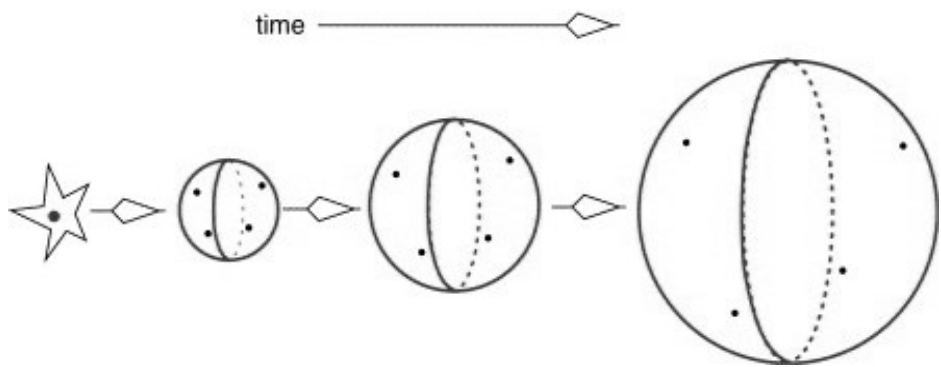

شکل (۳.۵) تحول شامه به صورت مقطع های زمانی نشان داده شده است. مشاهده می کنیم که انفجار بزرگ بصورت یک تکینگی در توده است و با گذشت زمان بصورت یک موج شوک نسبیتی منبسط می شود. در شکل قبل دیدیم که انفجار بزرگ یک تکینگی نورگونه است. پس هر نقطه روی شامه در هر لحظه باید اینگونه باشد و با سرعت نور حرکت کند [برگرفته از مرجع

[ A. Lue, Phys. Reports **423** (2006) 1

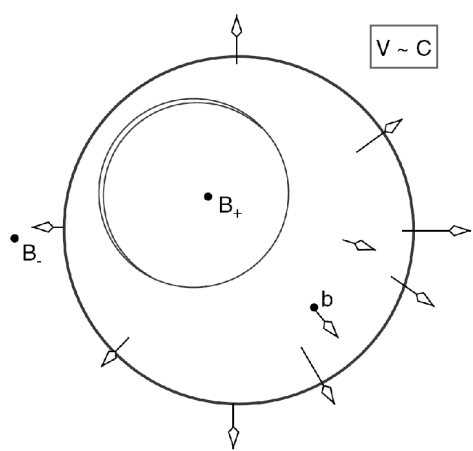

شکل (۳.۶) سطح شامه در فضای توده در این شکل نشان داده شده است. مشاهده می کنیم که از نقطه نظر ناظر داخل توده داخل شامه، شامه در حال دور شدن از او است. اما از نقطه نظر ناظر توده که بیرون از شامه قرار دارد، شامه در حال نزدیک شدن به او است. برای توضیح بیشتر به متن مراجعه کنید [برگرفته از مرجع 1 (2006) **423** A. Lue, Phys. Reports ]

شاخه FRW در واقع مربوط به قسمتی از توده است که در داخل فضای بسته شامه در حال انبساط قرار دارد. در این حالت کل توده دو کپی از این فضای داخلی است و ناظر در توده مشاهده می کند که شامه از او دور می شود. در این حالت یک راه سریعتر برای ارتباط بین دو نقطه در شامه استفاده از فضای توده خواهد بود. فضای توده در شاخه انبساط تندشونده در واقع مربوط به قسمت بیرونی فضای شامه است و کل فضای توده دو کپی از این فضاست که بوسیله شامه از هم جدا شده است. در این حالت ناظر در توده مشاهده می کند که شامه به او نزدیک می شود و وقتی شعاع این شامه قابل مقایسه با $r_0$ می شود مشاهده می کند که شامه با شتاب تندشونده شروع به نزدیک شدن به او می کند. در این حالت به خاطر هندسه خاص توده نزدیک ترین راه برای اتصال دو نقطه روی شامه این است که از روی خود شامه مسیر را بپیماییم. شکل (۳.۶) گفته های بالا را روشن تر می کند.

## ۳.۵ قید های مشاهداتی

در این بخش فاصله درخشندگی را برای مدل DGP بدست می آوریم و نتایج مدل را با نتایج حاصل از فرض وجود انرژی تاریک مقایسه می کنیم. همان طور که در مقدمه گفته شد انبساط تندشونده جهان را می توان با اضافه کردن یک چگالی انرژی با فشار منفی به سیستم توضیح داد. در این حالت معادله فریدمن بصورت زیر نوشته می شود

$$H^2 = \frac{8\pi G}{3}(\rho_M + \rho_D). \qquad (۳.۵.۱)$$

انرژی تاریک دارای معادله حالت $p_D = \omega \rho_D$ با تابع چگالی انرژی بصورت

$$\rho_D(\tau) = \rho_D^0 a^{-3(1+\omega)}, \qquad (۳.۵.۲)$$

است. در این معادلات $\omega$ مقداری ثابت است و $\omega = -1$ نتایج رصدی را به خوبی توضیح می دهد. معادله فریدمن در مدل DGP بصورت

$$H^2 \pm \frac{H}{r_0} = \frac{8\pi G}{3}\rho(\tau), \qquad (۳.۵.۳)$$

است که علامت منفی همان طور که گفته شد می تواند شتاب تندشونده جهان را توضیح دهد. می توانیم چگالی انرژی تاریک مؤثر در مدل DGP را بصورت زیر تعریف کنیم

$$\frac{8\pi G}{3} \rho_D^{eff} = \frac{H}{r_0}. \tag{3.5.4}$$

محاسبه فاصله درخشندگی در مدل DGP ابتدا در مراجع [30 و31] انجام شده است. از تعریف (3.5.4) این طور بر می آید که با گذشت زمان چگالی انرژی تاریک در این مدل باید کاهش بیابد. لذا این مدل باید مترادف با مقدار $\omega$ مؤثر بیشتر از $-1$ باشد. اگر فرض کنیم که جهان از نظر فضایی تخت است و پارامتر $\Omega_M$ را بصورت معمول

$$\Omega_M = \Omega_M^0 (1+z)^3, \tag{3.5.5}$$

تعریف کنیم که در آن

$$\Omega_M^0 = \frac{8\pi G \rho_M^0}{3H_0^2}, \tag{3.5.6}$$

و صفر به معنی مقدار کنونی کمیت است. معادله فریدمن (3.5.3) حکم می کند که داشته باشیم

$$1 = \Omega_M + \Omega_{r_0}, \tag{3.5.7}$$

که در آن مؤلفه مؤثر انرژی تاریک را بصورت

$$\Omega_{r_0} = \frac{1}{r_0 H}, \tag{3.5.8}$$

تعریف کرده ایم. اگر $\omega$ مؤثر تابع زمان است. اگر $p_D^{eff} = \omega_{eff}(z) \rho_D^{eff}$ بصورت تعریف کنیم با استفاده از معادلات (3.4.12) و (3.4.13) خواهیم داشت

$$\omega_{eff}(z) = -\frac{1}{1+\Omega_M}. \tag{3.5.9}$$

با تعاریف بالا معادله فریدمن را می توان بصورت

$$\frac{H(z)}{H_0} = \frac{1}{2}\left[\frac{1}{r_0 H_0} + \sqrt{\left(\frac{1}{r_0 H_0}\right)^2 + 4\Omega_M^0 (1+z)^3}\right], \quad (3.5.10)$$

نوشت که با استفاده از رابطه

$$r_0 H_0 = \frac{1}{1-\Omega_M^0}, \quad (3.5.11)$$

فقط تابع یک پارامتر آزاد است. فاصله درخشندگی بصورت زیر تعریف می شود

$$d_L^{DGP}(z) = (1+z)\int_0^z \frac{dz}{H(z)} = (1+z)\int_0^z \frac{2H_0^{-1}dz}{\frac{1}{r_0 H} + \sqrt{\left(\frac{1}{r_0 H}\right)^2 + 4\Omega_M^0 (1+z)^3}}, \quad (3.5.12)$$

که از (3.5.10) استفاده کرده ایم. فاصله درخشندگی برای مدل انرژی تاریک بصورت زیر است

$$d_L^\omega(z) = (1+z)\int_0^z \frac{H_0^{-1}dz}{\sqrt{4\Omega_M^\omega (1+z)^3 + \left(1-\Omega_M^\omega\right)(1+z)^{3(1+\omega)}}}. \quad (3.5.13)$$

در شکل (3.7) این دو فاصله را با هم مقایسه کرده ایم. مشاهده می کنیم که برای اهداف عملی این دو مدل در واقع از هم قابل تشخیص نیستند. با استفاده از روابط بالا و داده های رصدی می توان تخمینی برای مقدار $r_0$ بدست آورد.

مشاهدات رصدی مقدار $\Omega_M^0$ را بصورت زیر بدست می دهند

$$\Omega_M^0 = 0.18_{-0.06}^{+0.07}. \quad (3.5.14)$$

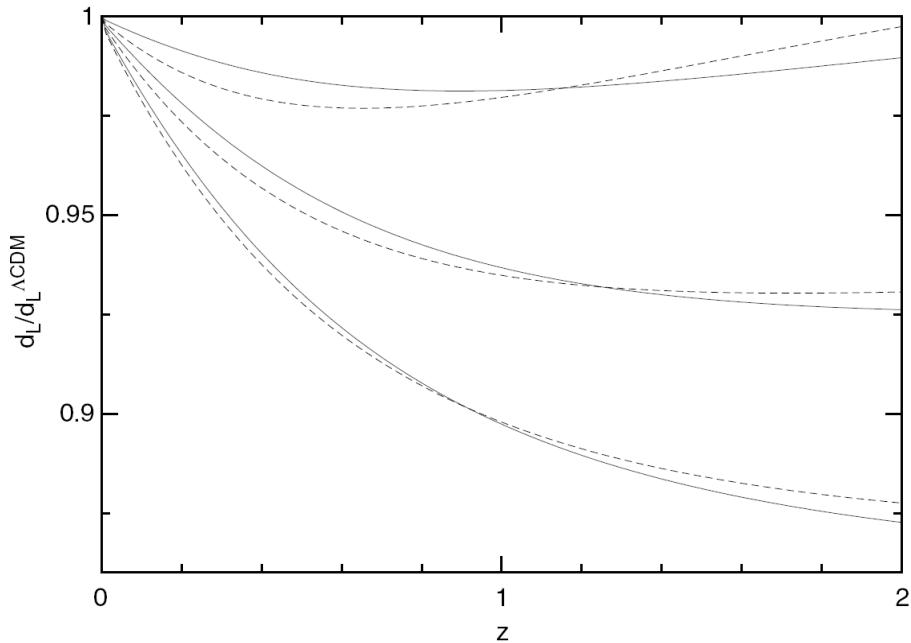

شکل (3.7) نمودارهای خط چین مربوط به مدل انرژی تاریک است. نمودارهای خط پر مربوط به پیش بینی مدل DGP برای فاصله درخشندگی است. هر دو نمودار بر واحد فاصله درخشندگی برای $\Lambda CDM$ کشیده شده اند.[برگرفته از مرجع 1  (2006) **423** A. Lue, Phys. Reports]

با استفاده از معادله (3.5.7) بدست می آوریم

$$r_0 = 1.21^{+0.09}_{-0.09} H_0^{-1}. \tag{3.5.15}$$

همان طور که از ابتدا انتظار داشتیم مقیاس فاصله $r_0$ هم مرتبه با شعاع هابل بدست آمده است. برای اینکه تخمینی از اندازه رابطه بالا ارائه دهیم فرض کنید که مقدار پارامتر هابل برابر $H_0 \sim 70\, km\, s^{-1} Mpc^{-1}$ باشد. در این صورت مقدار $r_0$ برابر با $r_0 \sim 5 Gpc$ خواهد بود.

## ۳.۶ بازیابی معادلات اینشتین

همان طور که دیدیم معادلات $DGP$ در مقیاس های خیلی کوچکتر از $r_0$ به یک نظریه چهار بعدی تبدیل می شوند. در این بخش متریک را برای یک منبع ایستا، فشرده و کروی بدست می آوریم و آن را در حد $\frac{r}{r_0} \to 0$ مطالعه خواهیم کرد.

المان طول را بصورت زیر انتخاب می کنیم

$$ds^2 = N^2(r,y)dt^2 - A^2(r,y)dr^2 - B^2(r,y)\left[d\theta^2 + \sin^2\theta d\varphi^2\right] - dy^2, \qquad (۳.۶.۱)$$

که شامل عمومی ترین فرم یک متریک ایستا با تقارن کروی روی شامه است. مولفه های تانسور اینشتین در توده بصورت زیر است

$$G^t_{\ t} = \frac{1}{B^2} - \frac{1}{A^2}\left[\frac{2B''}{B} - \frac{2A'B'}{AB} + \frac{B'^2}{B^2}\right] - \left[\frac{\ddot{A}}{A} + \frac{2\ddot{B}}{B} + 2\frac{\dot{A}\dot{B}}{AB} + \frac{\dot{B}^2}{B^2}\right], \qquad (۳.۶.۲)$$

$$G^r_{\ r} = \frac{1}{B^2} - \frac{1}{A^2}\left[\frac{N'B'}{NB} + \frac{B'^2}{B^2}\right] - \left[\frac{\ddot{N}}{N} + \frac{2\ddot{B}}{B} + 2\frac{\dot{N}\dot{B}}{NB} + \frac{\dot{B}^2}{B^2}\right], \qquad (۳.۶.۳)$$

$$G^\theta_{\ \theta} = G^\varphi_{\ \varphi} = -\frac{1}{A^2}\left[\frac{N''}{N} + \frac{B''}{B} - \frac{N'A'}{NA} + \frac{N'B'}{NB} - \frac{A'B'}{AB}\right]$$
$$- \left[\frac{\ddot{N}}{N} + \frac{\ddot{A}}{A} + \frac{\ddot{B}}{B} + \frac{\dot{N}\dot{A}}{NA} + \frac{\dot{N}\dot{B}}{NB} + \frac{\dot{A}\dot{B}}{AB}\right], \qquad (۳.۶.۴)$$

$$G^y_{\ y} = \frac{1}{B^2} - \frac{1}{A^2}\left[\frac{N''}{N} + \frac{2B''}{B} - \frac{N'A'}{NA} + 2\frac{N'B'}{NB} - 2\frac{A'B'}{AB} + \frac{B'^2}{B^2}\right]$$
$$- \left[\frac{\dot{N}\dot{A}}{NA} + 2\frac{\dot{N}\dot{B}}{NB} + 2\frac{\dot{A}\dot{B}}{AB} + \frac{\dot{B}^2}{B^2}\right], \qquad (۳.۶.۵)$$

$$G^y_{\ r} = -\left[\frac{\dot{N}'}{N} + \frac{2\dot{B}'}{B}\right] - \frac{\dot{A}}{A}\left[\frac{\dot{N}}{N} + 2\frac{\dot{B}}{B}\right], \qquad (۳.۶.۶)$$

۴۱

که در آن پریم نشان دهنده مشتق نسبت به $r$ و نقطه نشانه مشتق گیری نسبت به بعد اضافه است. حال فرض می کنیم که توده خالی است و تانسور انرژی ـ تکانه برای شامه بصورت

$$T^{\mu}_{\nu} = diag(\rho(r), p(r), p(r), p(r)), \qquad (3.6.7)$$

است و ثابت کیهان شناسی را نیز برابر با صفر در نظر می گیریم. با اعمال تقارن $Z_2$ و استفاده از پیمانه $r = B\big|_{y=0}$ و برابر قرار دادن ضریب $\delta(y)$ در دو طرف معادلات بالا به معادلات زیر برای شامه می رسیم

$$-\left(\frac{\dot{A}}{A} + \frac{2\dot{B}}{B}\right) = \frac{r_0}{A^2}\left[-\frac{2}{r}\frac{A'}{A} + \frac{1}{r^2}(1-A^2)\right] + \frac{8\pi r_0}{M_P^2}\rho(r), \qquad (3.6.8)$$

$$-\left(\frac{\dot{N}}{N} + \frac{2\dot{B}}{B}\right) = \frac{r_0}{A^2}\left[\frac{2}{r}\frac{N'}{N} + \frac{1}{r^2}(1-A^2)\right] - \frac{8\pi r_0}{M_P^2}p(r), \qquad (3.6.9)$$

$$-\left(\frac{\dot{N}}{N} + \frac{\dot{A}}{A} + \frac{\dot{B}}{B}\right) = \frac{r_0}{A^2}\left[\frac{N''}{N} + \frac{N'A'}{NA} + \frac{1}{r}\left(\frac{N'}{N} - \frac{A'}{A}\right)\right] - \frac{8\pi r_0}{M_P^2}p(r), \qquad (3.6.10)$$

که این سه معادله از رابطه های مربوط به $G_{tt}, G_{rr}, G_{\theta\theta}$ بدست آمده اند، و همه ی کمیت ها در $y=0$ محاسبه می شوند. مشاهده می کنیم که جملات اول طرف راست در معادلات بالا به خاطر بزرگی $r_0$ سهم ناصفر در معادلات خواهند داشت و این دلیل اصلی انحراف حل های معادلات بالا از حل های معمول اینشتینی است. برای ادامه فرض می کنیم که $\rho(r)$ توزیع ماده ایستا باشد. شعاع مؤثر شوارتزشیلد را بصورت زیر تعریف می کنیم

$$R_g(r) = \frac{8\pi}{m_3^2}\int_0^r r^2 \rho(r)dr, \qquad (3.6.11)$$

و قرار می دهیم $r_g = R_g(r \to \infty)$. در اینجا ما به منابع ضعیف ماده مند علاقه مند هستیم. همچنین به قسمت هایی از فضا ـ زمان نگاه خواهیم کرد که انحراف مؤلفه های متریک از مشابه مینکوفسکی آنها کوچک باشد. لذا توابع $\{n(r,y), a(r,y), b(r,y)\}$ را به صورت زیر تعریف می کنیم

۴۲

$$N(r,y) = 1 + n(r,y), \tag{3.6.12}$$

$$A(r,y) = 1 + a(r,y), \tag{3.6.13}$$

$$B(r,y) = r\left[1 + b(r,y)\right]. \tag{3.6.14}$$

چون مطالعه ما بر روی متریک شامه است لذا پیمانه ای را انتخاب می کنیم که در آن $b(r, y = 0) = 0$ باشد و متریک روی شامه به صورت زیر در آید

$$ds^2 = \left[1 + n(r)\right]^2 dt^2 - \left[1 + a(r)\right]^2 dr^2 - r^2 d\Omega, \tag{3.6.15}$$

که در آن کمیت ها را در $y = 0$ محاسبه کرده ایم یعنی $n(r) = n(r, y = 0)$. این همان متریک استاندارد با دو پتانسیل $n(r)$ و $a(r)$ مربوط به پتانسیل نیوتنی و پتانسیل گرویتومغناطیسی است. حال روی این دو پتانسیل بحث خواهیم کرد. از آنجایی که به دنبال حل های اختلالی نسبت به فضای تخت مینکوفسکی می گردیم، با فرض $n(r,y), a(r,y), b(r,y) \ll 1$ و قرار دادن معادلات (3.6.2)-(3.6.6) در معادلات (3.6.8)-(3.6.10) مربوط به شامه تا مرتبه اول اختلال بدست می آوریم

$$-(\dot{a} + 2\dot{b}) = r_0\left[-\frac{2a'}{r} - \frac{2a}{r^2}\right] + \frac{r_0}{r^2} R'_g(r), \tag{3.6.16}$$

$$-(\dot{n} + 2\dot{b}) = r_0\left[-\frac{2n'}{r} - \frac{2a}{r^2}\right], \tag{3.6.17}$$

$$-(\dot{n} + \dot{a} + \dot{b}) = r_0\left[n'' + \frac{n'}{r} - \frac{a'}{r^2}\right]. \tag{3.6.18}$$

جملات فشار و چگالی انرژی از معادلات (3.5.17)-(3.5.18) حذف شده اند چون همگی از مرتبه صفر هستند. معادله حاکم بر فشار و چگالی انرژی طبق معمول از پایستگی انرژی بدست می آید

$$p'_g = n' \rho_g. \tag{3.6.19}$$

۴۳

حال فرض کنید که $g_{AB} = \eta_{AB} + h_{AB}$ باشد که در آن $h_{AB}$ متریک اختلالی است. معادله حاکم بر $h_{\mu\nu}$ با تبدیل فوریه با معادله (۳.۳.۲۲) داده می شود. برای حل شوارتزشیلد این رابطه ایجاب می کند که برای $r \gg r_0$ داشته باشیم

$$h_{00} = -\frac{4}{3}\frac{r_0 r_g}{r^2},$$
$$h_{xx} = h_{yy} = h_{zz} = -\frac{2}{3}\frac{r_0 r_g}{r^2},$$
(۳.۶.۲۰)

و برای $r \ll r_0$ داشته باشیم

$$h_{00} = -\frac{4}{3}\frac{r_g}{r^2},$$
$$h_{xx} = h_{yy} = h_{zz} = -\frac{2}{3}\frac{r_g}{r^2}.$$
(۳.۶.۲۱)

اگر معادله آخر را برای پتانسیل ها بنویسیم بدست می آوریم

$$n(r) = -\frac{4}{3}\frac{r_g}{2r},$$
$$a(r) = +\frac{2}{3}\frac{r_g}{2r}.$$
(۳.۶.۲۲)

در واقع این پتانسیل ها مشابه پتانسیل های نظریه برنز- دیکی[20] هستند که در آن قرار داده ایم $\omega = 0$. برای نظریه گرانش معمول اینشتینی این پتانسیل ها به ترتیب برابرند با $-r_g/2r$ و $+r_g/2r$. همان طوری که در [۳۲] بدست آمده است با اینکه جملات غیر خطی در معادلات (۳.۶.۸) تا (۳.۶.۱۰) در حد $r \gg r_0$ قابل صرف نظر کردن هستند اما جملاتی مانند $\frac{A'}{A}\frac{N'}{N}$ در معادلات (۳.۶.۲) تا (۳.۶.۶) تنها وقتی قابل صرف نظر کردن هستند که داشته باشیم

$$r \gg r_\star \equiv \left(r_g r_0^2\right)^{\frac{1}{3}}.$$
(۳.۶.۲۳)

---
[20] Brans-Dicke

۴۴

به $r_\star$ شعاع واینشتاین[21] می‌گوییم. این شعاع جدید به ما کمک خواهد کرد که حد چهار بعدی مدل DGP را بهتر بشناسیم. در جدول زیر مقدار $r_\star$ برای زمین، خورشید و کهکشان راه شیری آمده است

جدول (3.1) مقادیر $r_\star$ برای زمین، خورشید و کهکشان راه شیری.

| | |
|---|---|
| Earth | $1.2\, pc$ |
| Sun | $150\, pc$ |
| Milky Way ($10^{12}\, M_\odot$) | $1.2\, Mpc$ |

در واقع معادلات (3.6.22) در حد $r_\star \ll r \ll r_0$ به دست آمده اند. حال حد $r \ll r_\star$ را در نظر می گیریم. در این حد فقط کافی است که جملات خاصی را در (3.6.2) تا (3.6.6) در نظر بگیریم [32 و 34]. این جملات در واقع آنهایی هستند که یکی از جملات $\dfrac{\dot{B}}{B}$ یا $\dfrac{\dot{A}}{A}$ در آنها وجود داشته باشد. اگر یک منبع جرم نقطه ای در نظر بگیریم بطوریکه $R_g = r_g = const.$ باشد پتانسیل های $n(r)$ و $a(r)$ را بصورت زیر بدست می آوریم

$$n(r) = -\frac{r_g}{2r} + \sqrt{\frac{r_g r}{2r_0^2}}, \qquad (3.6.24)$$

$$a(r) = +\frac{r_g}{2r} - \sqrt{\frac{r_g r}{8r_0^2}}. \qquad (3.6.25)$$

از آنجایی که برای این مسئله داریم $r_g \ll r$ براحتی مشاهده می کنیم که جملات دوم در معادلات بالا قابل صرف نظر کردن هستند و لذا این دو معادله در این حد همان پتانسیل های مربوط به نسبیت عام معمولی هستند. در پایان نتایج بدست آمده را

---
[21] Veinstein

خلاصه می کنیم. مشاهده کردیم که دو مقیاس فاصله ای برای مدل وجود دارد $r_0$ و $r_\star$. در حد $r \gg r_0$ مدل DGP یک مدل گرانشی در پنج بعد است، در حد میانی $r_\star \ll r \ll r_0$ یک مدل گرنش چهار بعدی اما بصورت نظریه برنز – دیکی با پارامتر $\omega = 0$ است و در حد $r \ll r_\star$ به گرانش چهار بعدی اینشتینی کاهش می یابد.

## ۳.۷ معادلات میدان برای فضای استاتیک با تقارن کروی

در این بخش معادلات میدان $DGP$ را برای یک فضا با تقارن کروی و استاتیک بدست می آوریم. در فصل بعد از این معادلات برای بدست آوردن قضیه ویریال در مدل $DGP$ استفاده خواهیم کرد. برای سادگی فرض می کنیم که تانسورهای انرژی ـ تکانه ی توده و شامه به ترتیب بصورت زیر باشند

$$\hat{T}^A{}_B = diag\left(-\rho_B, p_B, p_B, p_B, p_B\right), \qquad (۳.۷.۱)$$

$$T^\mu{}_\nu = diag\left(-\rho_b, p_b, p_b, p_b\right), \qquad (۳.۷.۲)$$

که در آن اندیس $B$ و b به ترتیب مربوط به توده و شامه هستند. المان طول پنج بعدی مورد نظر ما برای فضای توده بصورت زیر است

$$ds^2 = -e^{\nu(r,y)}dt^2 + e^{\mu(r,y)}dr^2 + r^2\left(d\theta^2 + \sin^2\theta d\varphi^2\right) + e^{\lambda(r,y)}dy^2. \qquad (۳.۷.۳)$$

همان طور که گفته شد متریک فضای شامه متریک القایی از توده است لذا برای شامه داریم

$$ds^2 = -e^{\nu_0(r)}dt^2 + e^{\mu_0(r)}dr^2 + r^2\left(d\theta^2 + \sin^2\theta d\varphi^2\right), \qquad (۳.۷.۴)$$

که در آن عبارت های با اندیس صفر همان عبارت های موجود در (۳.۲.۶) هستند که در مکان شامه $y=0$ محاسبه شده اند. همان طور که گفته شد در مدل $DGP$ فرض می کنیم که توده حول شامه دارای تقارن $Z_2$ است. برای اعمال این تقارن توجه

۴۶

می‌کنیم که اگر $f\left(|y|\right)$ و $g\left(|y|\right)$ دو تابع دلخواه در توده باشند (برای سادگی چهار متغیر اول مشترک با شامه حذف شده اند، همچنین توجه کنید که تمام توابع در توده نسبت به بعد پنجم به خاطر تقارن $Z_2$ فقط تابع $|y|$ هستند) [10] داریم

$$\frac{df}{dy} = \dot{f}\frac{d|y|}{dy} = \dot{f}\left[2\theta(y)-1\right], \tag{3.7.5}$$

$$\frac{d^2f}{dy^2} = \ddot{f} + 2\dot{f}\delta(y), \tag{3.7.6}$$

$$\left(\frac{df}{dy}\right)\left(\frac{dg}{dy}\right) = \dot{f}\dot{g}, \tag{3.7.7}$$

که در آن $\dot{f} = \frac{df}{d|y|}$ است. معادلات میدان در توده بعد از جایگذاری متریک‌های (3.7.3) و (3.7.4) در معادلات (3.2.7) و استفاده از روابط بالا بصورت زیر در می‌آیند

(3.7.8)

$$\frac{m_4^3}{4r^2}\Big[\left(4+4r\lambda'+2\lambda''r^2+\lambda'^2r^2-4r\mu'-\mu'\lambda'r^2\right)e^{-\mu}+\left(2\ddot{\mu}r^2+4\dot{\mu}r^2\delta(y)+\dot{\mu}^2r^2-\dot{\mu}\dot{\lambda}r^2\right)e^{-\lambda}-4\Big]$$
$$-\frac{m_3^2}{r^2}e^{-\mu_0}\Big[r\mu_0'+e^{\mu_0}-1\Big]\delta(y) = -(\rho_b+m_3^2\Lambda)\delta(y)-\rho_B,$$

$$\frac{m_4^3}{4r^2}\Big[\left(4+4r\nu'+4r\lambda'+\nu'\lambda'r^2\right)e^{-\mu}+\left(2\ddot{\nu}r^2+4\dot{\nu}r^2\delta(y)+\dot{\nu}^2r^2-\dot{\nu}\dot{\lambda}r^2\right)e^{-\lambda}-4\Big]$$
$$+\frac{m_3^2}{r^2}e^{-\mu_0}\Big[r\nu_0'-e^{\mu_0}+1\Big]\delta(y) = (p_b-m_3^2\Lambda)\delta(y)+p_B, \tag{3.7.9}$$

$$\frac{m_4^3}{4r}[\left(2\nu'-2\mu'+2\lambda'+2\nu''r+\nu'^2r+2\lambda''r+\lambda'^2r-\mu'\nu'r+\nu'\lambda'r-\mu'\lambda'r\right)e^{-\mu}$$
$$+\left(2\ddot{\nu}r+4\dot{\nu}r\delta(y)+\dot{\nu}^2r+2\ddot{\mu}r+4\dot{\mu}r\delta(y)+\dot{\mu}^2r+\dot{\nu}\dot{\mu}r-\dot{\nu}\dot{\lambda}r-\dot{\mu}\dot{\lambda}r\right)e^{-\lambda}] \tag{3.7.10}$$
$$+\frac{m_3^2}{4r}e^{-\mu_0}\Big[2\nu_0'-2\mu_0'-\nu_0'\mu_0'r+2\nu_0''r+\nu'^2r\Big]\delta(y) = (p_b-m_3^2\Lambda)\delta(y)+p_B,$$

۴۷

$$\frac{m_4^3}{4r^2}\Big[\big(4-4r\mu'+4r\nu'\mu'r^2+\nu'^2 r^2+2\nu'r^2\big)e^{-\mu}+\dot\nu\dot\mu r^2 e^{-\lambda}-4\Big]=p_B, \qquad (3.7.11)$$

$$\frac{m_4^3}{4r}\big(2\dot\nu'r+\nu'\dot\nu r-\dot\mu\nu'r-\lambda'\dot\nu r-4\dot\mu\big)\big(2\theta(y)-1\big)=0, \qquad (3.7.12)$$

که در آن مشتق نسبت به $r$ را با پریم نشان داده ایم. برای بدست آوردن معادلات میدان برای شامه ضرایب ($\delta(y)$) را در دو طرف معادلات (3.7.8)-(3.7.12) برابر قرار می دهیم. اگر تعریف کنیم

$$\mathcal{U}(r)=m_4^3\,\dot\mu\Big|_{y=0}e^{-\lambda_0}, \qquad (3.7.13)$$

$$\mathcal{P}(r)=-m_4^3\,\dot\nu\Big|_{y=0}e^{-\lambda_0}, \qquad (3.7.14)$$

خواهیم داشت

$$m_3^2 e^{-\mu_0}\left(\frac{\mu_0'}{r}-\frac{1}{r^2}+\frac{e^{\mu_0}}{r^2}\right)=\rho_b(r)+\mathcal{U}(r)+m_3^2\Lambda, \qquad (3.7.15)$$

$$m_3^2 e^{-\mu_0}\left(\frac{\nu_0'}{r}+\frac{1}{r^2}-\frac{e^{\mu_0}}{r^2}\right)=p_b(r)+\mathcal{P}(r)-m_3^2\Lambda, \qquad (3.7.16)$$

$$m_3^2 e^{-\mu_0}\left(\frac{\nu_0'}{2r}-\frac{\mu_0'}{2r}-\frac{\mu_0'\nu_0'}{4}+\frac{\nu_0'}{2}+\frac{\nu_0'^2}{4}\right)=p_b(r)+\big[\mathcal{P}(r)-\mathcal{U}(r)\big]-m_3^2\Lambda. \qquad (3.7.17)$$

در واقع معادلات (3.7.13) و (3.7.14) یک ماده جدید برای جهان تعریف می کند که صرفاً از بعد اضافه آمده است. معادلات پایستگی برای این مدل به صورت صفر شدن مشتق کواریانت طرف راست معادلات بالا خواهد بود. این معادلات به صورت زیر هستند

$$\nu_0'=-2\frac{p_b'+\left(\mathcal{P}'+\dfrac{\mathcal{U}'}{r}\right)}{(\rho_b+p_b)+(\mathcal{P}+\mathcal{U})}, \qquad (3.7.18)$$

۴۸

$$\mathcal{U} = 0. \tag{3.7.19}$$

شرط آخر، معادلات اینشتین روی شامه را بسیار ساده خواهد کرد. معادلات اینشتین در مدل $DGP$ روی شامه به صورت زیر خواهد بود

$$m_3^2 e^{-\mu_0}\left(\frac{\mu_0'}{r} - \frac{1}{r^2} + \frac{e^{\mu_0}}{r^2}\right) = \rho_b(r) + m_3^2 \Lambda, \tag{3.7.20}$$

$$m_3^2 e^{-\mu_0}\left(\frac{\nu_0'}{r} + \frac{1}{r^2} - \frac{e^{\mu_0}}{r^2}\right) = p_b(r) + \mathcal{P}(r) - m_3^2 \Lambda, \tag{3.7.21}$$

$$m_3^2 e^{-\mu_0}\left(\frac{\nu_0'}{2r} - \frac{\mu_0'}{2r} - \frac{\mu_0' \nu_0'}{4} + \frac{\nu_0''}{2} + \frac{\nu_0'^2}{4}\right) = p_b(r) + \mathcal{P}(r) - m_3^2 \Lambda. \tag{3.7.22}$$

از این معادلات در فصل چهار برای بدست آوردن قضیه ویریال در مدل $DGP$ استفاده خواهیم کرد. این معادلات نشان می دهند که در این مدل نیز با همان معادلات اینشتین معمولی سر و کار داریم با این تفاوت که در این مدل یک جمله انرژی – تکانه جدید به جهان اضافه شده است که منشأ آن بعد اضافه است. همان طور که از معادله (3.7.19) مشاهده می کنیم، این ماده فاقد مؤلفه چگالی انرژی است.

# فصل ۴

## قضیه ویریال

### ۴.۱ مسئله جرم ویریال

جرم ویریال یکی از مسائل مهم کیهان شناسی است و از آنجا که پیش بینی های نسبیت عام اینشتین مشاهدات تجربی را توضیح نمی دهد این مسئله به یک چالش برای نظریه پردازان نیز تبدیل شده است. یک راه آسان و در عین حال خوشایند برای کیهان شناسان این است که همچنان فرض کنیم نسبیت عام اینشتین نظریه درست بر برهم کنش های گرانشی درون سیستم مورد مطالعه ما است و مسئله جرم ویریال را با وارد کردن یک ماده اضافه (ماده تاریک) حل کنیم. راه دیگر استفاده از ثابت کیهان شناسی برای توضیح مسئله جرم ویریال است [۳]. در زیر مختصری از این روش و نواقص آن ارائه خواهد شد. اما راه دیگر که در دهه های اخیر بیشتر به آن پرداخته شده است یافتن نظریه جدیدی بر پایه نسبیت عام اینشتین است که بتواند رفتار و مقدار جرم ویریال را در خوشه های کهکشانی پیش بینی کند. مدل $DGP$ به عنوان یک چنین نظریه ای کاندیدای خوبی برای حل مسئله جرم ویریال خواهد بود. برای یافتن حل مسئله جرم ویریال در مدل های دیگر به مراجع [۳۶ و ۳۷] مراجعه کنید.

برای اندازه گیری جرم یک خوشه کهکشانی دو راه متفاوت وجود دارد، اما از آنجا که هر دو روش یک کمیت را اندازه می گیرند انتظار داریم که هر دو یک جواب برای هر دو پیش بینی کنند. اگر حرکت هر یک از اعضای یک خوشه کهکشانی را بدانیم، یک راه محاسبه جرم استفاده از قضیه ویریال خواهد بود (که در بخش بعد در مورد آن توضیح می دهیم). این جرم را $M_V$ بنامید. راه دیگر این است که جرم تک تک کهکشان ها در یک خوشه را محاسبه بعد آنها را با هم جمع بزنیم. جرم یک کهکشان را می توان با اندازه گیری درخشندگی یک کهکشان و ضرب آن در نسبت جرم به درخشندگی مناسب آن کهکشان بدست آورد. این جرم را $M$ می نامیم. تقریباً در تمامی خوشه های کهکشانی آزمایش شده $M_V$ به اندازه ۲۰ تا ۳۰ برابر بیشتر از $M$ است. این اساس مسئله جرم ویریال است. در این فصل نشان می دهیم که می توان این مسئله را با استفاده از مدل $DGP$ توضیح داد. این کار با

توجه به جمله جرمی که صرفاً از بعد اضافه آمده است انجام می شود [۳۹]. در این فصل قضیه ویریال را برای یک جهان با تقارن کروی با در نظر گرفتن ثابت کیهان شناسی بدست می آوریم. سپس مشابه این فرایند را برای مدل $DGP$ دنبال کرده و نشان خواهیم داد که جرم ویریال متناسب با جرم $\mathcal{N}(r)$ خواهد بود که صرفاً از بعد پنجم آمده است. سپس مقدار $\mathcal{N}(r)$ را با کمک داده های رصدی محاسبه کرده و نشان خواهیم داد که جرم ویریال بطور خطی با شعاع تغییر می کند. همچنین نشان می دهیم مقدار این کمیت با مقداری که انتظار داریم برای جرم ویریال بدست آوریم هم مرتبه است. در آخر پراکندگی شعاعی سرعت کهکشان ها را در مدل $DGP$ محاسبه خواهیم کرد.

## ۴.۲ قضیه ویریال در نسبیت عام/ جرم ویریال

در این بخش قضیه ویریال را در کیهان شناسی با در نظر گرفتن ثابت کیهان شناسی بدست می آوریم [۳]. المان طول را بصورت زیر تعریف می کنیم

$$ds^2 = -e^{\nu(r)}dt^2 + e^{\mu(r)}dr^2 + r^2 d\Omega^2. \tag{۴.۲.۱}$$

برای راحتی محاسبات بهتر است که از رویکرد tetrad استفاده کنیم. برای این منظور بردارهای متعامد زیر را تعریف می کنیم

$$e_\rho^{(0)} = e^{\frac{\nu}{2}}\delta_\rho^0, \qquad e_\rho^{(1)} = e^{\frac{\mu}{2}}\delta_\rho^1, \qquad e_\rho^{(2)} = r\delta_\rho^2, \qquad e_\rho^{(3)} = r\sin\theta\delta_\rho^3, \tag{۴.۲.۲}$$

که در آن $g^{\mu\nu}e_\mu^{(a)}e_\nu^{(b)} = \eta^{(a)(b)}$ و مختصات tetrad را داخل پرانتز قرار داده ایم. در واقع $e_\mu^{(a)}$ ها چهار بردار هموردا هستند یعنی

$$\begin{aligned} e_\mu^{(0)} &= (e^{\frac{\nu}{2}}, 0, 0, 0), \\ e_\mu^{(1)} &= (0, e^{\frac{\mu}{2}}, 0, 0), \\ e_\mu^{(2)} &= (0, 0, r, 0), \\ e_\mu^{(3)} &= (0, 0, 0, r\sin\theta). \end{aligned} \tag{۴.۲.۳}$$

۵۱

در این رویکرد ۴-بردار سرعت $u^\mu$ یک کهکشان با خاصیت $u^\mu u_\mu = -1$ بصورت زیر نوشته می شود

$$u^{(a)} = u^\mu e_\mu^{(a)}, \quad a = 0,1,2,3. \tag{۴.۲.۴}$$

با معادله بولتزمان در رویکرد tetrad شروع می کنیم. این معادله توسط Lindquist در [۴] بدست آمده است. فرض کنید $f(x^\mu, u^{(a)})$ تابع توزیع کهکشان ها در فضا-زمان باشد، اگر کهکشان ها را بصورت ذرات تمیزناپذیر و بدون برهم کنش در نظر بگیریم خواهیم داشت

$$u^{(a)} e_{(a)}^\rho \frac{\partial f}{\partial x^\rho} + \gamma_{(b)(c)}^{(a)} u^{(a)} u^{(b)} \frac{\partial f}{\partial u^{(a)}} = 0, \tag{۴.۲.۵}$$

که در آن $\gamma_{(b)(c)}^{(a)} = e_{\rho;\sigma}^{(a)} e_{(b)}^\rho e_{(c)}^\sigma$ ضرایب چرخش ریچی[22] هستند. با استفاده از (۴.۲.۱) و با فرض اینکه تابع توزیع فقط از طریق $r$ به مختصات وابسته است، معادله بولتزمان بصورت زیر در می آید

$$u_r \frac{\partial f}{\partial r} - \left( \frac{u_t^2}{2} v' - \frac{u_\theta^2 + u_\varphi^2}{r} \right) \frac{\partial f}{\partial u_r} - \frac{u_r}{r} \left( u_\theta \frac{\partial f}{\partial u_\theta} + u_\varphi \frac{\partial f}{\partial u_\varphi} \right)$$
$$- \frac{e^{\frac{\mu}{2}} u_\varphi}{r} \cot\theta \left( u_\theta \frac{\partial f}{\partial u_\varphi} - u_\varphi \frac{\partial f}{\partial u_\theta} \right) = 0, \tag{۴.۲.۶}$$

که در آن تعریف کرده ایم

$$u^{(0)} = u_t, \quad u^{(1)} = u_r, \quad u^{(2)} = u_\theta, \quad u^{(3)} = u_\varphi. \tag{۴.۲.۷}$$

چون جهان (۴.۲.۱) با توزیع کروی است معادله حاکم بر توزیع کهکشان ها نمی تواند وابسته به $\theta$ باشد، لذا جمله $\cot\theta$ در معادله (۴.۲.۶) باید صفر باشد یعنی باید داشته باشیم

$$u_\theta \frac{\partial f}{\partial u_\varphi} = u_\varphi \frac{\partial f}{\partial u_\theta}. \tag{۴.۲.۸}$$

---

[22] Ricci rotation coefficients

۵۲

با تعریف متغیرهای جدید

$$x = u_\theta^2 + u_\varphi^2, \quad y = u_\theta^2 - u_\varphi^2, \tag{4.2.9}$$

معادله (4.2.8) نتیجه می دهد

$$\frac{\partial f}{\partial y} = 0. \tag{4.2.10}$$

یعنی $f$ تنها به $r$، $u_r$ و $u_\theta^2 + u_\varphi^2$ بستگی خواهد داشت. لذا معادله (4.2.6) را می توان بصورت

$$u_r \frac{\partial f}{\partial r} - \left( \frac{u_t^2}{2} v' - \frac{u_\theta^2 + u_\varphi^2}{r} \right) \frac{\partial f}{\partial r} - \frac{2 u_r x}{r} \frac{\partial f}{\partial x} = 0, \tag{4.2.11}$$

نوشت. حال طرفین معادله بولتزمان را در $m u_r du$ که در آن $du = \frac{1}{u_t} du_r du_\theta du_\varphi$ المان ناوردای حجم در فضای سرعت ها و $m$ جرم کهکشان ها است، ضرب می کنیم و روی فضای سرعت ها انتگرال می گیریم. اگر فرض کنیم که وقتی سرعت ها به بینهایت می روند $f$ به قدر کافی سریع به صفر میل کند بدست می آوریم

$$r \frac{\partial}{\partial r} \left[ \rho \langle u_r^2 \rangle \right] + \frac{1}{2} \rho \left[ \langle u_t^2 \rangle + \langle u_r^2 \rangle \right] r v' - \rho \left[ \langle u_\theta^2 \rangle + \langle u_\varphi^2 \rangle - 2 \langle u_r^2 \rangle \right] = 0, \tag{4.2.12}$$

که در آن $\rho$ چگالی جرم در خوشه کهکشانی است و $\langle \ \rangle$ به معنی مقدار میانگین کمیتی است که در درون آن قرار گرفته است. حال (4.2.12) را در $4\pi r^2$ ضرب می کنیم و روی خوشه کهکشان ها انتگرال گیری می کنیم

$$\frac{1}{2} \int_0^R 4\pi r^3 v' \rho \left[ \langle u_t^2 \rangle + \langle u_r^2 \rangle \right] dr - \int_0^R 4\pi r^2 \rho \left[ \langle u_\theta^2 \rangle + \langle u_\varphi^2 \rangle + \langle u_r^2 \rangle \right] dr = 0. \tag{4.2.13}$$

اگر انرژی کل کهکشان ها را به صورت

$$K = \int_0^R 2\pi r^2 \rho \left[ \langle u_\theta^2 \rangle + \langle u_\varphi^2 \rangle + \langle u_r^2 \rangle \right] dr, \tag{4.2.14}$$

تعریف کنیم خواهیم داشت

$$2K = \frac{1}{2}\int_0^R 4\pi r^3 v' \rho \left[\langle u_t^2 \rangle + \langle u_r^2 \rangle\right] dr. \qquad (4.2.15)$$

برای ادامه باید مؤلفه های تانسور انرژی ـ تکانه را به تابع توزیع کهکشان ها در خوشه کهکشانی مربوط کنیم. این کار از طریق رابطه زیر انجام می شود

$$T_{\mu\nu} = \int fm u_\mu u_\nu du. \qquad (4.2.16)$$

اگر تانسور انرژی ـ تکانه را به صورت

$$T_\mu^{\ \nu} = diag(-\rho_b, p_b, p_b, p_b), \qquad (4.2.17)$$

تعریف کنیم نتیجه می گیریم

$$\rho_b = \rho \langle u_t^2 \rangle, \qquad p_b = \rho \langle u_r^2 \rangle = \rho \langle u_\theta^2 \rangle = \rho \langle u_\varphi^2 \rangle. \qquad (4.2.18)$$

معادلات اینشتین برای متریک (4.2.1) به صورت زیر است

$$e^{-\mu}\left(\frac{\mu'}{r} - \frac{1}{r^2} + \frac{e^\mu}{r^2}\right) = 8\pi G \rho_b + \Lambda, \qquad (4.2.19)$$

$$e^{-\mu}\left(\frac{v'}{r} + \frac{1}{r^2} - \frac{e^\mu}{r^2}\right) = 8\pi G p_b - \Lambda, \qquad (4.2.20)$$

$$e^{-\mu}\left(\frac{v'}{2r} + \frac{\mu'}{2r} - \frac{\mu' v'}{4} + \frac{v''}{2} + \frac{v'^2}{4}\right) = 8\pi G p_b - \Lambda. \qquad (4.2.21)$$

با جمع کردن (4.2.19) و (4.2.20) و کم کردن دو برابر معادله (4.2.21) از آنها بدست می آوریم

۵۴

$$e^{-\mu}\left(\frac{v'}{r}-\frac{\mu'v'}{4}+\frac{v'}{2}+\frac{v'^2}{4}\right)=4\pi G\left\langle u^2\right\rangle-\Lambda, \qquad (4.2.22)$$

که در آن تعریف کرده ایم $\left\langle u^2\right\rangle=\left\langle u_t^2\right\rangle+\left\langle u_r^2\right\rangle+\left\langle u_\theta^2\right\rangle+\left\langle u_\varphi^2\right\rangle$ . برای یک خوشه کهکشانی می توانیم فرض کنیم که $\mu(r)$ و $v(r)$ به قدری کوچک هستند که جملات مرتبه دوم آنها در (4.2.22) قابل صرف نظر کردن است. همچنین در مورد کهکشان ها در یک خوشه کهکشانی  می توانیم به خوبی تقریب زیر را بکار ببریم

$$\left\langle u_r^2\right\rangle,\left\langle u_\theta^2\right\rangle,\left\langle u_\varphi^2\right\rangle\ll\left\langle u_t^2\right\rangle\approx 1. \qquad (4.2.23)$$

با این تقریب ها معادله (4.2.22) بصورت

$$4\pi G\rho=\frac{1}{2}\frac{1}{r^2}\frac{\partial}{\partial r}\left(r^2 v'\right)+\Lambda, \qquad (4.2.24)$$

در می آید. با ضرب طرفین (4.2.24) در $r^2$ و انتگرال گیری از 0 تا r بدست می آوریم

$$GM(r)=\frac{1}{2}r^2 v'+\frac{1}{3}\Lambda r^3, \qquad (4.2.25)$$

که در آن تعریف کرده ایم

$$M(r)=4\pi\int_0^r\rho r'^2 dr'. \qquad (4.2.26)$$

دوباره معادله (4.2.25) را در $\frac{dM(r)}{r}$ ضرب می کنیم و این بار از 0 تا $R$ انتگرال گیری می کنیم. این کار ما را به قضیه ویریال برای یک جهان با تقارن کروی با ثابت کیهان شناسی می رساند

$$2K+W+\frac{1}{3}\Lambda I=0. \qquad (4.2.27)$$

که در آن انرژی پتانسیل گرانشی سیستم را به صورت

۵۵

$$W = -\int_0^R \frac{GM(r)}{r} dM(r), \tag{۴.۲.۲۸}$$

تعریف کرده ایم و $I$ گشتاور لختی سیستم است و به صورت

$$I = \int_0^R r^2 dM(r), \tag{۴.۲.۲۹}$$

تعریف می شود. همچنین از معادله (۴.۲.۱۴) نیز استفاده کرده ایم. بدون جمله ثابت کیهان شناسی این معادله همان قانون ویریال معمول است. این معادله می تواند به معادله ای برای جرم ویریال تبدیل شود. برای این کار شعاع های زیر را تعریف می کنیم

$$R_V = \frac{M^2}{\int_0^R \frac{M(r)}{r} dM(r)}, \tag{۴.۲.۳۰}$$

$$R_I^2 = \frac{\int_0^R r^2 dM(r)}{M(r)}. \tag{۴.۲.۳۱}$$

جرم ویریال را می توان توسط قانون نیوتن به صورت زیر تعریف کرد

$$2K = \frac{GM_V^2}{R_V}. \tag{۴.۲.۳۲}$$

اگر معادلات (۴.۲.۳۰) و (۴.۲.۳۱) را به صورت

$$W = -\frac{GM^2}{R_V}, \quad I = MR_I^2, \tag{۴.۲.۳۳}$$

بنویسیم و از قضیه ویریال استفاده کنیم نسبت جرم ویریال به جرم کهکشان ها به صورت زیر بدست می آید

$$\left(\frac{M_V}{M}\right)^2 = 1 - \frac{\Lambda}{3G} \frac{R_I^2 R_V}{M}. \tag{۴.۲.۳۴}$$

۵۶

مشاهده می کنیم که برای ارضاء مشاهدات رصدی چون طرف اول بزرگتر از ۱ است باید داشته باشیم $\Lambda < 0$. اگر $M_V > 3M$ باشد، که برای اکثر کهکشان ها درست است می توانیم از عدد ۱ در (۴.۲.۳۴) صرف نظر کنیم. در واقع این تقریب معادل است با اینکه داشته باشیم $W \ll \frac{1}{3} \Lambda I$. در این صورت می توان معادله (۴.۲.۳۴) را به صورت زیر نوشت

$$\log \frac{M_V}{M} = -\frac{1}{2} \log \left( \frac{3M}{4\pi R_V^3} \right) + \frac{1}{2} \log \left( \frac{-\Lambda \beta^2}{4\pi G} \right), \qquad (۴.۲.۳۵)$$

که در آن فرض کرده ایم $R_I = \beta R_V$. $\beta$ پارامتری هم مرتبه با ۱ است که مقدار دقیق آن بستگی به چگالی جرمی خوشه کهکشانی دارد اما از هر کهکشان به کهکشان دیگر تغییر زیادی نمی کند. مشاهده می کنیم که نمودار $\log \frac{M_V}{M}$ نسبت به $\log \left( \frac{3M}{4\pi R_V^3} \right)$ باید خطی و با شیب $-0.5$ باشد. این نتیجه را با نتایج رصدی در شکل (۴.۱) مقایسه کرده ایم. مشاهده می کنیم که این مدل می تواند مسئله جرم ویریال را با در نظر گرفتن ثابت کیهان شناسی منفی برای خوشه های کهکشانی که از تقارن کروی برخوردار باشند توضیح دهد. مثالی از یک خوشه کهکشانی با این خاصیت، خوشه کهکشانی کما[23] است. اما این مدل نمی تواند همزمان با حل مسئله جرم ویریال عمر عالم را نیز به درستی محاسبه کند و این به خاطر منفی در نظر گرفتن ثابت کیهان شناسی است. برای مطالعه این موضوع ابتدا معادلات فریدمن را برای یک جهان با پارامتر انحنای $k$ می نویسیم

$$8\pi G \rho = \frac{3k}{a^2} + 3 \left( \frac{\dot{a}}{a} \right)^2 - \lambda, \qquad (۴.۲.۳۶)$$

$$8\pi G p = -\frac{k}{a^2} - 2\frac{\ddot{a}}{a} - \left( \frac{\dot{a}}{a} \right)^2 + \lambda. \qquad (۴.۲.۳۷)$$

---

[23] Coma cluster

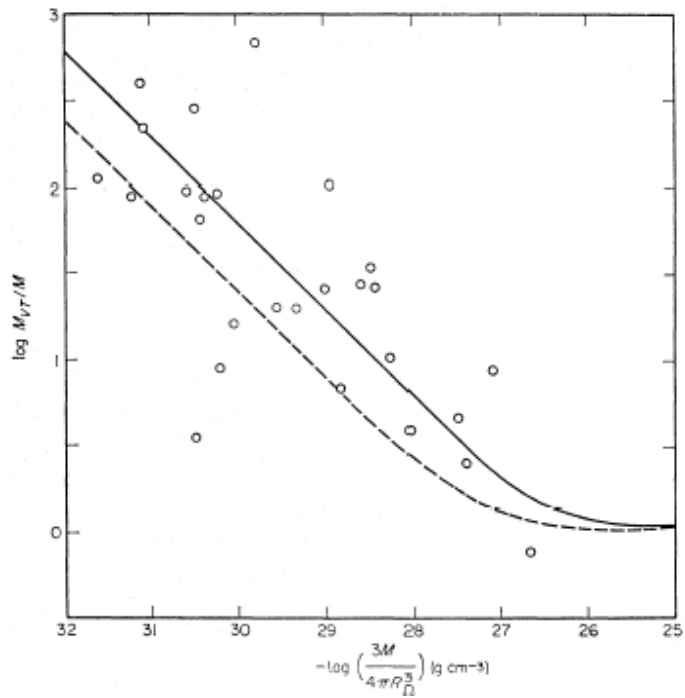

شکل (۴.۱) رسم معادله (۴.۲.۳۵). نمودار خط پر رسم برای $\Lambda = -2.9 \times 10^{-33}$

و نمودار خط چین برای $\Lambda = -5 \times 10^{-34}$ است. نقطه ها داده های تجربی اند

[برگرفته از مرجع 2 (1970) **148** Jackson, Mon. Not. Roy. Astro. Soc.]

با فرض منفی بودن $\lambda$ از این معادلات نتیجه می گیریم که $k = -1$. اگر این اتفاق نیافتد از معادله (۴.۲.۳۶) داریم

$$8\pi G \rho > -\lambda, \qquad (\text{۴.۲.۳۸})$$

و با استفاده از معادله (۴.۲.۳۴) می توانیم بنویسیم

$$\frac{M_V}{M} < \left(1 + 2\frac{\rho}{\bar{\rho}}\right)^{\frac{1}{2}}, \qquad (\text{۴.۲.۳۹})$$

که در آن تعریف کرده ایم

$$\bar{\rho} = \frac{M}{\frac{4}{3}\pi R_V R_I^{\,2}}. \qquad (4.2.40)$$

در واقع $\bar{\rho}$ با چگالی جرم متوسط خوشه کهکشانی متناسب است و ضریب تناسب برای بیشتر خوشه های کهکشانی از مرتبه 1 است [3]. از آنجایی که خوشه کهکشانی طبق تعریف جایی است که چگالی آن به طور قابل ملاحظه ای از $\bar{\rho}$ بیشتر است از معادله (4.2.39) نتیجه می گیریم که نسبت جرم ویریال به جرم مشاهده شده خوشه باید اندکی از 1 بیشتر باشد و این تناقض است. لذا $k = -1$. برای بدست آوردن عمر جهان تعریف می کنیم

$$q_0 = -\frac{1}{H_0^{\,2}} \frac{\ddot{a}_0}{a_0}, \qquad (4.2.41)$$

$$\lambda_0 = \frac{1}{3} \frac{\lambda}{H_0^{\,2}}, \qquad (4.2.42)$$

که در آن $q_0$ پارامتر کند شدن شتاب[24] است و صفر یعنی مقدار کمیت در زمان حال. اگر فرض کنیم $p = \rho = 0$ داریم

$$q_0 = -\lambda_0. \qquad (4.2.43)$$

حل معادلات بالا به صورت زیر خواهد بود

$$a(t) = \left(-\frac{1}{3}\lambda\right)^{\frac{1}{2}} \sin\left[\left(-\frac{1}{3}\lambda\right)^{\frac{1}{2}} t\right], \qquad (4.2.44)$$

$$H_0 t_0 = q_0^{-\frac{1}{2}} \cot^{-1}\left(q_0^{-\frac{1}{2}}\right), \qquad (4.2.45)$$

که در آن $t_0$ عمر کنونی جهان است. با استفاده از مقادیر شکل (4.1) بدست می آوریم

---

[24] Deceleration parameter

$$q_0 = 92.3 \qquad\qquad H_0 t_0 = 0.153,$$
$$t_0 = 1.5 \times 10^9 \ \ years \ \ if \ \ H_0 = 100 \, km \, s^{-1} \, Mpc^{-1}, \qquad (4.2.46)$$
$$t_0 = 3 \times 10^9 \ \ \ \ years \ \ if \ \ H_0 = 50 \ km \, s^{-1} \, Mpc^{-1}.$$

مشاهده می کنید که این مقدار برای عمر عالم بسیار کم است. در بخش بعد مسئله جرم ویریال را با استفاده از مدل DGP توضیح خواهیم داد. از آنجایی که در حل DGP احتیاجی به وارد کردن قیدی روی پارامترهای مسئله مانند ثابت کیهان شناسی نداریم حل DGP مشکل کوتاه پیش بینی کردن عمر جهان را نخواهد داشت.

## 4.3 جرم ویریال در مدل DGP

در بخش قبل قضیه ویریال را برای یک جهان با تقارن کروی با ثابت کیهان شناسی بدست آوردیم [39]. در این بخش راهکار قبل را برای مدل DGP تکرار می کنیم. در معادلات (3.7.20) تا (3.7.22) بخش آخر از فصل قبل معادلات DGP را برای یک جهان کروی بر روی شامه بدست آوردیم. اگر در این معادلات جمله $\mathcal{P}(r)$ مربوط به بعد اضافه را حذف کنیم و از رابطه $m_3^{-2} = 8\pi G$ استفاده نماییم، مشاهده می کنیم که این معادلات همان معادلات اینشتین در بخش قبل هستند. لذا می توان حدس زد که قضیه ویریال باید به همان شکل بخش قبل به اضافه یک جمله مربوط به بعد اضافه باشد. برای بدست آوردن قضیه ویریال در مدل DGP مانند بخش قبل معادلات (3.6.20) و (3.6.21) را با هم جمع کرده و دو برابر معادله (3.6.22) را از آنها کم می کنیم. اگر از فرض های بخش قبل در مورد سرعت کهکشان ها در خوشه کهکشانی استفاده کنیم خواهیم داشت

$$\rho = m_3^2 \frac{1}{r^2} \frac{\partial}{\partial r}\left(r^2 \nu_0'\right) + 2 m_3^2 \Lambda - 3 \mathcal{P}(r), \qquad (4.3.1)$$

و با ضرب $r^2$ در آن و انتگرال گیری بدست می آوریم

$$m_3^2 r^2 \nu_0' - \frac{1}{4\pi} M(r) + \frac{2}{3} m_3^2 \Lambda r^3 - 4\pi \mathcal{N}(r) = 0, \qquad (4.3.2)$$

که در آن $M(r)$ جرم کهکشان ها (4.2.26) است و تعریف کرده ایم

$$\mathcal{N}(r) = 4\pi \int_0^r 3\mathcal{P}(r')r'^2 dr'. \tag{۴.۳.۳}$$

با ضرب $\frac{dM(r)}{r}$ در معادله قبل و انتگرال گیری از آن بین $0$ تا $R$ قضیه ویریال در مدل $DGP$ را بدست می آوریم

$$2K + W + \frac{1}{3}\Lambda I + \mathcal{W}_B = 0, \tag{۴.۳.۴}$$

که در آن تعریف کرده ایم

$$W = -\frac{1}{8\pi m_3^2} \int_0^R \frac{M(r)}{r} dM(r), \tag{۴.۳.۵}$$

$$\mathcal{W}_B = -\frac{2\pi}{m_3^2} \int_0^R \rho r \mathcal{N}(r) dr. \tag{۴.۳.۶}$$

معادله (۴.۳.۵) در واقع همان (۴.۲.۲۸) است. گشتاور لختی سیستم را هم در (۴.۲.۲۹) تعریف کرده ایم. برای بدست آوردن رابطه جرم ویریال و جرم جدیدی که مدل $DGP$ به ما معرفی کرده است شعاع مربوط به بعد اضافه را به صورت زیر تعریف می کنیم

$$\mathcal{R} = -\frac{1}{8\pi m_3^2} \frac{\mathcal{N}^2}{\mathcal{W}_B}. \tag{۴.۳.۷}$$

اگر از تعریف های بخش قبل در اینجا نیز استفاده کنیم رابطه بین جرم ویریال و جرم $\mathcal{N}(r)$ به صورت زیر بدست می آید

$$\left(\frac{M_V}{M}\right)^2 = 1 - \frac{8\pi m_3^2 \Lambda}{3} \frac{R_I^2 R_V}{M} + \left(\frac{\mathcal{N}}{M}\right)^2 \left(\frac{R_V}{\mathcal{R}}\right). \tag{۴.۳.۸}$$

در اینجا به خاطر جمله آخر که مربوط به بعد اضافه است احتیاجی به فرض $\Lambda < 0$ نداریم. اگر فرض کنیم $M_V > 3M$ باز هم می توانیم از عدد ۱ در این معادله صرفنظر کنیم. همچنین برای اینکه بتوانیم اثرات بعد اضافه را به تنهایی مطالعه کنیم در ادامه فرض می کنیم $\Lambda = 0$ . در این صورت معادله (۴.۳.۸) را می توان به صورت

$$M_V(r) \simeq \mathcal{N}(r)\sqrt{\frac{R_V}{\mathcal{R}}} \qquad (4.3.9)$$

نوشت. مشاهده می کنیم که جرم ویریال مستقیماً توسط $\mathcal{N}(r)$، جرمی که بوسیله بعد اضافه ایجاد شده است، تعیین می شود. در بخش بعد تخمینی برای مقدار این کمیت ارائه خواهیم کرد.

## ۴.۴ محاسبه جرم $\mathcal{N}(r)$

در این بخش ما جرم $\mathcal{N}(r)$ را با استفاده از داده های رصدی محاسبه خواهیم کرد [۳۹]. طبق ملاحظات فصل قبل برای این کار کافی است $\mathcal{P}(r)$ را محاسبه کنیم. ابتدا توجه کنید که بیشتر جرم باریونیک داخل کهکشان ها به صورت گازهای داخل خوشه کهکشانی است. این گازها را می توان به وسیله چگالی انرژی

$$\rho_g(r) = \rho_0\left(1+\frac{r^2}{r_c^2}\right)^{-\frac{3\beta}{2}}, \qquad (4.4.1)$$

توصیف کرد [۲۴]. این معادله با داده های تجربی به خوبی همخوانی دارد و در آن $r_c$ شعاع هسته خوشه کهکشانی است و $\beta$ و $\rho_0$ ثابت هایی وابسته به خوشه کهکشانی می باشند. یک سیستم استاتیک با تقارن کروی که از ذرات بدون برهم کنش تشکیل شده و در حال تعادل ترمودینامیکی باشد را می توان بوسیله معادله جینز[25] توصیف کرد. این معادله به صورت زیر نوشته می شود

$$\frac{d}{dr}\left[\rho_g \sigma_r^2\right] + \frac{2\rho_g(r)}{r}\left(\sigma_r^2 - \sigma_{\theta,\varphi}^2\right) = -\rho_g(r)\frac{d\Phi}{dr}, \qquad (4.4.2)$$

که در آن $\Phi(r)$ پتانسیل گرانشی است و $\sigma_r$ و $\sigma_{\theta,\varphi}$ به ترتیب پراکندگی های سرعت در جهت های شعاعی و مماسی هستند. این معادله در [۲۵] بدست آمده است. اگر فرض کنیم که گاز بطور همسانگرد داخل خوشه کهکشانی توزیع شده است می توانیم بنویسیم $\sigma_r = \sigma_{\theta,\varphi}$. همچنین فشار گاز با پراکندگی سرعت و چگالی انرژی گاز از طریق معادله زیر به هم مربوط هستند

---

[25] Jean's equation

$$p_g = \rho_g \sigma_r^2. \qquad (4.4.3)$$

فرض می کنیم که میدان گرانشی ضعیف باشد بطوری که بتوانیم از معادله معمول پواسون

$$2m_3^2 \nabla^2 \Phi \approx \rho_{tot}, \qquad (4.4.4)$$

برای توصیف $\Phi(r)$ استفاده کنیم، که در آن $\rho_{tot}$ چگالی انرژی کل خوشه کهکشانی است و شامل ماده باریونی $\rho_g$ و حالت های دیگر ماده مانند ماده تاریک و غیره است. با این فرضیات می توان معادله جینز را بصورت

$$\frac{dp_g(r)}{dr} = -\frac{1}{8\pi m_3^2} \frac{M_{tot}}{r^2} \rho_g(r), \qquad (4.4.5)$$

نوشت که در آن $M_{tot}$ کل جرم داخل شعاع $r$ است. برای ادامه باید یک معادله حالت برای گاز موجود در خوشه کهکشانی فرض کنیم. گسیل های پرتوی X مشاهده شده از گازهای یونیزه شده داخل خوشه کهکشانی معمولاً به صورت گاز در حال تعادل همدما در نظر گرفته می شوند. ما فرض می کنیم که معادله حالت این گاز بصورت

$$p_g(r) = \frac{k_B T_g}{\mu m_p} \rho_g(r), \qquad (4.4.6)$$

باشد که در آن $\mu = 0.61$ وزن اتمی میانگین ذرات گاز داخل خوشه کهکشانی و $m_p$ جرم پروتون است. همچنین دمای معمول گاز داخل خوشه کهکشانی در حدود $k_B T_g \approx 5 KeV$ است. با استفاده از معادله حالت (4.4.6) معادله جینز را می توان برای $M_{tot}$ حل کرد

$$M_{tot} = -8\pi m_3^2 \frac{k_B T_g}{\mu m_p} r^2 \frac{d}{dr} \ln \rho_g = 8\pi m_3^2 \frac{3 k_B T_g}{\mu m_p} \beta \frac{r^3}{r^2 + r_c^2}. \qquad (4.4.7)$$

برای اینکه بتوانیم رابطه ای برای $\mathcal{P}(r)$ به دست بیاوریم باید رابطه ای برای $M_{tot}$ از داخل نظریه داشته باشیم. این رابطه توسط معادلات (4.3.3) و (4.2.26) داده می شود و به صورت زیر است

۶۳

$$\frac{dM_{tot}}{dr} = 4\pi\rho_g r^2 + 3\mathcal{P}r^2. \qquad (4.4.8)$$

از حل دو معادله (4.4.7) و (4.4.8) برای $\mathcal{P}(r)$ بدست می آوریم

$$\mathcal{P}(r) = 8\pi m_3^2 \frac{k_B T_g}{\mu m_p} \beta \frac{3r_c^2 + r^2}{\left(r^2 + r_c^2\right)^2} - \frac{4\pi}{3}\rho_0\left(1 + \frac{r^2}{r_c^2}\right)^{-\frac{3\beta}{2}}. \qquad (4.4.9)$$

حال توجه خود را به محدوده مرزی خوشه کهکشانی معطوف می کنیم. در این حد داریم $r_c \ll r$. پس در معادله (4.4.9) می توان از $r_c$ در جمله اول و از 1 در جمله دوم صرف نظر کرد. لذا داریم

$$\mathcal{P}(r) = 8\pi m_3^2 \frac{k_B T_g}{\mu m_p} \beta \frac{1}{r^2} - \frac{4\pi}{3}\rho_0\left(\frac{r}{r_c}\right)^{-3\beta} = \left[8\pi m_3^2 \frac{k_B T_g}{\mu m_p} \beta - \frac{4\pi}{3}\rho_0 r_c^{3\beta} r^{2-3\beta}\right]\frac{1}{r^2}. \qquad (4.4.10)$$

جدول (4.1) اطلاعات مربوط به خوشه های کهکشانی را نشان می دهد. برای مثال مشاهده می کنید که برای بیشتر کهکشان ها داریم $\beta > \frac{2}{3}$. در این حالت جمله دوم در (4.4.10) قابل صرفنظر کردن است و لذا داریم

$$\mathcal{P}(r) = 8\pi m_3^2 \frac{k_B T_g}{\mu m_p} \beta \frac{1}{r^2}. \qquad (4.4.11)$$

که با استفاده از تعریف جرم $\mathcal{N}(r)$ خواهیم داشت

$$\mathcal{N}(r) = 8\pi m_3^2 \frac{3k_B T_g}{\mu m_p} \beta\, r. \qquad (4.4.12)$$

این معادله رفتار خطی بر حسب $r$ را برای جرم $\mathcal{N}(r)$ نشان می دهد که در توافق با مشاهدات برای جرم ویریال است. برای اینکه تخمینی از این رابطه بزنیم ابتدا تخمینی برای شعاع ماکزیمم خوشه کهکشانی در این مدل ارائه می دهیم. برای تعریف شعاع یک خوشه کهکشانی معمولاً فاصله ای را در نظر می گیرند که در آن مقدار چگالی انرژی کل خوشه کهکشانی 200 یا 500 برابر چگالی بحرانی انرژی جهان یعنی $\rho_c = \frac{3H^2}{8\pi G} = 4.09 \times 10^{-30} \frac{gr}{cm^3}$ باشد. این فواصل را با $r_{200}$ و $r_{500}$ و مقدار جرمی را که

داخل این فواصل وجود دارند با $M_{200}$ و $M_{500}$ نشان می دهیم. مقادیر این کمیت ها برای خوشه های کهکشانی مختلف در جدول (۴.۱) آمده است. مشاهده می کنید که مقدار $r_{200}$ در حدود $2 Mpc$ برای خوشه های کهکشانی است. معمولاً شعاع ویریال را برابر $r_{200}$ و جرم ویریال را برابر $M_{200}$ در نظر می گیرند. برای جرم $\mathcal{N}(r)$ که از بعد اضافه آمده است نیز فرض می کنیم که شعاع آن جایی باشد که چگالی انرژی آن $\mathcal{P}(r)$ به اندازه $200\rho_c$ کاهش یابد [۳۹]. در این حالت داریم

$$\mathcal{P}(r) = 200\rho_c = 9.38 \times 10^{-28} \frac{gr}{cm^3}, \qquad (4.4.13)$$

که نتیجه می دهد

$$r = 4.28 \beta^{\frac{1}{2}} \left(\frac{k_B T_g}{5 KeV}\right)^{\frac{1}{2}} Mpc. \qquad (4.4.14)$$

در این حالت داریم

$$\mathcal{N}(r) = \frac{3 k_B T_g}{\mu m_p G} \beta r = 32.72 \times 10^{14} \beta^{\frac{3}{2}} \left(\frac{k_B T_g}{5 KeV}\right)^{\frac{3}{2}} M_\odot. \qquad (4.4.15)$$

مشاهده می کنیم که مقدارهای بالا، در حد $M_{200}$ و $r_{200}$ برای خوشه های کهکشانی است و لذا می تواند مقدار جرم و شعاع ویریال را توضیح دهد. برای حالت $\beta = \frac{2}{3}$ داریم

$$\mathcal{P}(r) = \left[8\pi m_3^2 \frac{k_B T_g}{\mu m_p} \beta - \frac{4\pi}{3} \rho_0 r_c^2\right] \frac{1}{r^2}, \qquad (4.4.16)$$

که از آن جرم ویریال به صورت

$$\mathcal{N}(r) = 3 \left[8\pi m_3^2 \frac{k_B T_g}{\mu m_p} \beta - \frac{4\pi}{3} \rho_0 r_c^2\right] r, \qquad (4.4.17)$$

۶۵

به دست می آید. در حالت $\beta > \frac{2}{3}$ جرم $\mathcal{N}(r)$ دارای یک ماکزیمم در شعاع

$$r_{\max} = \left( \frac{6m_3^2 k_B T_g}{\mu m_p \rho_0 r_c^{3\beta}} \right)^{\frac{1}{2-3\beta}},\qquad(4.4.18)$$

خواهد بود و مقدار ماکزیمم $\mathcal{N}(r)$ برابر است با

$$\mathcal{N}(r_{\max}) = \left( \frac{24\pi m_3^2 k_B T_g}{\mu m_p} \frac{2-3\beta}{3-3\beta} \right) r_{\max}.\qquad(4.4.19)$$

## ۴.۵ پراکندگی سرعت شعاعی در خوشه های کهکشانی

در این بخش پراکندگی سرعت شعاعی خوشه های کهکشانی را در مدل DGP به دست می آوریم [۳۹]. اگر فرض هایی را که بعد از معادله (۴.۲.۲۲) کرده بودیم روی معادله (۴.۲.۱۲) اعمال کنیم این معادله به صورت زیر نوشته می شود

$$\frac{d}{dr}\left(\rho \sigma_r^2\right) + \frac{1}{2}\rho v_0' = 0.\qquad(4.5.1)$$

فرض می کنیم توزیع سرعت در خوشه کهکشانی همسانگرد باشد به طوری که

$$\langle u^2 \rangle = \langle u_r^2 \rangle + \langle u_\theta^2 \rangle + \langle u_\varphi^2 \rangle = 3\langle u_r^2 \rangle = 3\sigma_r^2.\qquad(4.5.2)$$

CLUSTER PROPERTIES

| Cluster (1) | $\beta$ (2) | $r_c$ (3) | $T_X$ (4) | $M_{500}$ (5) | $r_{500}$ (6) | $M_{200}$ (7) | $r_{200}$ (8) | $M_A$ (9) | Ref. (10) |
|---|---|---|---|---|---|---|---|---|---|
| A0085 | $0.532^{+0.004}_{-0.004}$ | $83^{+3}_{-3}$ | $6.90^{+0.40}_{-0.40}$ | $6.84^{+0.66}_{-0.66}$ | $1.68^{+0.05}_{-0.06}$ | $10.80^{+1.12}_{-1.04}$ | $2.66^{+0.09}_{-0.09}$ | 12.21 | 1 |
| A0119 | $0.675^{+0.026}_{-0.023}$ | $501^{+28}_{-26}$ | $5.60^{+0.30}_{-0.30}$ | $6.23^{+0.92}_{-0.76}$ | $1.63^{+0.08}_{-0.07}$ | $10.76^{+1.50}_{-1.39}$ | $2.66^{+0.11}_{-0.13}$ | 12.24 | 1 |
| A0133 | $0.530^{+0.004}_{-0.004}$ | $45^{+2}_{-2}$ | $3.80^{+2.00}_{-0.90}$ | $2.78^{+2.51}_{-0.95}$ | $1.24^{+0.30}_{-0.16}$ | $4.41^{+4.00}_{-1.52}$ | $1.97^{+0.47}_{-0.27}$ | 6.71 | 9 |
| NGC 507 | $0.444^{+0.005}_{-0.005}$ | $19^{+1}_{-1}$ | $1.26^{+0.07}_{-0.07}$ | $0.41^{+0.04}_{-0.04}$ | $0.66^{+0.02}_{-0.02}$ | $0.64^{+0.07}_{-0.06}$ | $1.04^{+0.04}_{-0.04}$ | 1.86 | 2 |
| A0262 | $0.443^{+0.018}_{-0.017}$ | $42^{+12}_{-10}$ | $2.15^{+0.06}_{-0.09}$ | $0.90^{+0.10}_{-0.09}$ | $0.86^{+0.03}_{-0.03}$ | $1.42^{+0.15}_{-0.13}$ | $1.35^{+0.05}_{-0.04}$ | 3.17 | 2 |
| A0400 | $0.534^{+0.014}_{-0.013}$ | $154^{+9}_{-9}$ | $2.31^{+0.14}_{-0.14}$ | $1.28^{+0.17}_{-0.15}$ | $0.96^{+0.04}_{-0.04}$ | $2.07^{+0.30}_{-0.25}$ | $1.53^{+0.08}_{-0.06}$ | 4.10 | 2 |
| A0399 | $0.713^{+0.137}_{-0.095}$ | $450^{+132}_{-100}$ | $7.00^{+0.40}_{-0.40}$ | $10.00^{+3.73}_{-2.48}$ | $1.91^{+0.21}_{-0.18}$ | $16.64^{+6.61}_{-4.32}$ | $3.07^{+0.36}_{-0.30}$ | 16.24 | 1 |
| A0401 | $0.613^{+0.010}_{-0.010}$ | $246^{+11}_{-10}$ | $8.00^{+0.40}_{-0.40}$ | $10.27^{+1.08}_{-0.93}$ | $1.92^{+0.07}_{-0.05}$ | $16.59^{+1.62}_{-1.40}$ | $3.07^{+0.09}_{-0.10}$ | 16.21 | 1 |
| A3112 | $0.576^{+0.006}_{-0.006}$ | $61^{+3}_{-3}$ | $5.30^{+0.70}_{-1.00}$ | $5.17^{+1.17}_{-1.45}$ | $1.53^{+0.11}_{-0.16}$ | $8.22^{+1.79}_{-2.31}$ | $2.43^{+0.16}_{-0.25}$ | 10.16 | 1 |
| Fornax | $0.804^{+0.098}_{-0.084}$ | $174^{+17}_{-15}$ | $1.20^{+0.04}_{-0.04}$ | $0.87^{+0.22}_{-0.16}$ | $0.84^{+0.07}_{-0.06}$ | $1.42^{+0.36}_{-0.27}$ | $1.35^{+0.11}_{-0.09}$ | 3.20 | 2 |
| 2A 0335 | $0.575^{+0.004}_{-0.003}$ | $33^{+1}_{-1}$ | $3.01^{+0.07}_{-0.07}$ | $2.21^{+0.10}_{-0.09}$ | $1.15^{+0.02}_{-0.02}$ | $3.51^{+0.16}_{-0.15}$ | $1.83^{+0.03}_{-0.03}$ | 5.76 | 2 |
| Zw III 54 | $0.887^{+0.320}_{-0.151}$ | $289^{+124}_{-73}$ | $(2.16^{+2.22}_{-0.30})$ | $2.36^{+2.22}_{-0.90}$ | $1.18^{+0.29}_{-0.17}$ | $3.93^{+3.83}_{-1.54}$ | $1.89^{+0.48}_{-0.29}$ | 6.32 | 11 |
| A3158 | $0.661^{+0.025}_{-0.022}$ | $269^{+20}_{-19}$ | $5.77^{+0.10}_{-0.05}$ | $7.00^{+0.52}_{-0.42}$ | $1.69^{+0.04}_{-0.03}$ | $11.29^{+0.95}_{-0.79}$ | $2.69^{+0.08}_{-0.06}$ | 12.61 | 3 |
| A0478 | $0.613^{+0.004}_{-0.004}$ | $98^{+2}_{-2}$ | $8.40^{+0.80}_{-1.40}$ | $11.32^{+1.78}_{-2.81}$ | $1.99^{+0.10}_{-0.18}$ | $17.89^{+2.95}_{-4.35}$ | $3.13^{+0.18}_{-0.27}$ | 17.12 | 1 |
| NGC 1550 | $0.554^{+0.049}_{-0.037}$ | $45^{+15}_{-11}$ | $1.43^{+0.04}_{-0.03}$ | $0.69^{+0.12}_{-0.09}$ | $0.78^{+0.04}_{-0.04}$ | $1.09^{+0.20}_{-0.14}$ | $1.23^{+0.07}_{-0.06}$ | 2.64 | 5 |
| EXO 0422 | $0.722^{+0.104}_{-0.071}$ | $142^{+40}_{-30}$ | $2.90^{+0.90}_{-0.60}$ | $2.89^{+2.39}_{-1.14}$ | $1.26^{+0.28}_{-0.19}$ | $4.63^{+3.84}_{-1.82}$ | $2.00^{+0.44}_{-0.31}$ | 6.96 | 9 |
| A3266 | $0.796^{+0.020}_{-0.019}$ | $564^{+21}_{-20}$ | $8.00^{+0.50}_{-0.50}$ | $14.17^{+1.94}_{-1.84}$ | $2.14^{+0.09}_{-0.10}$ | $23.76^{+3.23}_{-2.91}$ | $3.45^{+0.15}_{-0.15}$ | 20.47 | 1 |
| A0496 | $0.484^{+0.003}_{-0.003}$ | $30^{+1}_{-1}$ | $4.13^{+0.08}_{-0.08}$ | $2.76^{+0.11}_{-0.11}$ | $1.24^{+0.02}_{-0.02}$ | $4.35^{+0.18}_{-0.17}$ | $1.96^{+0.03}_{-0.03}$ | 6.66 | 2 |
| A3376 | $1.054^{+0.101}_{-0.083}$ | $755^{+69}_{-60}$ | $4.00^{+0.40}_{-0.40}$ | $6.32^{+2.11}_{-1.59}$ | $1.64^{+0.17}_{-0.15}$ | $11.96^{+3.82}_{-2.91}$ | $2.75^{+0.26}_{-0.25}$ | 13.20 | 1 |
| A3391 | $0.579^{+0.026}_{-0.024}$ | $234^{+24}_{-22}$ | $5.40^{+0.60}_{-0.60}$ | $5.18^{+1.31}_{-1.08}$ | $1.53^{+0.12}_{-0.11}$ | $8.41^{+2.13}_{-1.81}$ | $2.44^{+0.19}_{-0.19}$ | 10.35 | 1 |
| A3395s | $0.964^{+0.275}_{-0.167}$ | $604^{+173}_{-118}$ | $5.00^{+0.30}_{-0.30}$ | $8.82^{+4.79}_{-2.61}$ | $1.83^{+0.29}_{-0.20}$ | $15.34^{+8.79}_{-4.74}$ | $2.99^{+0.49}_{-0.35}$ | 15.42 | 1 |
| A0576 | $0.825^{+0.432}_{-0.185}$ | $394^{+221}_{-125}$ | $4.02^{+0.07}_{-0.07}$ | $5.36^{+4.42}_{-1.66}$ | $1.55^{+0.34}_{-0.18}$ | $8.96^{+8.01}_{-2.91}$ | $2.50^{+0.60}_{-0.31}$ | 10.86 | 3 |
| A0754 | $0.698^{+0.027}_{-0.024}$ | $239^{+17}_{-16}$ | $9.50^{+0.70}_{-0.40}$ | $16.37^{+2.91}_{-1.84}$ | $2.25^{+0.13}_{-0.09}$ | $26.19^{+4.45}_{-2.95}$ | $3.57^{+0.18}_{-0.15}$ | 21.94 | 1 |
| Hydra A | $0.573^{+0.003}_{-0.003}$ | $50^{+1}_{-1}$ | $4.30^{+0.40}_{-0.40}$ | $3.76^{+0.58}_{-0.55}$ | $1.38^{+0.07}_{-0.07}$ | $5.94^{+0.91}_{-0.84}$ | $2.17^{+0.11}_{-0.10}$ | 8.21 | 1 |
| A1060 | $0.607^{+0.040}_{-0.034}$ | $94^{+15}_{-13}$ | $3.24^{+0.06}_{-0.06}$ | $2.66^{+0.34}_{-0.28}$ | $1.23^{+0.05}_{-0.04}$ | $4.24^{+0.55}_{-0.47}$ | $1.95^{+0.08}_{-0.08}$ | 6.54 | 2 |
| A1367 | $0.695^{+0.035}_{-0.032}$ | $383^{+24}_{-22}$ | $3.55^{+0.08}_{-0.08}$ | $3.34^{+0.36}_{-0.32}$ | $1.32^{+0.05}_{-0.04}$ | $5.69^{+0.63}_{-0.56}$ | $2.14^{+0.08}_{-0.07}$ | 8.08 | 2 |
| MKW 4 | $0.440^{+0.004}_{-0.005}$ | $11^{+1}_{-1}$ | $1.71^{+0.09}_{-0.09}$ | $0.64^{+0.06}_{-0.06}$ | $0.76^{+0.03}_{-0.03}$ | $1.00^{+0.10}_{-0.09}$ | $1.20^{+0.04}_{-0.03}$ | 2.51 | 2 |
| Zw Cl 1215 | $0.819^{+0.038}_{-0.034}$ | $431^{+28}_{-25}$ | $(5.58^{+0.89}_{-0.78})$ | $8.79^{+3.00}_{-2.29}$ | $1.83^{+0.19}_{-0.18}$ | $14.52^{+4.92}_{-3.67}$ | $2.93^{+0.30}_{-0.27}$ | 14.91 | 11 |
| NGC 4636 | $0.491^{+0.032}_{-0.027}$ | $6^{+3}_{-2}$ | $0.76^{+0.01}_{-0.01}$ | $0.22^{+0.03}_{-0.02}$ | $0.53^{+0.02}_{-0.02}$ | $0.35^{+0.04}_{-0.04}$ | $0.85^{+0.03}_{-0.03}$ | 1.24 | 4 |
| A3526 | $0.495^{+0.011}_{-0.010}$ | $37^{+5}_{-4}$ | $3.68^{+0.06}_{-0.06}$ | $2.39^{+0.15}_{-0.13}$ | $1.18^{+0.02}_{-0.02}$ | $3.78^{+0.23}_{-0.18}$ | $1.87^{+0.04}_{-0.03}$ | 6.07 | 2 |
| A1644 | $0.579^{+0.111}_{-0.074}$ | $300^{+128}_{-92}$ | $4.70^{+0.90}_{-0.70}$ | $4.10^{+2.64}_{-1.41}$ | $1.42^{+0.26}_{-0.18}$ | $6.73^{+4.54}_{-2.38}$ | $2.27^{+0.43}_{-0.31}$ | 8.98 | 10 |
| A1650 | $0.704^{+0.131}_{-0.081}$ | $281^{+104}_{-71}$ | $6.70^{+0.80}_{-0.80}$ | $9.62^{+4.91}_{-2.92}$ | $1.88^{+0.28}_{-0.21}$ | $15.60^{+8.08}_{-4.85}$ | $3.01^{+0.45}_{-0.35}$ | 15.56 | 1 |
| A1651 | $0.643^{+0.014}_{-0.013}$ | $181^{+10}_{-10}$ | $6.10^{+0.40}_{-0.40}$ | $7.45^{+1.00}_{-0.95}$ | $1.73^{+0.07}_{-0.08}$ | $11.91^{+1.60}_{-1.52}$ | $2.75^{+0.12}_{-0.13}$ | 13.01 | 1 |
| Coma | $0.654^{+0.019}_{-0.021}$ | $344^{+22}_{-21}$ | $8.38^{+0.34}_{-0.34}$ | $11.99^{+1.28}_{-1.29}$ | $2.03^{+0.08}_{-0.08}$ | $19.38^{+2.08}_{-1.97}$ | $3.22^{+0.11}_{-0.11}$ | 18.01 | 2 |
| NGC 5044 | $0.524^{+0.002}_{-0.003}$ | $11^{+1}_{-1}$ | $1.07^{+0.01}_{-0.01}$ | $0.41^{+0.01}_{-0.01}$ | $0.66^{+0.01}_{-0.01}$ | $0.65^{+0.01}_{-0.01}$ | $1.04^{+0.01}_{-0.01}$ | 1.87 | 2 |
| A1736 | $0.542^{+0.147}_{-0.092}$ | $374^{+178}_{-130}$ | $3.50^{+0.40}_{-0.40}$ | $2.19^{+1.23}_{-0.74}$ | $1.15^{+0.18}_{-0.15}$ | $3.78^{+2.41}_{-1.34}$ | $1.87^{+0.34}_{-0.25}$ | 6.22 | 1 |
| A3558 | $0.580^{+0.006}_{-0.005}$ | $224^{+5}_{-5}$ | $5.50^{+0.40}_{-0.40}$ | $5.37^{+0.70}_{-0.64}$ | $1.55^{+0.07}_{-0.06}$ | $8.64^{+1.12}_{-1.03}$ | $2.46^{+0.10}_{-0.10}$ | 10.56 | 1 |
| A3562 | $0.472^{+0.006}_{-0.006}$ | $99^{+5}_{-5}$ | $5.16^{+0.16}_{-0.16}$ | $3.68^{+0.24}_{-0.23}$ | $1.37^{+0.03}_{-0.03}$ | $5.83^{+0.38}_{-0.36}$ | $2.16^{+0.05}_{-0.04}$ | 8.10 | 3 |
| A3571 | $0.613^{+0.010}_{-0.010}$ | $181^{+7}_{-7}$ | $6.90^{+0.20}_{-0.20}$ | $8.33^{+0.56}_{-0.53}$ | $1.79^{+0.04}_{-0.04}$ | $13.31^{+0.90}_{-0.85}$ | $2.85^{+0.06}_{-0.06}$ | 14.04 | 1 |
| A1795 | $0.596^{+0.003}_{-0.002}$ | $78^{+1}_{-1}$ | $7.80^{+1.00}_{-1.00}$ | $9.75^{+2.01}_{-1.90}$ | $1.89^{+0.12}_{-0.14}$ | $15.39^{+3.17}_{-2.92}$ | $2.99^{+0.19}_{-0.20}$ | 15.46 | 1 |
| A3581 | $0.543^{+0.024}_{-0.022}$ | $35^{+5}_{-4}$ | $1.83^{+0.04}_{-0.04}$ | $0.96^{+0.09}_{-0.09}$ | $0.87^{+0.02}_{-0.03}$ | $1.52^{+0.16}_{-0.13}$ | $1.38^{+0.05}_{-0.04}$ | 3.30 | 5 |
| MKW 8 | $0.511^{+0.098}_{-0.059}$ | $107^{+70}_{-42}$ | $3.29^{+0.23}_{-0.22}$ | $2.10^{+0.86}_{-0.52}$ | $1.14^{+0.13}_{-0.10}$ | $3.33^{+1.45}_{-0.83}$ | $1.79^{+0.24}_{-0.17}$ | 5.60 | 5 |
| A2029 | $0.582^{+0.004}_{-0.004}$ | $83^{+2}_{-2}$ | $9.10^{+1.00}_{-1.00}$ | $11.82^{+2.14}_{-1.99}$ | $2.01^{+0.11}_{-0.12}$ | $18.79^{+3.40}_{-3.17}$ | $3.20^{+0.18}_{-0.19}$ | 17.62 | 1 |
| A2052 | $0.526^{+0.005}_{-0.005}$ | $37^{+2}_{-2}$ | $3.03^{+0.04}_{-0.04}$ | $1.95^{+0.07}_{-0.07}$ | $1.10^{+0.02}_{-0.01}$ | $3.10^{+0.09}_{-0.11}$ | $1.75^{+0.01}_{-0.02}$ | 5.30 | 3 |
| MKW 3s | $0.581^{+0.008}_{-0.007}$ | $66^{+3}_{-3}$ | $3.70^{+0.20}_{-0.20}$ | $3.06^{+0.32}_{-0.30}$ | $1.29^{+0.05}_{-0.04}$ | $4.84^{+0.51}_{-0.47}$ | $2.03^{+0.07}_{-0.07}$ | 7.16 | 1 |
| A2065 | $1.162^{+0.734}_{-0.282}$ | $690^{+360}_{-186}$ | $5.50^{+0.40}_{-0.40}$ | $13.44^{+16.12}_{-5.17}$ | $2.10^{+0.63}_{-0.31}$ | $23.37^{+29.87}_{-9.42}$ | $3.43^{+1.09}_{-0.54}$ | 20.21 | 1 |
| A2063 | $0.561^{+0.011}_{-0.011}$ | $110^{+7}_{-6}$ | $3.68^{+0.11}_{-0.11}$ | $2.84^{+0.23}_{-0.19}$ | $1.25^{+0.04}_{-0.03}$ | $4.54^{+0.36}_{-0.31}$ | $1.99^{+0.06}_{-0.04}$ | 6.86 | 2 |
| A2142 | $0.591^{+0.006}_{-0.006}$ | $154^{+6}_{-6}$ | $9.70^{+1.50}_{-1.10}$ | $13.29^{+3.45}_{-2.41}$ | $2.10^{+0.17}_{-0.14}$ | $21.04^{+5.46}_{-3.69}$ | $3.31^{+0.26}_{-0.20}$ | 19.05 | 1 |
| A2147 | $0.444^{+0.071}_{-0.046}$ | $238^{+103}_{-65}$ | $4.91^{+0.28}_{-0.28}$ | $2.99^{+0.92}_{-0.63}$ | $1.28^{+0.12}_{-0.10}$ | $4.84^{+1.64}_{-1.03}$ | $2.03^{+0.21}_{-0.15}$ | 7.21 | 2 |
| A2163 | $0.796^{+0.030}_{-0.028}$ | $519^{+31}_{-29}$ | $13.29^{+4.24}_{-0.64}$ | $31.85^{+4.96}_{-3.74}$ | $2.81^{+0.12}_{-0.11}$ | $51.99^{+6.96}_{-6.13}$ | $4.49^{+0.19}_{-0.18}$ | 34.18 | 3 |
| A2199 | $0.655^{+0.019}_{-0.021}$ | $139^{+10}_{-10}$ | $4.10^{+0.08}_{-0.08}$ | $4.21^{+0.33}_{-0.29}$ | $1.43^{+0.04}_{-0.03}$ | $6.73^{+0.52}_{-0.51}$ | $2.27^{+0.06}_{-0.06}$ | 8.92 | 2 |
| A2204 | $0.597^{+0.008}_{-0.008}$ | $67^{+3}_{-3}$ | $7.21^{+0.25}_{-0.25}$ | $8.67^{+0.67}_{-0.57}$ | $1.82^{+0.05}_{-0.04}$ | $13.79^{+0.96}_{-1.00}$ | $2.89^{+0.06}_{-0.06}$ | 14.34 | 3 |
| A2244 | $0.607^{+0.016}_{-0.015}$ | $126^{+11}_{-10}$ | $7.10^{+5.00}_{-2.20}$ | $8.65^{+11.47}_{-3.89}$ | $1.82^{+0.59}_{-0.33}$ | $13.78^{+18.02}_{-6.20}$ | $2.89^{+0.92}_{-0.52}$ | 14.33 | 10 |
| A2256 | $0.914^{+0.054}_{-0.047}$ | $587^{+40}_{-37}$ | $6.60^{+0.40}_{-0.40}$ | $12.83^{+2.38}_{-2.00}$ | $2.07^{+0.12}_{-0.11}$ | $21.81^{+4.07}_{-3.54}$ | $3.36^{+0.19}_{-0.20}$ | 19.34 | 1 |
| A2255 | $0.797^{+0.033}_{-0.030}$ | $593^{+35}_{-32}$ | $6.87^{+0.20}_{-0.20}$ | $10.90^{+1.15}_{-0.95}$ | $1.96^{+0.07}_{-0.05}$ | $18.65^{+2.01}_{-1.67}$ | $3.18^{+0.11}_{-0.09}$ | 17.54 | 3 |
| A3667 | $0.541^{+0.008}_{-0.008}$ | $279^{+10}_{-10}$ | $7.00^{+0.60}_{-0.60}$ | $6.88^{+1.08}_{-1.02}$ | $1.68^{+0.08}_{-0.09}$ | $11.19^{+1.76}_{-1.65}$ | $2.69^{+0.13}_{-0.14}$ | 12.50 | 1 |
| S1101 | $0.639^{+0.006}_{-0.007}$ | $56^{+2}_{-2}$ | $3.00^{+1.20}_{-0.88}$ | $2.58^{+1.76}_{-0.88}$ | $1.22^{+0.23}_{-0.16}$ | $4.08^{+2.78}_{-1.39}$ | $1.92^{+0.36}_{-0.25}$ | 6.38 | 9 |
| A2589 | $0.596^{+0.013}_{-0.012}$ | $118^{+8}_{-7}$ | $3.70^{+2.20}_{-1.10}$ | $3.14^{+3.44}_{-1.35}$ | $1.29^{+0.37}_{-0.22}$ | $5.01^{+5.41}_{-2.15}$ | $2.06^{+0.56}_{-0.35}$ | 7.33 | 9 |
| A2597 | $0.633^{+0.008}_{-0.008}$ | $58^{+2}_{-2}$ | $4.40^{+0.40}_{-0.70}$ | $4.52^{+0.72}_{-1.11}$ | $1.47^{+0.07}_{-0.14}$ | $7.14^{+1.14}_{-1.74}$ | $2.31^{+0.11}_{-0.20}$ | 9.27 | 1 |
| A2634 | $0.640^{+0.051}_{-0.043}$ | $364^{+44}_{-39}$ | $3.70^{+0.28}_{-0.28}$ | $3.15^{+0.78}_{-0.60}$ | $1.29^{+0.10}_{-0.09}$ | $5.35^{+1.34}_{-1.04}$ | $2.10^{+0.17}_{-0.14}$ | 7.77 | 2 |
| A2657 | $0.556^{+0.008}_{-0.007}$ | $119^{+5}_{-5}$ | $3.70^{+0.30}_{-0.30}$ | $2.83^{+0.43}_{-0.39}$ | $1.25^{+0.06}_{-0.06}$ | $4.52^{+0.68}_{-0.62}$ | $1.99^{+0.10}_{-0.09}$ | 6.84 | 1 |
| A4038 | $0.541^{+0.009}_{-0.008}$ | $59^{+4}_{-4}$ | $3.15^{+0.03}_{-0.03}$ | $2.16^{+0.09}_{-0.08}$ | $1.14^{+0.02}_{-0.02}$ | $3.41^{+0.13}_{-0.14}$ | $1.80^{+0.03}_{-0.01}$ | 5.67 | 3 |
| A4059 | $0.582^{+0.010}_{-0.010}$ | $90^{+5}_{-5}$ | $4.40^{+0.30}_{-0.30}$ | $3.95^{+0.52}_{-0.48}$ | $1.40^{+0.06}_{-0.06}$ | $6.30^{+0.83}_{-0.76}$ | $2.22^{+0.09}_{-0.09}$ | 8.52 | 1 |

| Cluster (1) | $\beta$ (2) | $r_c$ (3) | $T_X$ (4) | $M_{500}$ (5) | $r_{500}$ (6) | $M_{200}$ (7) | $r_{200}$ (8) | $M_A$ (9) | Ref. (10) |
|---|---|---|---|---|---|---|---|---|---|
| Clusters from the Extended Sample Not Included in HIFLUGCS ||||||||||
| A2734 | $0.624^{+0.034}_{-0.029}$ | $212^{+26}_{-23}$ | $(3.85^{+0.62}_{-0.54})$ | $3.49^{+1.25}_{-0.89}$ | $1.34^{+0.15}_{-0.12}$ | $5.67^{+1.98}_{-1.48}$ | $2.14^{+0.22}_{-0.21}$ | 7.97 | 11 |
| A2877 | $0.566^{+0.029}_{-0.025}$ | $190^{+19}_{-17}$ | $3.50^{+2.20}_{-1.10}$ | $2.61^{+3.32}_{-1.24}$ | $1.22^{+0.39}_{-0.23}$ | $4.24^{+5.28}_{-2.00}$ | $1.95^{+0.60}_{-0.38}$ | 6.57 | 10 |
| NGC 499 | $0.722^{+0.034}_{-0.030}$ | $24^{+2}_{-2}$ | $0.72^{+0.03}_{-0.02}$ | $0.36^{+0.05}_{-0.04}$ | $0.63^{+0.03}_{-0.02}$ | $0.58^{+0.08}_{-0.06}$ | $1.00^{+0.04}_{-0.04}$ | 1.73 | 4 |
| AWM 7 | $0.671^{+0.027}_{-0.025}$ | $173^{+18}_{-15}$ | $3.75^{+0.09}_{-0.09}$ | $3.79^{+0.38}_{-0.32}$ | $1.38^{+0.05}_{-0.04}$ | $6.08^{+0.62}_{-0.52}$ | $2.19^{+0.08}_{-0.06}$ | 8.35 | 2 |
| Perseus | $0.540^{+0.006}_{-0.004}$ | $64^{+2}_{-2}$ | $6.79^{+0.12}_{-0.12}$ | $6.84^{+0.29}_{-0.26}$ | $1.68^{+0.02}_{-0.02}$ | $10.80^{+0.46}_{-0.41}$ | $2.66^{+0.04}_{-0.04}$ | 12.20 | 2 |
| S405 | $0.664^{+0.263}_{-0.133}$ | $459^{+262}_{-159}$ | $(4.21^{+0.67}_{-0.59})$ | $3.91^{+3.56}_{-1.57}$ | $1.40^{+0.33}_{-0.22}$ | $6.75^{+6.80}_{-2.81}$ | $2.27^{+0.60}_{-0.37}$ | 9.09 | 11 |
| 3C 129 | $0.601^{+0.260}_{-0.131}$ | $318^{+178}_{-107}$ | $5.60^{+0.70}_{-0.60}$ | $5.68^{+5.58}_{-2.29}$ | $1.58^{+0.40}_{-0.25}$ | $9.30^{+9.51}_{-3.85}$ | $2.53^{+0.67}_{-0.42}$ | 11.08 | 9 |
| A0539 | $0.561^{+0.020}_{-0.018}$ | $148^{+13}_{-12}$ | $3.24^{+0.09}_{-0.09}$ | $2.33^{+0.21}_{-0.19}$ | $1.18^{+0.03}_{-0.03}$ | $3.74^{+0.35}_{-0.34}$ | $1.87^{+0.05}_{-0.06}$ | 6.04 | 2 |
| S540 | $0.641^{+0.073}_{-0.051}$ | $130^{+38}_{-29}$ | $(2.40^{+0.38}_{-0.34})$ | $1.83^{+0.83}_{-0.54}$ | $1.08^{+0.21}_{-0.12}$ | $2.93^{+1.34}_{-0.87}$ | $1.72^{+0.23}_{-0.19}$ | 5.13 | 11 |
| A0548w | $0.666^{+0.194}_{-0.111}$ | $198^{+90}_{-62}$ | $(1.20^{+0.19}_{-0.17})$ | $0.63^{+0.48}_{-0.23}$ | $0.76^{+0.16}_{-0.11}$ | $1.06^{+0.84}_{-0.41}$ | $1.23^{+0.26}_{-0.19}$ | 2.64 | 11 |
| A0548e | $0.480^{+0.013}_{-0.013}$ | $118^{+12}_{-11}$ | $3.10^{+0.10}_{-0.10}$ | $1.74^{+0.15}_{-0.15}$ | $1.07^{+0.03}_{-0.03}$ | $2.77^{+0.27}_{-0.23}$ | $1.68^{+0.06}_{-0.05}$ | 4.95 | 3 |
| A3395n | $0.981^{+0.619}_{-0.244}$ | $672^{+383}_{-203}$ | $5.00^{+0.30}_{-0.30}$ | $8.70^{+9.53}_{-3.19}$ | $1.82^{+0.51}_{-0.26}$ | $15.47^{+18.82}_{-6.07}$ | $2.99^{+0.92}_{-0.46}$ | 15.55 | 1 |
| UGC 03957 | $0.740^{+0.133}_{-0.086}$ | $142^{+45}_{-33}$ | $(2.58^{+0.43}_{-0.36})$ | $2.51^{+1.50}_{-0.83}$ | $1.20^{+0.22}_{-0.15}$ | $4.02^{+2.41}_{-1.33}$ | $1.91^{+0.33}_{-0.23}$ | 6.35 | 11 |
| PKS 0745 | $0.608^{+0.006}_{-0.006}$ | $71^{+2}_{-2}$ | $7.21^{+0.11}_{-0.11}$ | $8.88^{+0.35}_{-0.28}$ | $1.83^{+0.03}_{-0.01}$ | $14.12^{+0.56}_{-0.53}$ | $2.91^{+0.04}_{-0.04}$ | 14.58 | 3 |
| A0644 | $0.700^{+0.011}_{-0.011}$ | $203^{+7}_{-7}$ | $7.90^{+0.80}_{-0.80}$ | $12.50^{+2.79}_{-2.11}$ | $2.06^{+0.12}_{-0.12}$ | $19.83^{+3.79}_{-3.23}$ | $3.24^{+0.21}_{-0.17}$ | 18.33 | 1 |
| S636 | $0.752^{+0.217}_{-0.123}$ | $344^{+130}_{-86}$ | $(1.18^{+0.44}_{-0.17})$ | $0.61^{+0.68}_{-0.22}$ | $0.75^{+0.10}_{-0.10}$ | $1.16^{+0.90}_{-0.44}$ | $1.26^{+0.27}_{-0.18}$ | 2.93 | 11 |
| A1413 | $0.660^{+0.017}_{-0.015}$ | $179^{+12}_{-11}$ | $7.32^{+0.26}_{-0.24}$ | $10.20^{+0.93}_{-0.82}$ | $1.92^{+0.05}_{-0.05}$ | $16.29^{+1.49}_{-1.31}$ | $3.05^{+0.09}_{-0.08}$ | 16.03 | 3 |
| M49 | $0.592^{+0.007}_{-0.007}$ | $11^{+1}_{-1}$ | $0.95^{+0.02}_{-0.02}$ | $0.41^{+0.02}_{-0.01}$ | $0.66^{+0.01}_{-0.01}$ | $0.65^{+0.04}_{-0.03}$ | $1.04^{+0.02}_{-0.01}$ | 1.87 | 4 |
| A3528n | $0.621^{+0.034}_{-0.030}$ | $178^{+17}_{-14}$ | $3.40^{+1.66}_{-0.64}$ | $2.89^{+2.84}_{-0.94}$ | $1.26^{+0.32}_{-0.15}$ | $4.65^{+4.54}_{-1.48}$ | $2.00^{+0.51}_{-0.23}$ | 7.00 | 8 |
| A3528s | $0.463^{+0.013}_{-0.012}$ | $101^{+9}_{-8}$ | $3.15^{+0.89}_{-0.59}$ | $1.69^{+0.87}_{-0.50}$ | $1.05^{+0.16}_{-0.12}$ | $2.70^{+1.39}_{-0.80}$ | $1.67^{+0.25}_{-0.19}$ | 4.86 | 8 |
| A3530 | $0.773^{+0.114}_{-0.085}$ | $421^{+75}_{-61}$ | $3.89^{+0.27}_{-0.25}$ | $4.52^{+1.52}_{-1.05}$ | $1.47^{+0.15}_{-0.13}$ | $7.64^{+2.72}_{-1.80}$ | $2.36^{+0.26}_{-0.20}$ | 9.82 | 7 |
| A3532 | $0.653^{+0.034}_{-0.029}$ | $282^{+27}_{-24}$ | $4.58^{+0.19}_{-0.17}$ | $4.77^{+0.86}_{-0.52}$ | $1.49^{+0.07}_{-0.05}$ | $7.79^{+1.16}_{-0.91}$ | $2.38^{+0.12}_{-0.10}$ | 9.88 | 7 |
| A1689 | $0.690^{+0.011}_{-0.011}$ | $163^{+7}_{-6}$ | $9.23^{+0.28}_{-0.28}$ | $15.49^{+1.18}_{-1.00}$ | $2.20^{+0.05}_{-0.05}$ | $24.68^{+1.70}_{-1.76}$ | $3.50^{+0.07}_{-0.08}$ | 21.13 | 3 |
| A3560 | $0.566^{+0.033}_{-0.029}$ | $256^{+30}_{-27}$ | $(3.16^{+0.51}_{-0.44})$ | $2.16^{+0.79}_{-0.56}$ | $1.14^{+0.12}_{-0.11}$ | $3.59^{+1.30}_{-0.95}$ | $1.84^{+0.20}_{-0.18}$ | 5.92 | 11 |
| A1775 | $0.673^{+0.026}_{-0.023}$ | $260^{+19}_{-18}$ | $3.69^{+0.20}_{-0.34}$ | $3.61^{+0.50}_{-0.34}$ | $1.36^{+0.06}_{-0.06}$ | $5.91^{+0.83}_{-0.77}$ | $2.17^{+0.09}_{-0.07}$ | 8.21 | 3 |
| A1800 | $0.766^{+0.308}_{-0.139}$ | $392^{+223}_{-132}$ | $(4.02^{+0.64}_{-0.56})$ | $4.75^{+4.64}_{-1.85}$ | $1.49^{+0.38}_{-0.23}$ | $7.97^{+8.31}_{-3.17}$ | $2.39^{+0.65}_{-0.37}$ | 10.08 | 11 |
| A1914 | $0.751^{+0.018}_{-0.017}$ | $231^{+11}_{-10}$ | $10.53^{+0.51}_{-0.50}$ | $21.43^{+2.39}_{-2.16}$ | $2.46^{+0.09}_{-0.08}$ | $33.99^{+4.06}_{-3.43}$ | $3.88^{+0.16}_{-0.13}$ | 26.20 | 3 |
| NGC 5813 | $0.766^{+0.179}_{-0.103}$ | $25^{+9}_{-6}$ | $(0.52^{+0.08}_{-0.07})$ | $0.24^{+0.17}_{-0.08}$ | $0.55^{+0.11}_{-0.07}$ | $0.38^{+0.27}_{-0.13}$ | $0.87^{+0.16}_{-0.12}$ | 1.32 | 11 |
| NGC 5846 | $0.599^{+0.016}_{-0.015}$ | $7^{+1}_{-1}$ | $0.82^{+0.01}_{-0.01}$ | $0.33^{+0.02}_{-0.02}$ | $0.61^{+0.01}_{-0.01}$ | $0.53^{+0.03}_{-0.03}$ | $0.97^{+0.02}_{-0.01}$ | 1.63 | 4 |
| A2151w | $0.564^{+0.014}_{-0.013}$ | $68^{+5}_{-5}$ | $2.40^{+0.06}_{-0.06}$ | $1.52^{+0.12}_{-0.10}$ | $1.02^{+0.03}_{-0.02}$ | $2.42^{+0.18}_{-0.18}$ | $1.61^{+0.03}_{-0.04}$ | 4.51 | 3 |
| A3627 | $0.555^{+0.056}_{-0.044}$ | $299^{+56}_{-49}$ | $6.02^{+0.08}_{-0.08}$ | $5.63^{+0.95}_{-0.68}$ | $1.57^{+0.09}_{-0.06}$ | $9.20^{+1.61}_{-1.14}$ | $2.51^{+0.14}_{-0.10}$ | 11.03 | 3 |
| Triangulum | $0.610^{+0.010}_{-0.010}$ | $279^{+11}_{-10}$ | $9.60^{+0.60}_{-0.60}$ | $13.42^{+1.70}_{-1.36}$ | $2.10^{+0.09}_{-0.07}$ | $21.54^{+2.73}_{-2.36}$ | $3.34^{+0.14}_{-0.11}$ | 19.35 | 1 |
| Ophiuchus | $0.747^{+0.035}_{-0.032}$ | $279^{+23}_{-22}$ | $10.26^{+0.32}_{-0.32}$ | $20.25^{+2.51}_{-2.10}$ | $2.41^{+0.10}_{-0.08}$ | $32.43^{+4.05}_{-3.38}$ | $3.83^{+0.16}_{-0.13}$ | 25.32 | 2 |
| Zw Cl 1742 | $0.717^{+0.073}_{-0.053}$ | $232^{+46}_{-38}$ | $(5.23^{+0.84}_{-0.73})$ | $6.88^{+3.06}_{-1.96}$ | $1.68^{+0.22}_{-0.18}$ | $11.05^{+4.93}_{-3.16}$ | $2.67^{+0.35}_{-0.28}$ | 12.42 | 11 |
| A2319 | $0.591^{+0.013}_{-0.012}$ | $285^{+15}_{-14}$ | $8.80^{+0.50}_{-0.50}$ | $11.16^{+1.39}_{-1.20}$ | $1.97^{+0.08}_{-0.07}$ | $18.07^{+2.12}_{-2.06}$ | $3.16^{+0.11}_{-0.13}$ | 17.17 | 1 |
| A3695 | $0.642^{+0.259}_{-0.117}$ | $399^{+254}_{-149}$ | $(5.29^{+0.85}_{-0.74})$ | $5.57^{+5.29}_{-2.16}$ | $1.57^{+0.39}_{-0.24}$ | $9.32^{+9.56}_{-3.74}$ | $2.53^{+0.67}_{-0.40}$ | 11.12 | 11 |
| Zw II 108 | $0.662^{+0.167}_{-0.097}$ | $365^{+159}_{-105}$ | $(3.44^{+0.55}_{-0.48})$ | $2.96^{+2.09}_{-1.02}$ | $1.27^{+0.26}_{-0.16}$ | $5.04^{+3.60}_{-1.80}$ | $2.06^{+0.40}_{-0.28}$ | 7.47 | 11 |
| A3822 | $0.639^{+0.150}_{-0.093}$ | $351^{+160}_{-111}$ | $(4.90^{+0.78}_{-0.69})$ | $4.97^{+3.30}_{-1.75}$ | $1.51^{+0.28}_{-0.20}$ | $8.26^{+5.64}_{-3.34}$ | $2.43^{+0.46}_{-0.34}$ | 10.29 | 11 |
| A3827 | $0.989^{+0.410}_{-0.192}$ | $593^{+248}_{-149}$ | $(7.08^{+1.13}_{-0.99})$ | $16.35^{+17.02}_{-6.76}$ | $2.25^{+0.60}_{-0.32}$ | $27.44^{+29.53}_{-11.46}$ | $3.62^{+0.99}_{-0.60}$ | 22.44 | 11 |
| A3888 | $0.928^{+0.084}_{-0.066}$ | $401^{+46}_{-40}$ | $(8.84^{+1.41}_{-1.24})$ | $22.00^{+9.28}_{-6.28}$ | $2.48^{+0.31}_{-0.26}$ | $35.74^{+15.07}_{-10.38}$ | $3.96^{+0.49}_{-0.44}$ | 26.85 | 11 |
| A3921 | $0.762^{+0.036}_{-0.030}$ | $328^{+26}_{-23}$ | $5.73^{+0.24}_{-0.23}$ | $8.46^{+1.13}_{-0.96}$ | $1.80^{+0.08}_{-0.07}$ | $13.80^{+1.87}_{-1.59}$ | $2.89^{+0.12}_{-0.12}$ | 14.37 | 3 |
| HCG 94 | $0.514^{+0.007}_{-0.006}$ | $86^{+4}_{-4}$ | $3.45^{+0.30}_{-0.30}$ | $2.28^{+0.36}_{-0.34}$ | $1.17^{+0.06}_{-0.06}$ | $3.62^{+0.56}_{-0.51}$ | $1.84^{+0.09}_{-0.09}$ | 5.90 | 6 |
| RXJ 2344 | $0.807^{+0.033}_{-0.030}$ | $301^{+20}_{-18}$ | $(4.73^{+0.76}_{-0.66})$ | $6.91^{+2.30}_{-1.69}$ | $1.68^{+0.17}_{-0.14}$ | $11.27^{+3.74}_{-2.80}$ | $2.69^{+0.27}_{-0.25}$ | 12.58 | 11 |

جدول (۴.۱) ستون اول نام خوشه های کهکشانی است. ستون دوم پارامتر $\beta$ برای خوشه است. ستون سوم شعاع هسته خوشه کهکشانی در واحد $Kpc$ است. ستون چهارم دمای تابش X خوشه است. ستون های پنجم و هفتم مربوط به جرم های $M_{200}$ و $M_{500}$ در واحد $10^{14} M_\odot$ هستند. ستون های ششم و هشتم مربوط به شعاع های $r_{200}$ و $r_{500}$ در واحد $Mpc$ هستند. ستون نهم جرم کل خوشه کهکشانی تا شعاع آبل در واحد $10^{14} M_\odot$ است.

[برگرفته از مرجع T. H. Reiprich and H. Bringer, Astrophys. J. **567** (2002) 716]

۶۸

از معادلات (3.7.20) تا (3.7.22) داریم

$$m_3^2\left(\frac{2v_0'}{r}+v_0''\right)=3\mathcal{P}(r)+\rho(r). \tag{4.5.3}$$

با انتگرال گیری از این معادله به دست می آوریم

$$m_3^2 r^2 v_0' = \mathcal{N}(r)+\frac{1}{4\pi}M(r)+C_1, \tag{4.5.4}$$

که در آن $C_1$ ثابت انتگرال گیری است. از معادلات (4.5.1) و (4.5.4) به دست می آوریم

$$2m_3^2\frac{d}{dr}\left(\rho\sigma_r^2\right)=-\frac{\mathcal{N}(r)}{r^2}\rho(r)-\frac{M(r)}{4\pi r^2}\rho(r)-\frac{C_1}{r^2}\rho(r). \tag{4.5.5}$$

اگر از روابطی که برای $\rho(r)$ و $\mathcal{N}(r)$ در (4.4.1) و (4.4.12) داریم استفاده کنیم به دست می آوریم

$$\sigma_r^2 = \frac{4\pi k_B T_g}{\mu m_p}-\frac{\rho_0 r^2}{4(\gamma-3)(\gamma-1)m_3^2}\left(\frac{r}{r_c}\right)^{-\gamma}+\frac{C_1}{2(\gamma+1)m_3^2}\frac{1}{r}+\frac{C_2}{2m_3^2\rho_0}\left(\frac{r}{r_c}\right)^{\gamma} \quad \gamma\neq 3, \tag{4.5.6}$$

و

$$\sigma_r^2 = \frac{4\pi k_B T_g}{\mu m_p}+\frac{\rho_0 r_c^3}{8m_3^2}\frac{\ln r}{r}+\frac{C_3}{r}+\frac{C_2}{2m_3^2\rho_0}\left(\frac{r}{r_c}\right)^3 \quad \gamma=3, \tag{4.5.7}$$

که در آن برای سادگی تعریف کرده ایم $3\beta=\gamma$. معمولاً برای پراکندگی سرعت یک خوشه کهکشانی فرم ساده

$$\sigma_r^2 = \frac{B}{r+b}, \tag{4.5.8}$$

را برای انطباق بر نمودارهای تجربی انتخاب می کنند. همان طور که مشاهده می کنید معادله (4.5.6) دو ثابت قابل تنظیم دارد که می تواند آن را بر نمودارهای تجربی منطبق کند. همچنین می توان از این معادله برای مقایسه این مدل با مدل های دیگری که جرم ویریال را اندازه می گیرند استفاده کرد. در اینجا ما از ذکر جزئیات آن صرفنظر می کنیم. در آخر فرمول (4.5.6) را برای

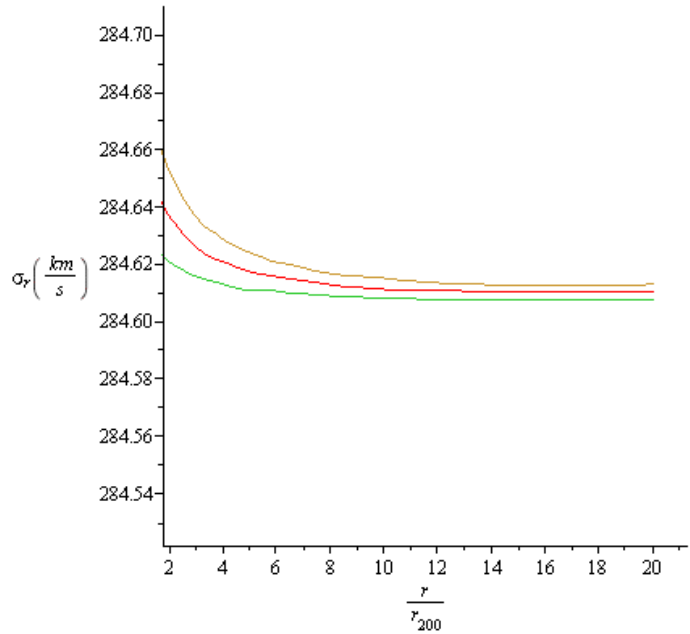

شکل (۴.۲) رسم پراکندگی سرعت شعاعی برای خوشه کهکشانی $NGC\,5813$. مقادیر ثابت های مسئله عبارتند از
$C_1 = 4.2 \times 10^{-8}, 2.9 \times 10^{-8}, 1.4 \times 10^{-8} M_\odot$ و $C_2 = 0.0390, 0.0260, 0.0130 M_\odot^2 / Kpc^4$.

پراکندگی سرعت شعاعی در خوشه کهکشانی $NGC\,5813$ می آزماییم. برای $\beta = 0.766$ است و $r_{200} = 0.87 Mpc, r_c = 25 Kpc, k_B T_g = 0.52 KeV$. با انتخاب ثابت های مسئله به صورت $C_1 = 4.2 \times 10^{-8}, 2.9 \times 10^{-8}, 1.4 \times 10^{-8} M_\odot$ و $C_2 = 0.0390, 0.0260, 0.0130 M_\odot^2 / Kpc^4$ به نمودارهای شکل (۴.۲) می رسیم. پراکندگی سرعت شعاعی برای این کهکشان در حدود $240 km/s$ است [۳۸].

**جمع بندی**

در این پایان نامه مسئله جرم ویریال در مدل شامه ای DGP بررسی شد. برای این منظور معادلات میدان در این مدل را برای یک شامه ایستا با تقارن کروی در معادلات (۳.۷.۲۰)-(۳.۷.۲۲) بدست آوردیم. با استفاده از این معادلات و معادله نسبیتی بولتزمن که در معادله (۴.۲.۵) داده شده است، قضیه ویریال را در کیهان شناسی DGP در معادله (۴.۳.۴) بدست آوردیم. یکی از نتایج این معادله ظاهر شدن یک جمله اضافه است که صرفا منشاء هندسی دارد و مربوط به بعد پنجم است. با استفاده از قضیه ویریال، جرم ویریال برای یک خوشه کهکشانی با تقارن کروی در معادله (۴.۳.۹) بدست آمده است. این معادله نشان می دهد که می توان مسئله جرم ویریال را در مدل DGP به جرمی مربوط کرد که منشاء هندسی دارد. مقدار این جرم در (۴.۴.۱۵) بدست آمده است، و این مقدار برابر با همان مقداری است که برای جرم ویریال انتظار داریم. این تحقیق نشان می دهد که می توانیم برای حل مسئله جرم ویریال از مدل DGP استفاده کنیم. در پایان پراکندگی شعاعی سرعت در خوشه های کهکشانی در (۴.۵.۶) بدست آمده است، که ما این فرمول را برای یک خوشه کهکشانی دلخواه در شکل (۴.۲) رسم کرده ایم. نمودارهای بدست آمده در توافق با مشاهدات است، و لذا می توان نتیجه گرفت که این مدل می تواند برای پراکندگی سرعت شعاعی خوشه های کهکشانی پیش بینی قابل قبولی داشته باشد. نتیجه آخر اینکه می توان مدل DGP را به عنوان جایگزینی برای آن قسمت از ماده تاریک که وظیفه توضیح مسئله جرم ویریال را دارد در نظر گرفت.

مراجع

**Abstract**

In this proposal we study the problem of the virial mass discrepancy in the context of DGP brane gravity. DGP model is a kind of brane-world model such that the corrections to the usual gravity occurred in the large distance limit relative to the distance scale of the model $r_0$ defined as a ratio of the brane Planck scale to the bulk Planck scale. The extra dimension of this model is noncompact. This model is composed with an Einstein-Hilbert action in 5 dimensions plus an induced 4D action guarantees the existence of gravity on the brane. The importance of this model is that it can explain the self-acceleration of our universe without use of any type of dark energy. The virial mass discrepancy is an important problem in cosmology and it can be explained by dark matter. This is due to our various ways in measurement of the mass of the galaxy clusters. One of the ways we can measure the mass of a cluster of galaxies is to measure the galaxy masses and then add them up to obtain the cluster mass. Another way is to use the virial theorem. Almost in all clusters the virial mass is bigger than 20-30 times from the observed mass. In this proposal we show that the virial mass discrepancy can be addressed by the DGP model.

**Key words:** extra dimension, virial mass, DGP brane cosmology.

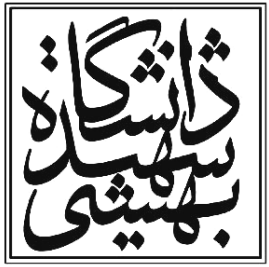

# Shahid Beheshti University
# Department of Physics

MSc. Proposal in

Gravitation and cosmology

# The Virial mass in DGP cosmology

Supervisor

Hamid Reza Sepangi

Advisor

Mehrdad Frarhoudi

Author

Shahab Shahidi

1388